\documentclass[12pt]{article}
\usepackage[dvips]{color}
\usepackage{amsmath}
\usepackage{graphicx}
\usepackage{subfigure}
\usepackage{epsfig}
\usepackage{amsmath}
\usepackage{cite}
\usepackage{color}
\usepackage{subfigure}

\begin{document}
\begin{center}
\Large{\bf Thermodynamic Topology and Photon Spheres in the Hyperscaling violating black holes}\\
 \small \vspace{0.3cm}
 {\bf Jafar Sadeghi $^{\star}$\footnote {Email:~~~pouriya@ipm.ir}}, \quad
 {\bf Mohammad Ali S. Afshar $^{\star}$\footnote {Email:~~~m.a.s.afshar@gmail.com}}\quad
\small \vspace{0.3cm}{\bf Saeed Noori Gashti$^{\dag,\star}$\footnote {Email:~~~saeed.noorigashti@stu.umz.ac.ir}}\\
{\bf Mohammad Reza Alipour $^{\star}$\footnote {Email:~~~mr.alipour@stu.umz.ac.ir}}, \quad\\
\vspace{0.2cm}$^{\star}${Department of Physics, Faculty of Basic
Sciences,\\
University of Mazandaran
P. O. Box 47416-95447, Babolsar, Iran}\\
\vspace{0.2cm}$^{\dag}${School of Physics, Damghan University, P. O. Box 3671641167, Damghan, Iran}
\small \vspace{0.5cm}
\end{center}
\begin{abstract}
It was shown that a standard ring of light can be imagined outside the event horizon for stationary rotating four-dimensional black holes with axial symmetry using the topological method\cite{001,002}. Based on this concept, in this paper, we investigate the topological charge and the conditions of existence of the photon sphere (PS) for a hyperscaling violation (HSV) black hole with various values of the parameters of this model. Then, after carrying out a detailed analysis, we show the conventional topological classes viz $Q=-1$ for the mentioned black hole and $Q=0$ for the naked singularities. \textit{Also, we propose a new topological class for naked singularities (Q=+1) with respect to $z\geq1$. We also determined that $z\geq2$, it either shows a naked singularity form with total topological charge $+1$ or has no solution. Therefore, we have the black hole solution only in $1\leq z<2$}. Then, we will use two different methods, namely the temperature (Duan's topological current $\Phi$-mapping theory) and the generalized Helmholtz free energy method, to study the topological classes of our black hole. By considering the black hole mentioned, we discuss the critical and zero points (topological charges and topological numbers) for different parameters of hyperscaling violating black holes, such as ($z, \overline{\theta}$) and other free parameters, and study their thermodynamic topology. We observe that for a given value of the parameters $1\leq z<2$, $\overline{\theta}$, and other free parameters, there exist two total topological charges $(Q_{total}=-1, 0)$ with the same phase structure for the $T$ method and total topological numbers $(W=+1)$ for the generalized Helmholtz free energy method. Additionally, we summarize the results for each study as photon sphere, temperature, and generalized Helmholtz free energy in some figures and tables. Finally, we compare our findings with other related studies in the literature.\\
Keywords: Hyperscaling violating black hole, Thermodynamic topological class, The photon sphere
\end{abstract}
PACS numbers: 04.70.Dy, 04.60.-m, 05.70.Ce
\tableofcontents

\section{Introduction}
As we know the investigation of black holes with the existence of horizons from a thermodynamics point of view can offer a beautiful insight for researchers to understand the structure and nature of gravitational degrees of freedom\cite{1,2}. Recently,  the thermodynamics of black holes from the topology point of view is discussed by several researchers in various gravity structures.  Therefore, we investigate the topological classification of Hyperscaling violating black holes from temperature and Helmholtz free energy methods.\cite{3,4,5}
Hyperscaling violating black holes have asymptotic geometries that do not obey the usual conformal field theory, because of the presence of both a dynamical critical exponent z and a “HSV” exponent $\theta$. This metric is not scale invariant,  the $\theta$ parameter breaks the conformal symmetry. Such parameter play an important role in AdS/QCD at the Quark Gluon Plasma phase in QCD\cite{6,7,8,9,10}. By adding some different matter fields in the corresponding system with the above-mentioned background, one can provide some gravitational duality for certain condensed matter systems. We also note here, the non-trivial relation between energy density and temperature leads us to account for some non-relativistic symmetries such as the Schr$\ddot{o}$dinger or the Lifshitz symmetries.
The study of hyperscaling violating black holes with different horizon topologies as spherical and hyperbolic can reveal new insights into the dual QFTs.\cite{7,8,9,10} In that case, on the gravitational side, we obtain some critical points and phase transitions which correspond to the gauge theory side(QCD) as confinement or non-confinement phases. Also, we know that the horizon topology affects the properties and behavior of hyperscaling violating black holes, such as their thermodynamics, stability, and phase transition. Therefore, thermodynamic topology is a way of studying the phase transitions and critical phenomena of black holes by using the topological properties of their thermodynamic state space.\cite{3,4,5,6,7,8,9,10,a,b,c,d}.\\\\
The photon sphere is a region of space where light is trapped in closed orbits around a black hole or other compact object. Any photon that can reach us from a black hole or its surroundings will undoubtedly, bring with it unique information about the history, structure, and behavior of the black hole. With this interpretation, there is no doubt that the study of light rings and photon spheres can play an essential role in our understanding of the history and structure of black holes\cite{001,002,10004}. For example, during black hole mergers, radiative gravitational waves are studied and understood based on quasi-normal modes that are related to light rings and photon spheres in the eikonal limit, or, in the formation of the shadow of a black hole, the role of light rings and photon spheres are important because the formation of the shadow depends on their existence\cite{001,002,10004}. Astronomical observation and the study of relativistic effects on light sources placed behind a very massive structure are strongly related to them. The study of light rings and photon spheres, or in other words, the study of null geodesics for photons around a black hole can be done based on different methods such as the use of killing vector or the killing equation\cite{001,002,10004}. But researchers based on Brouwer's mapping method established a new approach for the topological study of light rings, which they initially proposed for ultra-compact objects, but soon extended it to stationary 4-dimensional black holes with axial symmetry\cite{001,10002}. They showed that at least one standard light ring can always be found outside the horizon of the black hole by calculating the winding numbers of the vector field created by the effective potential. An ultra-compact object without a horizon will always show an even number of non-degenerate light rings\cite{001,002,10004}. Following the introduction of the topological method for rotating black holes this method expanded for non-rotating black holes with spherical symmetry, which was still based on Duan's topological mapping method (briefly introduced in the next sections), and we will also use the same method for our study\cite{002,11,12,10003}.\\\\In summary the topological current is nonzero only at the points where the vector field that determines the location of the photon sphere vanishes. The topological charge is an integer that measures how many times the vector field winds around the origin. So, can be shown that each photon sphere can be assigned a topological charge. The total topological charge -1 calculated in this case indicates that there must be at least one unstable photon sphere in the outer region of the black hole horizon. They also apply to dyonic black holes, which are black holes with both electric and magnetic charges. Moreover, for a naked singularity, which is a hypothetical object that has no horizon and exposes its singularity to the outside world, the topological charge is zero. This indicates that black holes and naked singularities are in different topological classes. They conclude that this novel topological argument could provide an insightful idea on the study of black hole photon spheres or light rings, and further cast new light on black hole astronomical phenomena\cite{002}.\\\\
Recently, the analysis of the black holes thermodynamics from a topology point of view has been carried out by two methods: Temperature (T) and generalized Helmholtz free energy $(F).$ The $T$ and $F$ method is a way of studying the thermodynamics and phase transitions of black holes for analyzing the topological properties of their thermodynamic state space with respect to Duan's $\phi$-mapping theory. With these methods, the topological charges can be used to classify the black holes into different classes, which have different properties and implications for the phase transition of the black hole.
The method can be applied to various types of black holes with different metric functions which are given by Ref.s \cite{14,15,16,17,18,19,600,601,602,6020,20,21,22,23,24,25,255,26,266,2666,27,277,2777,27777,277777}.
In this work, we are going to study photon sphere structures and thermodynamic topological properties of HSV black holes. Therefore, in the following sections, we choose the HSV black hole and use Duan's $\phi$-mapping theory. In such cases, we determine its classification and discuss its topological features.\\\\
So, based on the above explanations, we arrange the article as follows:\\
In section 2, we give a review of the corresponding black hole and study its thermodynamics. In section 3, we study the topological perspective of black hole thermodynamics and introduce some concepts: photon sphere, temperature-entropy, and the generalized Helmholtz free energy method. Then, by using Duan's $\phi$-mapping theory, we apply these methods to our black hole and investigate its topological thermodynamics properties in sections 4, 5, and 6. Finally, we summarize our results in section 7.\\\\
\section{Hyperscaling violation black holes}
As we know, Hyperscaling violating black holes are solutions of Einstein-Maxwell-Dilaton theories with additional fields that break the usual scaling symmetries of the AdS space. They have different horizon topologies such as spherical, planar, or hyperbolic with charge and rotating\cite{3,4,5,6}.
The motivation for studying hyperscaling violating black holes is twofold. On one hand, they provide a rich class of gravitational solutions that can test various aspects of holography, such as thermodynamics, phase transitions, entanglement entropy, transport coefficients, and holographic renormalization. On the other hand, they can be used to explore novel phenomena in condensed matter physics that arise from quantum criticality and non-relativistic symmetries, such as Schrödinger or Lifshitz invariance. The investigation of thermodynamics and phase transitions of the hyperscaling violating black holes can be complicated by the presence of non-trivial scalar fields and multiple charges. In that case, in order to have meaningful results,  one may need to use holographic renormalization techniques or extended thermodynamics frameworks\cite{5,6,7}. Also here we note that the exploring holographic duals of the corresponding black holes may not be well-defined or well-understood in some cases, because some hyperscaling violating black holes may have negative specific heat or negative entropy. So, in that case, it shows a breakdown of the holographic correspondence. Therefore, it needs a UV completion which is shown by Ref.s.\cite{3,4,5,6,7,8,9,10}.\\\\
The HSV parameter $\overline{\theta}$ is a measure of how the system deviates from the standard scaling behavior expected from relativistic conformal field theory. In the simplest scaling framework, HSV can be characterized by a single nonzero exponent $\overline{\theta}$, so that in a spatially isotropic state in $d$ spatial dimensions, the specific heat scales with temperature as $T^{(d-\overline{\theta})/z}$, and the optical conductivity scales with frequency as $\omega^{(d-\overline{\theta}-2)/z}$ for $\omega\gg T$, where $z$ is the dynamic critical exponent. The HSV parameter $\mathcal{\overline{\theta}}$ can also be related to the fractal dimension of the system, as $d_f=d-\mathcal{\overline{\theta}}$, where $d_f$ is the number of degrees of freedom per unit volume\cite{6,7,8,9,10}.\\\\
HSV can arise from various physical mechanisms, such as quantum fluctuations, disorder, interactions, or topology. For example, in quantum critical systems near a quantum phase transition, HSV can be induced by quantum fluctuations that modify the effective dimensionality of the system. In disordered systems, HSV can be caused by rare regions or Griffiths effects that lead to a broad distribution of energy scales. In strongly interacting systems, HSV can be a consequence of emergent gauge fields or fermionic excitations that affect the low-energy physics. In topological systems, HSV can be related to the presence of topological defects or fractionalized quasiparticles that change the scaling properties of the system\cite{7,8,9,10}.
HSV is an important concept in condensed matter physics, as it can reveal novel phenomena and exotic phases of matter that are beyond the conventional paradigms. HSV can also have implications for holography, as it can provide new ways to construct gravitational duals of non-relativistic systems that exhibit anomalous scaling behavior\cite{6,7,8,9,10}.\\\\
Observing HSV experimentally is a challenging task, because it requires probing quantum field theories (QFTs) that are strongly coupled and have non-relativistic scaling symmetries. However, there are some possible candidates for such QFTs in condensed matter physics and quantum information theory, where HSV can manifest itself in various physical quantities, such as specific heat, entropy, conductivity and entanglement entropy\cite{7,8,9}.
For example, one class of QFTs that can exhibit HSV are the so-called strange metals, which are materials that show anomalous transport and thermodynamic properties at low temperatures. Some examples of strange metals are high-temperature superconductors, heavy fermion compounds, and graphene. These materials do not obey the conventional Fermi liquid theory, which predicts that the resistivity should be proportional to the temperature squared at low temperatures. Instead, they show a linear dependence of resistivity on temperature, which suggests that they have a non-Fermi liquid behavior with a non-trivial scaling symmetry. Another class of QFTs that can exhibit HSV are the quantum critical points (QCPs), which are points in the phase diagram of a system where a continuous phase transition occurs at zero temperature\cite{7,8,9}. At a QCP, the system is tuned by an external parameter, such as pressure or magnetic field, to a critical value where the correlation length and the relaxation time diverge. This means that the system becomes scale invariant and strongly coupled at low energies\cite{6,7,8,9,10}. QCPs can also show non-relativistic scaling symmetries with HSV, depending on the dimensionality and the nature of the order parameter of the system\cite{6,7,8,9,10}.\\\\
One way to observe HSV experimentally is to measure the scaling behavior of physical quantities near a QCP or in a strange metal. Another way to observe HSV experimentally is to measure the entanglement entropy of a subsystem of a quantum system. Entanglement entropy is a measure of quantum correlations between two parts of a system that are separated by a boundary. According to holography, the entanglement entropy of a QFT can be computed by using the area of a minimal surface in a dual gravity theory. If the gravity theory has hyperscaling violating asymptotics, then the entanglement entropy will also show HSV. For example, for a strip-shaped subsystem of width $l$ in a $d$-dimensional QFT with HSV, one expects that the entanglement entropy per unit area $S_A$ scales as $S_A \sim l^{(d-\overline{\theta})/z}$. Therefore, by measuring $S_A$ for different widths and fitting it to a power law, one can also extract $\overline{\theta}$ and $z$ from the exponent\cite{3,4,5,6,7,8,9,10, a,b,c,d}.\\\\
As mentioned before the hyperscaling violating black holes are solutions of generalized Einstein-Maxwell-Dilaton theory with an additional vector field, which is given by,
\begin{equation}\label{1}
ds^2 = \frac{r}{r_{F}}^{-\frac{2\overline{\theta}}{d}}\bigg(-(\frac{r}{\ell})^{2z}f(r)dt^2 + \frac{dr^2}{r^2f(r)} + r^2d\Omega_{d,k}^2\bigg),
\end{equation}
where $d\Omega_{d,k}^2$ is the metric of a $(d+1)$-dimensional space with constant curvature $k=0,\pm 1$, and $f(r)$ is a function that determines the location and nature of the horizon. The parameter $z$ is the Lifshitz exponent, which controls the anisotropic scaling between time and space, and $\overline{\theta}$ is the HSV parameter, which controls the deviation from the area law of entropy. Two corresponding parameter as $z$ and $\overline{\theta}$ are dynamical and HSV parameters respectively.  The metric function $f(r)$ is given by,
\begin{equation}\label{2}
f(r)=1+\frac{k\ell^{2}}{r^2}\frac{(d-1)^2}{(z+d-\overline{\theta}-2)^2}-\frac{M}{r^{z+d-\overline{\theta}}}+
\frac{Q^2}{r^{2(z+d-\overline{\theta}-1)}}.
\end{equation}
By using the $T=\frac{r_{H}^{z+1}}{4\pi}|f'(r_{H})|$, one can obtain the Hawking temperature as,
\begin{equation}\label{3}
T=\frac{r_H^{z}}{4\pi\ell^{z+1}}\left((z+d-\overline{\theta})+k\frac{(d-1)^2}{(z+d-\overline{\theta}-2)}\frac{\ell^2}{r_{H}^{2}}
-\frac{(z+d-\overline{\theta}-2)Q^2}{r_{H}^{2(z+d-\overline{\theta}-1)}}\right).
\end{equation}
The entropy and the pressure are given by,
\begin{equation}\label{4}
S=\frac{\omega_{k,d}}{4G}r_{h}^{d-\overline{\theta}}r_{F}^{\overline{\theta}},
\end{equation}
\begin{equation}\label{5}
P=\frac{1}{16\pi G}\frac{(d-\overline{\theta}+z-1)(d-\overline{\theta}+z)}{\ell^{2}r_{F}^{2\overline{\theta}/d}e^{\lambda_{0}\phi_{0}}}.
\end{equation}
Also, the  total electric charge and associated potential of above  mentioned black hole are expressed as,
\begin{equation}\label{6}
Q=\frac{\omega_{k,d}}{16\pi G}Z_{0}\rho_{3}\ell^{z-1}r_{F}^{\overline{\theta}-2\overline{\theta}/d},
\end{equation}
\begin{equation}\label{7}
\Phi=\frac{qr_{F}^{d/\overline{\theta}}}{\ell^{2}e^{\lambda_{3}\phi_{0}/2}r^{d-\overline{\theta}+z-2}}\sqrt{\frac{2(d-\overline{\theta})}{Z_{0}(d-\overline{\theta}+z-2)}},
\end{equation}
where $Z_{0}$ is a positive parameter, $\lambda_{0}=2\overline{\theta}/\gamma d$, $\gamma\equiv\sqrt{2(d-\overline{\theta})(z-1-\overline{\theta}/d)}$, and $\lambda_{3}=\frac{\gamma}{d-\overline{\theta}}$. The charge parameter $q$ is related to the black hole charge $Q$ and $\ell$, which can be regarded as a generalization of the AdS radius. The $r_{F}$ is correspond to  the UV physics. Now we are ready to study the thermodynamics of black hole from a topological perspective. In that case we identify the topological class  and analyze its photon sphere of black hole. In the following we do some calculation  and compare our results with other works in the literature.

\section{Thermodynamics from topology perspective}
Topology of black hole thermodynamics is a new approach to studying the critical points and phase transitions of black holes by using Duan's topological current $\phi$-mapping theory. This theory assigns a topological charge to each critical or zero point, which can be used to classify different types of black holes and their thermodynamic behaviors. For example, the charged $AdS$ black hole and the Born-Infeld $AdS$ black hole have different topological charges. Here, we note that they are in different topological classes from thermodynamic point of view. Duan's topological current $\phi$-mapping theory is a mathematical tool to study the topological properties of a vector field in a manifold. It defines a topological current as the exterior derivative of a topological potential, which is a function that maps the vector field to a unit sphere. The topological charge is then the integral of the topological current over a closed surface. The topological charge is an integer that measures the winding number of the vector field around the zero point\cite{14,15,16,17,18,19,600,601,602,20,21,22,23,24,25,26,27}.\\\\
The applications of topology in black hole thermodynamics are to classify and characterize the critical points and phase transitions of black holes by using their topological charges. A critical point is a point where the first and second derivatives of a thermodynamic potential vanish. A phase transition is a process where a thermodynamic system changes from one state to another with different properties. By using Duan's topological current, one can assign a topological charge to each critical point and use it to distinguish different types of black holes and their thermodynamic behaviors. The motivation for the creation of such a method is to explore the topological properties of various physical systems, such as black holes, photon spheres, phase transitions, etc. The reason for its creation is to provide a novel and powerful way to classify and characterize different types of solutions and phenomena. Now we are going to write the topological current equation, which is given by\cite{14,15,16,17,18,19,600,601,602,20,21,22,23,24,25,26,27},
\begin{equation}\label{8}
J^\mu=\frac{1}{2\pi}\epsilon^{\mu\nu\rho}\varepsilon_{ab}\partial_\nu n^a\partial_\rho n^b.
\end{equation}
This equation defines the topological current as the exterior derivative of the topological potential, which is a function that maps the vector field to a unit sphere. The vector field $\phi$ is defined by the gradient of a scalar function $\Phi$, which is a function of some physical variables, such as the entropy $S$, the temperature $T$ and the spatial coordinates $x^i$. The vector field $\phi$ can be decomposed into two components, $\phi^S$ and $\phi^\theta$ as,
\begin{equation}\label{9}
\phi=(\phi^{S}, \phi^{\theta}),
\end{equation}
where
\begin{equation}\label{10}
\phi^{S}=(\partial_{S}\Phi)_{\theta, x^{i}},\hspace{1cm}\phi^{\theta}=(\partial_{\theta}\Phi)_{S, x^{i}}.
\end{equation}
The zero points of $\phi$ are the points where the vector field vanishes, i.e., $\phi^a(x^i)=0$, where $a=S,\theta$. These points are important because they are related to some physical quantities of interest, such as the photon sphere radius or the critical temperature. The zero points of $\phi$ are always located in $\theta=\pi/2$, which is a property of the scalar function $\Phi$. The main feature of Duan's $\phi$-mapping theory is the existence of a topological current $J^\mu$. So, we consider a superpotential such as \cite{002,14},
$$V^{\mu\nu}=\frac{1}{2\pi} \epsilon^{\mu\nu\rho} \epsilon_{ab}n^a\partial_\rho n^b,\hspace{0.5cm}\mu,\nu,\rho=0,1,2$$,
where $x^{\mu}=(t,r,\theta).$ It is clear that the superpotential is an antisymmetric tensor,
$$V^{\mu\nu}=-V^{\mu\nu}.$$
We can introduce a topological current by using the superpotential. So, we have,
$$j^{\mu}=\partial_{\nu}V^{\mu\nu}=\frac{1}{2\pi}\epsilon^{\mu\nu\rho}\epsilon_{ab}\partial_{\nu}n^a \partial_\rho n^b.$$
We find that this topological current satisfies,
$$\partial_\mu j^\mu=0.$$
The topological current $J^\mu$ has a non-zero contribution only from the zero points of $\phi$, and it can be used to define a topological charge $Q_t$, which is the integral of $J^\mu$ over a closed surface. The topological charge $Q_t$ is an integer that measures the winding number of the vector field $\phi$ around the zero point. The topological charge can also be expressed as a sum of products of two factors, $\beta_i$ and $\eta_i$, for each zero point $z_i$. The factor $\beta_i$ is called the Hopf index, which measures the number of loops of the vector field around the zero point. The factor $\eta_i$ is called the Brouwer degree, which measures the orientation of the vector field around the zero point. The product $\omega_i=\beta_i\eta_i$ is called the winding number, which measures how many times the vector field wraps around the zero point. Also, the charge can be considered as\cite{14,15,16,17,18,19,600,601,602,20,21,22,23,24,25,26,27},
\begin{equation}\label{11}
Q_{t}=\int_{\Sigma}\Sigma_{i=1}^{N}\beta_{i}\eta_{i}\delta^{2}(\overrightarrow{x}-\overrightarrow{z}_{i})d^{2}x=\Sigma_{i=1}^{N}\beta_{i}\eta_{i}=\Sigma_{i=1}^{N}\omega_{i}.
\end{equation}
This equation defines the topological charge as the integral of the topological current over a closed surface. The topological charge is an integer that measures the winding number of the vector field around the zero point.
A winding number is a mathematical concept that measures how many times a curve wraps around a point in a plane. It is an integer that can be positive, negative or zero, depending on the direction and number of turns of the curve. For example, a curve that makes two counterclockwise turns around a point has a winding number of two, while a curve that makes one clockwise turn around a point has a winding number of minus one. A curve that does not wrap around a point at all has a winding number of zero. Let $C_i$ be a closed curve that is smooth, positive oriented, and encloses only the i-th zero point of $\phi$, while the other zero points are outside of it. So the winding number can be calculated by the following formula,
\begin{equation}\label{12}
\omega_i=\frac{1}{2\pi}\int_{C_{i}}d\Omega,
\end{equation}
then the total charge will be,
\begin{equation}\label{13}
Q_{t}=\sum_{i}\omega_{i}.
\end{equation}
Another important point in this regard is that Duan's $\phi$-mapping theory can be generalized to higher dimensions and different physical systems, such as vortices, solitons, cosmic strings, etc.
As we mentioned before, one method that uses topology to study black hole thermodynamics is called Duan's topological current$\phi$ -mapping theory. This method assigns a topological charge to each critical point in the phase diagram, which measures how many times the system winds around the critical point. The topological charge can be positive or negative, depending on the direction of the winding. The method also defines a topological number for the whole phase diagram, which is the sum of all the topological charges. The topological number can tell us how many different classes of black holes there are in the phase diagram, and how they are related to each other.
Another method that uses topology to study black hole thermodynamics is called thermodynamic topological defects. This method treats black holes as defects in the generalized off-shell free energy landscape, which is a surface that represent the free energy of the system as a function of its parameters. The free energy is a measure of how much useful work can be extracted from the system. The method identifies different types of defects, such as winding numbers, vortices and monopoles, which correspond to different types of critical points and phase transitions in the phase diagram. The method also relates the local topological properties of the spacetime around the black hole to the global topological properties of the free energy landscape.
These methods are examples of how topology can provide new insights and perspectives on black hole thermodynamics, and reveal some hidden symmetries and patterns in their phase diagrams. They also show how black hole thermodynamics can be connected to other area of physics, such as condensed matter physics and quantum field theory, where similar methods have been applied.
\section{The investigation of photon sphere}
In this section we are going to study  photon sphere and light ring on the  hyperscaling violating black holes.  We will first give some definitions and then use the relevant equations for the corresponding black hole.
Usually, the black hole horizon's radius $r_h$ is the biggest solution of $g(r_h)=0$ or $f(r_h)=0$. By solving the null geodesics, one can also get the radial motion on the equatorial plane\cite{001},
\begin{equation}\label{14}
\begin{split}
\dot{r}^{2}+V_{eff}=0.
\end{split}
\end{equation}
So, we will have,
\begin{equation}\label{15}
\begin{split}
V_{eff}=g(r)\bigg(\frac{L^2}{h(r)}-\frac{E^2}{f(r)}\bigg),
\end{split}
\end{equation}
where $E$ and $L$ are the photon's energy and angular momentum, which are connected to the Killing vector fields $\partial_t$ and $\partial_\phi$, respectively. This solution has spherical symmetry, so there is a photon sphere at $r_{ps}$, which is given by,
\begin{equation}\label{16}
\begin{split}
V_{eff}=0,\hspace{1cm}\partial_{r}V_{eff}=0.
\end{split}
\end{equation}
So, with respect the above equation, one can calculate,
\begin{equation}\label{17}
\begin{split}
\bigg(\frac{f(r)}{h(r)}\bigg)^{'}_{r=r_{ps}}=0.
\end{split}
\end{equation}
The prime means the derivative with respect to $r$. Also, $\partial_{r,r}V_{eff}(r_{ps}) <(>)0$ means the photon sphere is unstable (stable). By taking the derivative, the above equation  will be as,
\begin{equation}\label{18}
\begin{split}
f(r)h(r)'-f(r)'h(r)=0.
\end{split}
\end{equation}
We should note that the first term disappears at the horizon where $f(r_h)=0$, but the second term is usually nonzero. So $r_{ps}$ and $r_h$ are not the same. But when a black hole has more than one horizon, there is the extremal black hole case, where two horizons are coincide. For the extremal black hole case, we have $f(r_h) = 0$ and $f(r_h)'= 0$.Then, in that case the  above condition holds. So the photon sphere and the extremal black hole horizon are the same in case of   extremal. To study the topology of the photon sphere, we use the regular potential function everywhere,
\begin{equation}\label{19}
\begin{split}
H(r, \theta)=\sqrt{\frac{-g_{tt}}{g_{\varphi\varphi}}}=\frac{1}{\sin\theta}\bigg(\frac{f(r)}{h(r)}\bigg)^{1/2}
\end{split}
\end{equation}
Clearly, the photon sphere radius is at the root of $\partial_{r}H=0$. So, we can use a vector field $\phi=(\phi^r,\phi^\theta)$,
\begin{equation}\label{20}
\begin{split}
&\phi^r=\frac{\partial_rH}{\sqrt{g_{rr}}}=\sqrt{g(r)}\partial_{r}H\\
&\phi^\theta=\frac{\partial_\theta H}{\sqrt{g_{\theta\theta}}}=\frac{\partial_\theta H}{\sqrt{h(r)}}.
\end{split}
\end{equation}
The circular photon orbit for a spherically symmetric black hole is a photon sphere, which does not depend on the coordinate $\theta$. \\\\Here we aim to investigate the topological property of the circular photon orbit, so we keep $\theta$ in our discussions. We can also write the vector as $\phi=||\phi||e^{i\Theta}$, where, $||\phi||=\sqrt{\phi^a\phi^a}$. However, at $\phi=0$, we have a PS. This implies that $\phi$ in $\phi=||\phi||e^{i\Theta}$ is not well-defined for the PS, so we consider the vector as $\phi = \phi^r + i\phi^\theta$. So the normalized vectors are defined as $n^a=\frac{\phi^a}{||\phi||}$, where $a=1,2$, $(\phi^1=\phi^r)$ and $(\phi^2=\phi^\theta)$.\\\\ Now, according to the above explanations, we apply the mentioned items to our black hole. Therefore, by considering equations (19) and (20), we draw diagrams and  analyze the results for different values of free parameters.\\\\
\textbf{Analysis pattern for photon sphere figures:}\\\\
We show the behavior of vector $n$ in Figs (1-7) at the plane of $(r, \theta)$.
Considering that the general structure of the description of the shapes in the Photon Sphere section follows the same structural pattern, also, with respect to the numbers of the shapes, to avoid confusion and better understanding, we will express the pattern of interpretation and classify it for a model. The rest will follow the same pattern. In general, when the contour includes a winding:\\
1. If the field lines diverge around the zero point (like positive electric charges), its topological charge is +1 (purple loop in Figure 1c). According to the previous definitions, this state will correspond to the naked singularity\cite{002}.\\
2.  If the field lines converge around the zero point (like negative electric charges), its topological charge is -1 (blue loop in Figure 1b). This state will correspond to a black hole with unstable PS \cite{002}.\\
3. If the contour does not include a winding, its charge will be zero, like (the purple loop in Figure 1b).
Finally, the total charges for each shape are the algebraic sum of the charges, which can be displayed by drawing a contour that includes all the zero points. For example, the total charge is zero for Figure 1c, -1 for Figure 1b, and +1 for Figures 7a and 7b. Based on this pattern, total charges for each figure have been calculated and included in the tables.

\subsection{z=1 $\&$ $\overline{\theta}=0$}

 \begin{figure}[h!]
 \begin{center}
 \subfigure[]{
 \includegraphics[height=4.5cm,width=5.5cm]{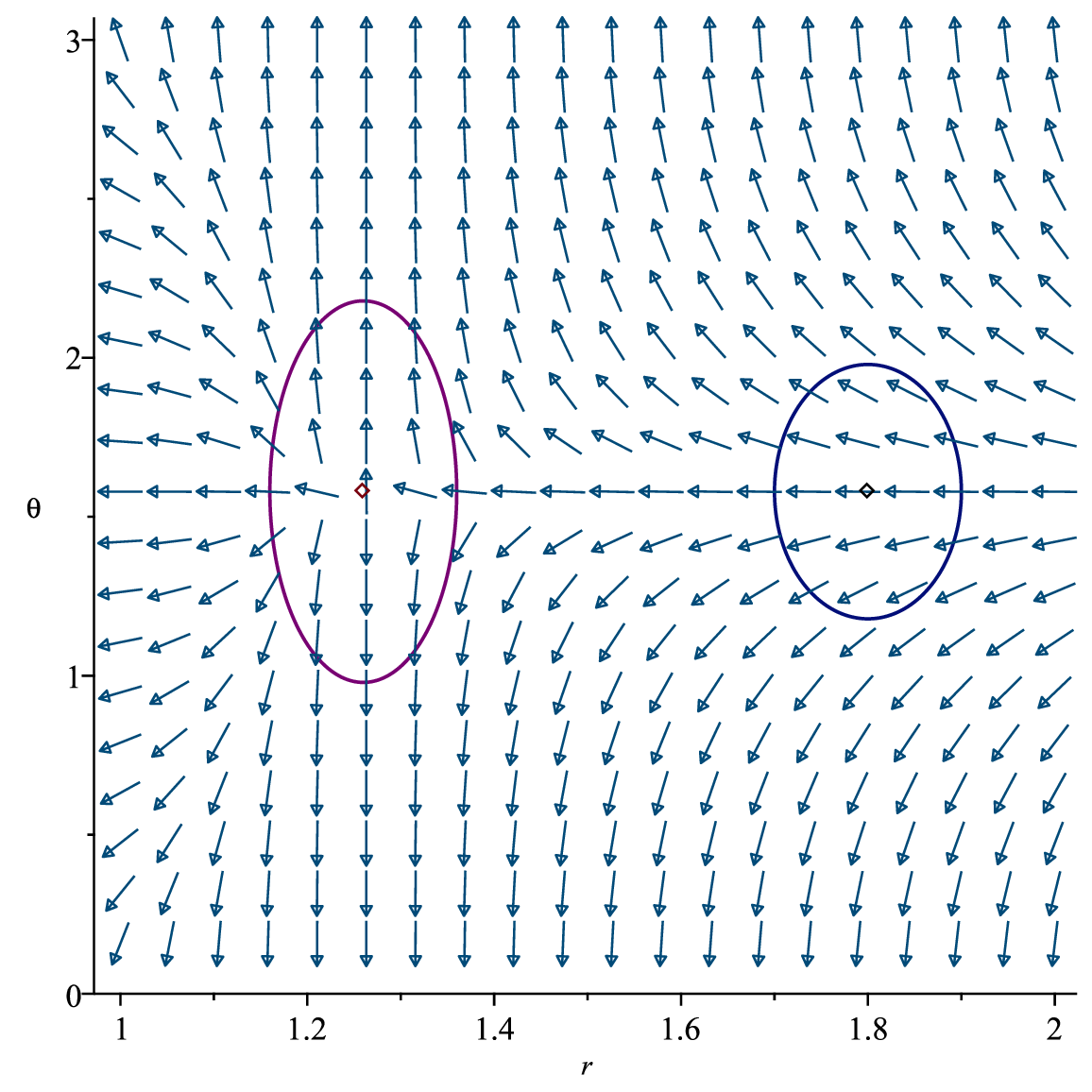}
 \label{1a}}
 \subfigure[]{
 \includegraphics[height=4.5cm,width=5.5cm]{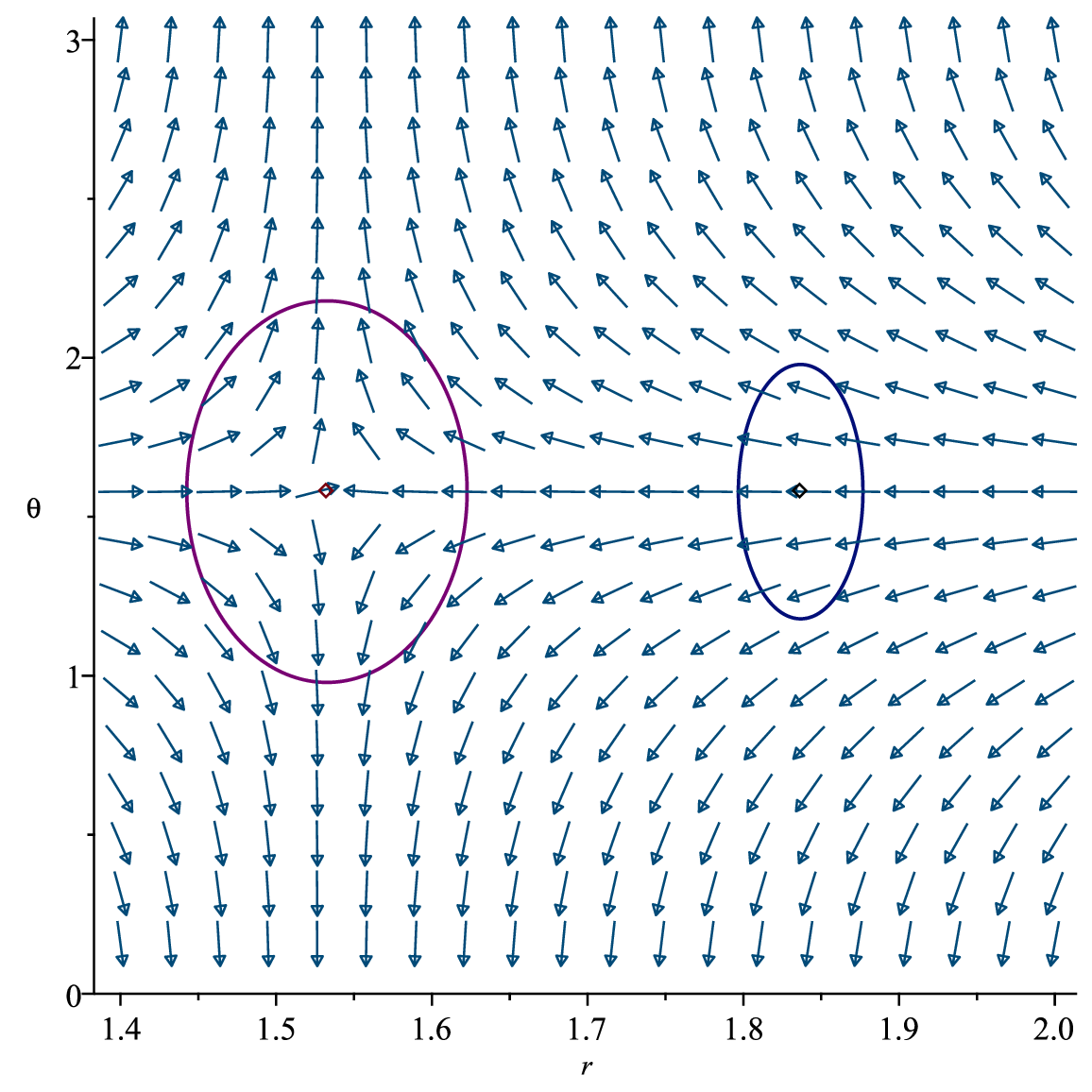}
 \label{1b}}
 \subfigure[]{
 \includegraphics[height=4.5cm,width=5.5cm]{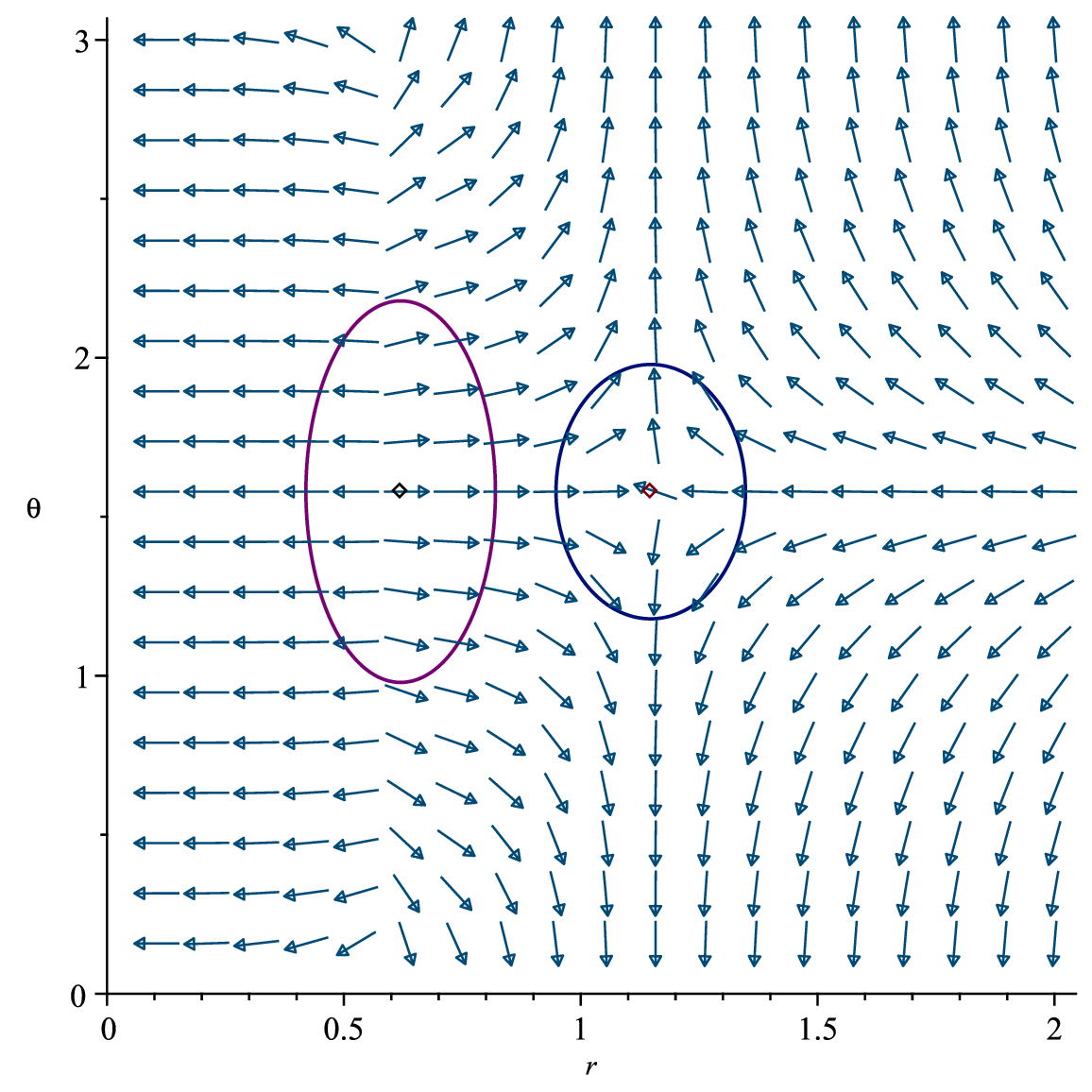}
 \label{1c}}
  \caption{\small{Figure 1 shows the vector field $n$ on a part of the $(r-\theta)$ plane. The blue arrows indicate the direction of $n$. The PSs is located at $(r,\theta)=(1.532618865,1.57)$ for fig 1b with respect to $(q=1, z=1, M=1.6, \overline{\theta}=0)$, $(q=0.6, z=1, M=1.6, \overline{\theta}=0)$ and $(r,\theta)=(0.6198,1.57)$ and $(r,\theta)=(1.147730106,1.57)$ for fig 1c with respect to $(q=0.3, z=1, M=0.7, \overline{\theta}=0)$. The contours $C_1$ (purple loop) and $C_2$ (blue loop) are two closed curves, where for 1b $C_1$ encircles the PS but $C_2$ does not and for fig 1c $C_1$ and $C_{2}$ encircles the PSs.}}
 \label{1}
 \end{center}
 \end{figure}

 \begin{figure}[h!]
 \begin{center}
 \subfigure[]{
 \includegraphics[height=4.5cm,width=5.5cm]{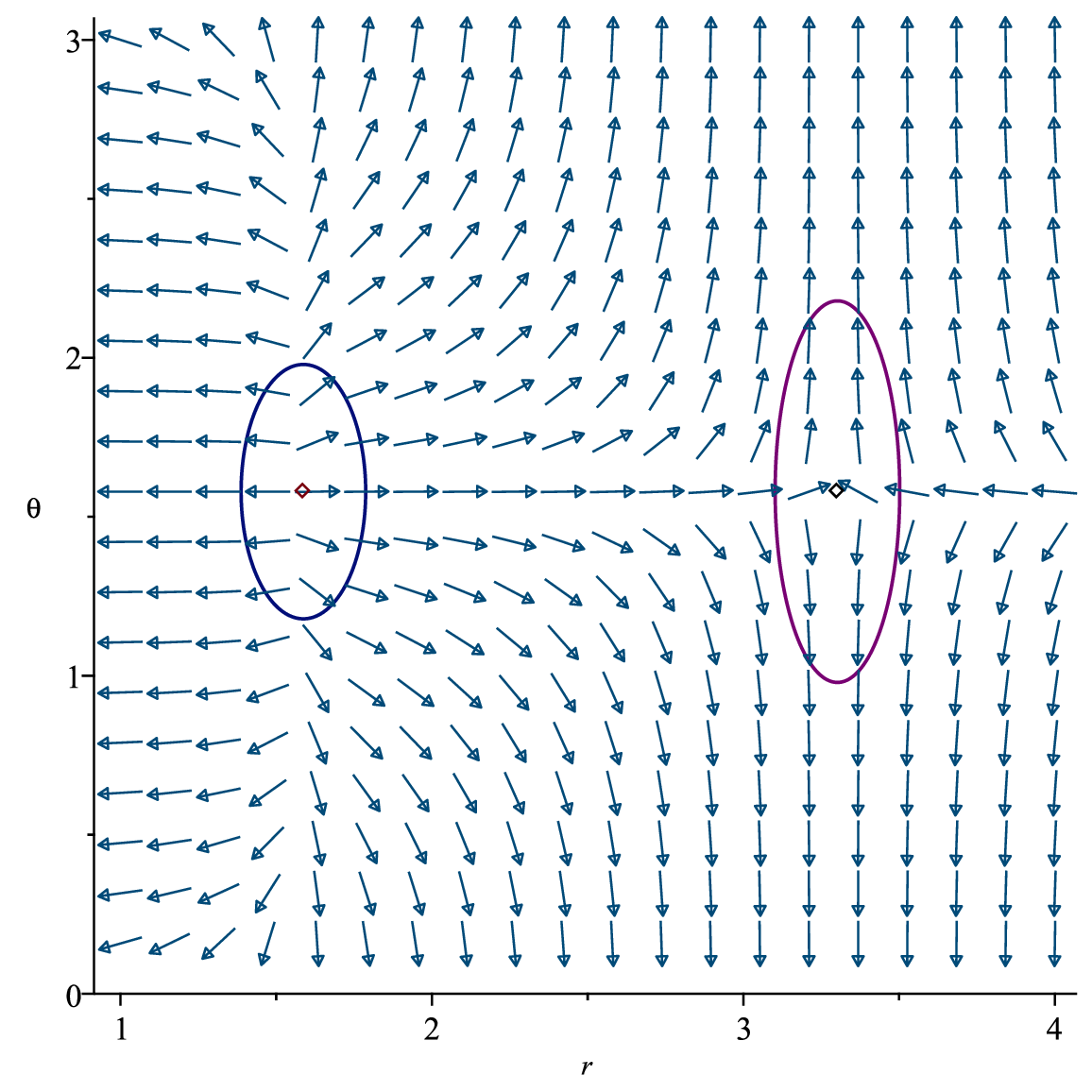}
 \label{2a}}
 \subfigure[]{
 \includegraphics[height=4.5cm,width=5.5cm]{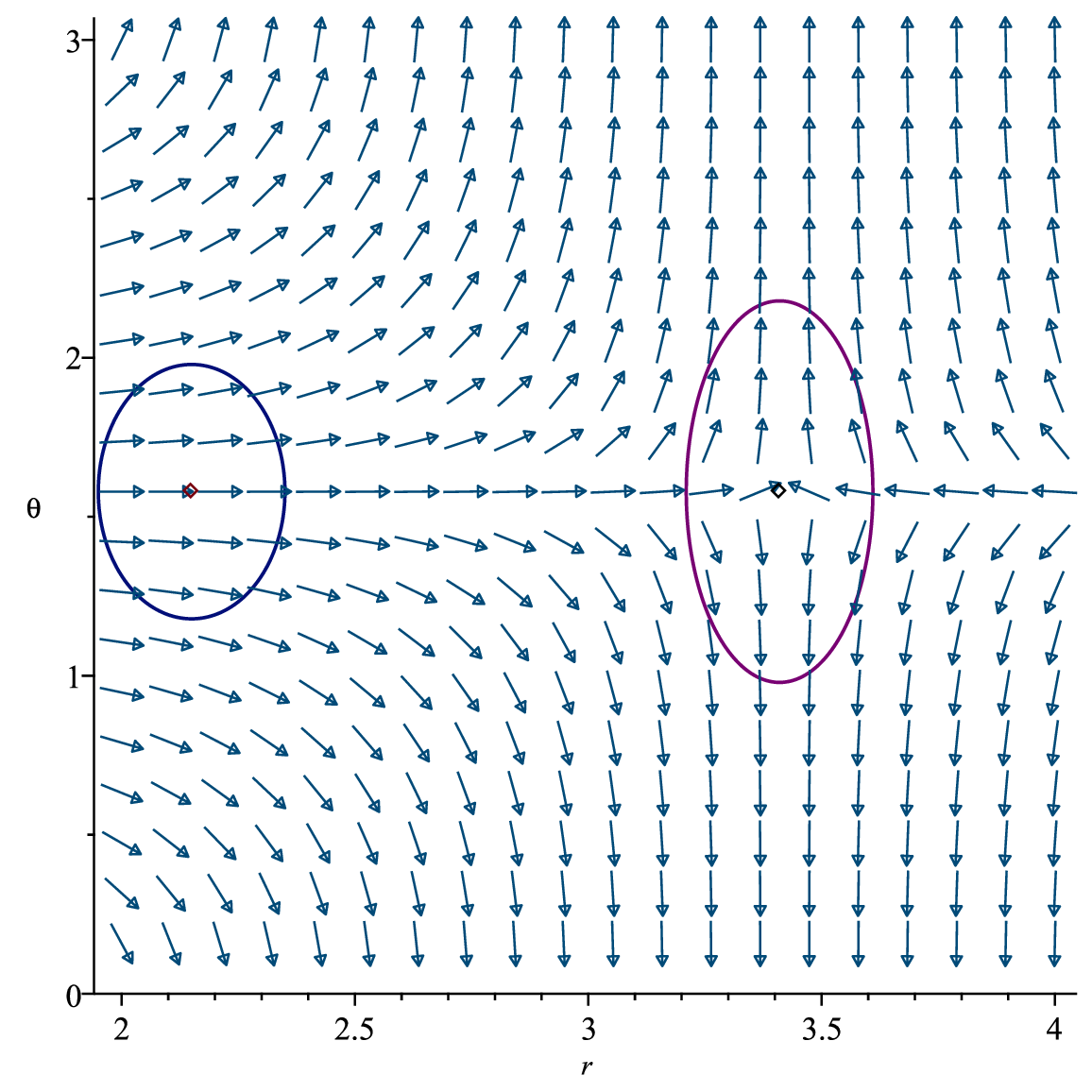}
 \label{2b}}
  \caption{\small{The PSs is located at $(r,\theta)=(1.5874,1.57)$ and $(r,\theta)=(3.3019,1.57)$ for fig 2a with respect to $(q=6, z=1, M=16, \overline{\theta}=0)$ and $(r,\theta)=(3.41,1.57)$  considering $(q=0.2, z=1, M=16, \overline{\theta}=0)$  for fig 2b. The contours $C_2$ (purple loop) and $C_1$ (blue loop) are two closed curves, where for fig 2b $C_2$ encircles the PS but $C_1$ does not and for fig 2a $C_1$ and $C_{2}$ encircles the PSs}}
 \label{2}
 \end{center}
 \end{figure}
\begin{center}
\begin{table}
  \centering
\begin{tabular}{|p{3cm}|p{6cm}|p{3cm}|p{3cm}|}
  \hline
  % after \\: \hline or \cline{col1-col2} \cline{col3-col4} ...
  \centering{Hyperscaling violating black hole}  & \centering{Conditions} &\centering{Topological Charge}& Total Topological Charge\\[3mm]
   \hline
  \centering{Case 1:} & $q=1, z=1, M=1.6, \overline{\theta}=0$ & \centering{0} & 0\\[3mm]
   \hline
  \centering{Case 2:} & $q=0.6, z=1, M=1.6, \overline{\theta}=0$ & \centering{$-1$} & $-1$ \\[3mm]
   \hline
   \centering{Case 3:} & $q=0.3, z=1, M=0.7, \overline{\theta}=0$ & \centering{+1, -1} & $0$ \\[3mm]
   \hline
   \centering{Case 4:} & $q=6, z=1, M=16, \overline{\theta}=0$ & \centering{+1, -1} & $0$ \\[3mm]
   \hline
   \centering{Case 5:} & $q=0.2, z=1, M=16, \overline{\theta}=0$ & \centering{-1} & $-1$ \\[3mm]
   \hline
\end{tabular}
\caption{Summary of the results for PSs (z=1 $\&$ $\overline{\theta}=0$)}\label{1}
\end{table}
 \end{center}

\subsection{z=1 $\&$ $\overline{\theta}<0$}
 \begin{figure}[h!]
 \begin{center}
 \subfigure[]{
 \includegraphics[height=4cm,width=5.5cm]{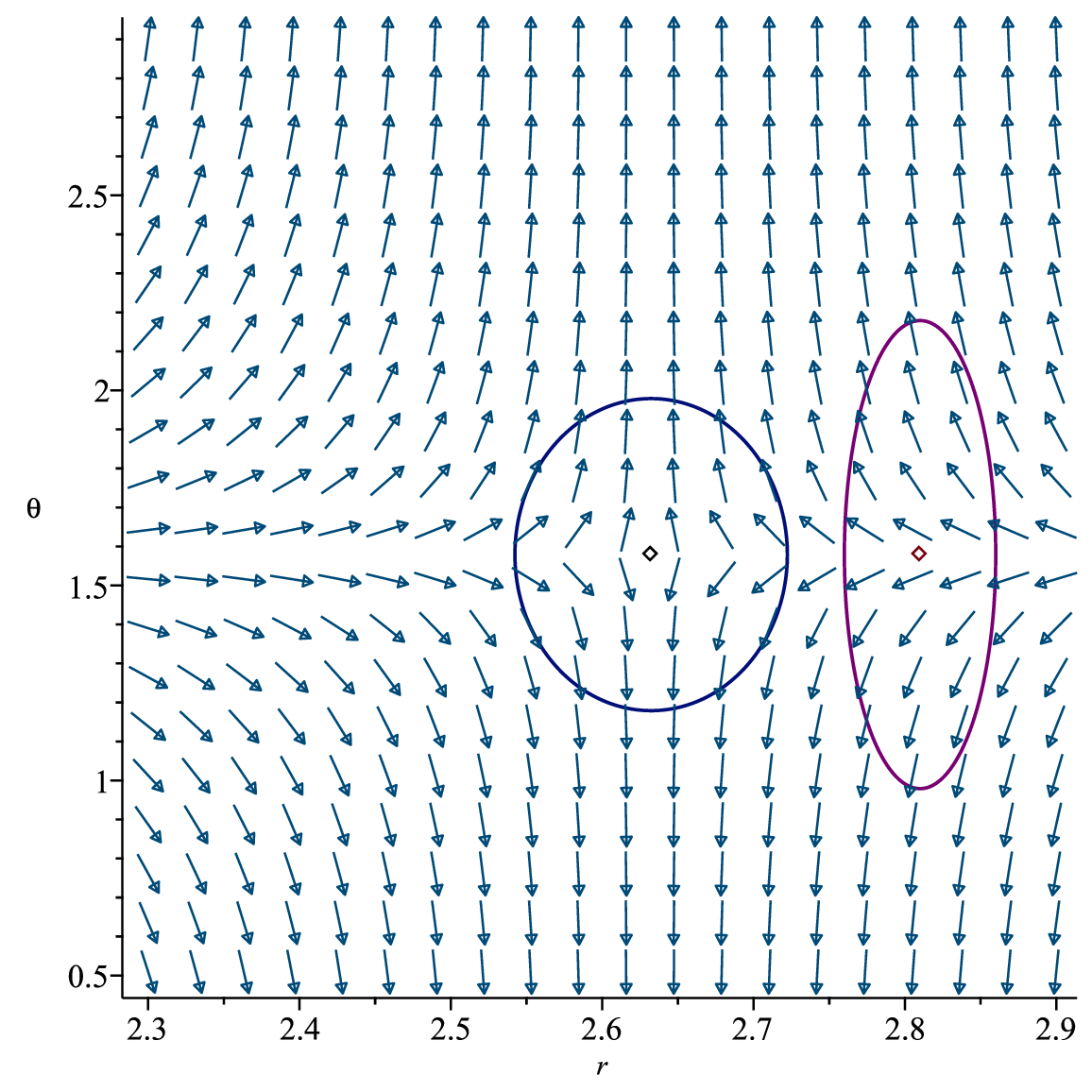}
 \label{3a}}
 \subfigure[]{
 \includegraphics[height=4cm,width=5.5cm]{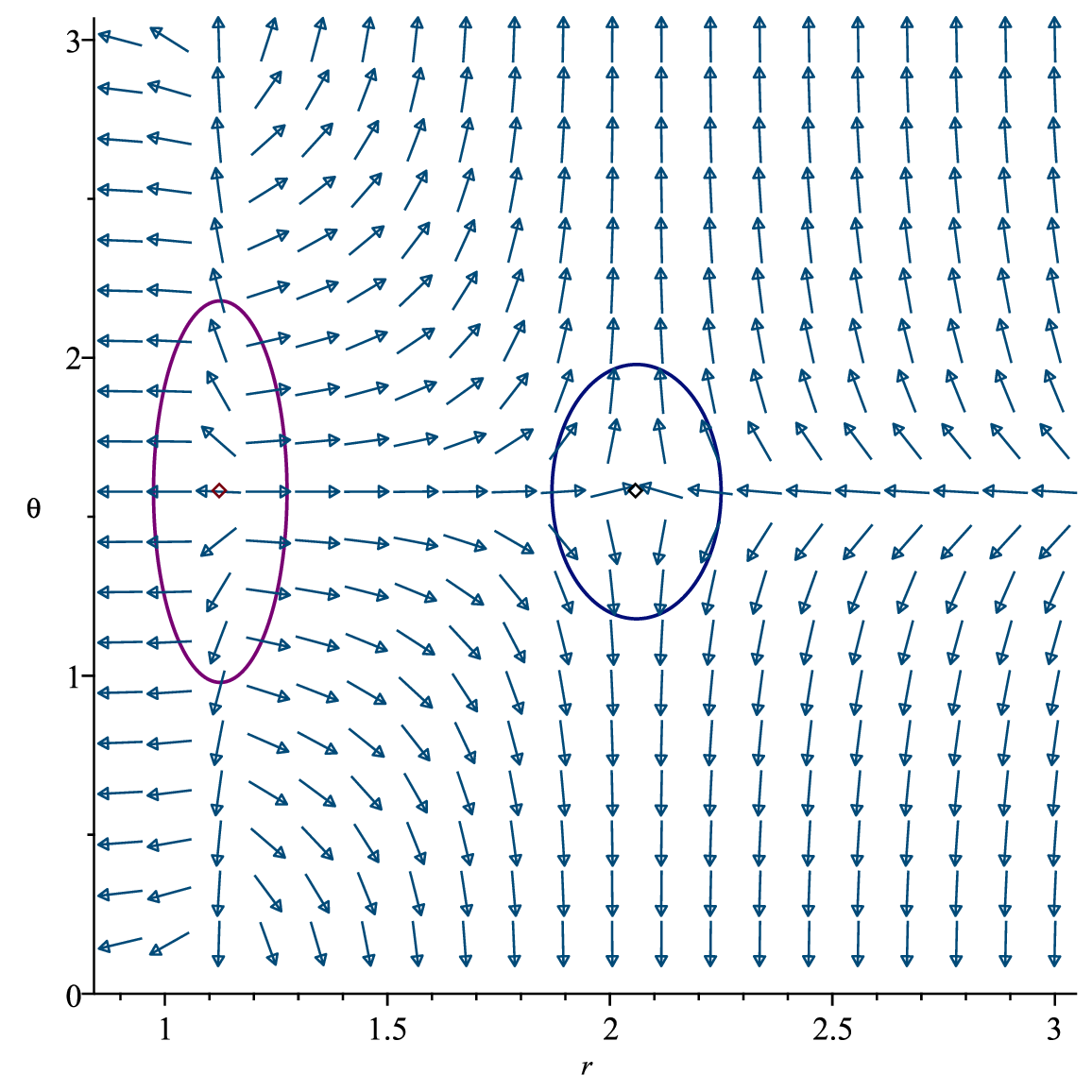}
 \label{3b}}
 \subfigure[]{
 \includegraphics[height=4cm,width=5.5cm]{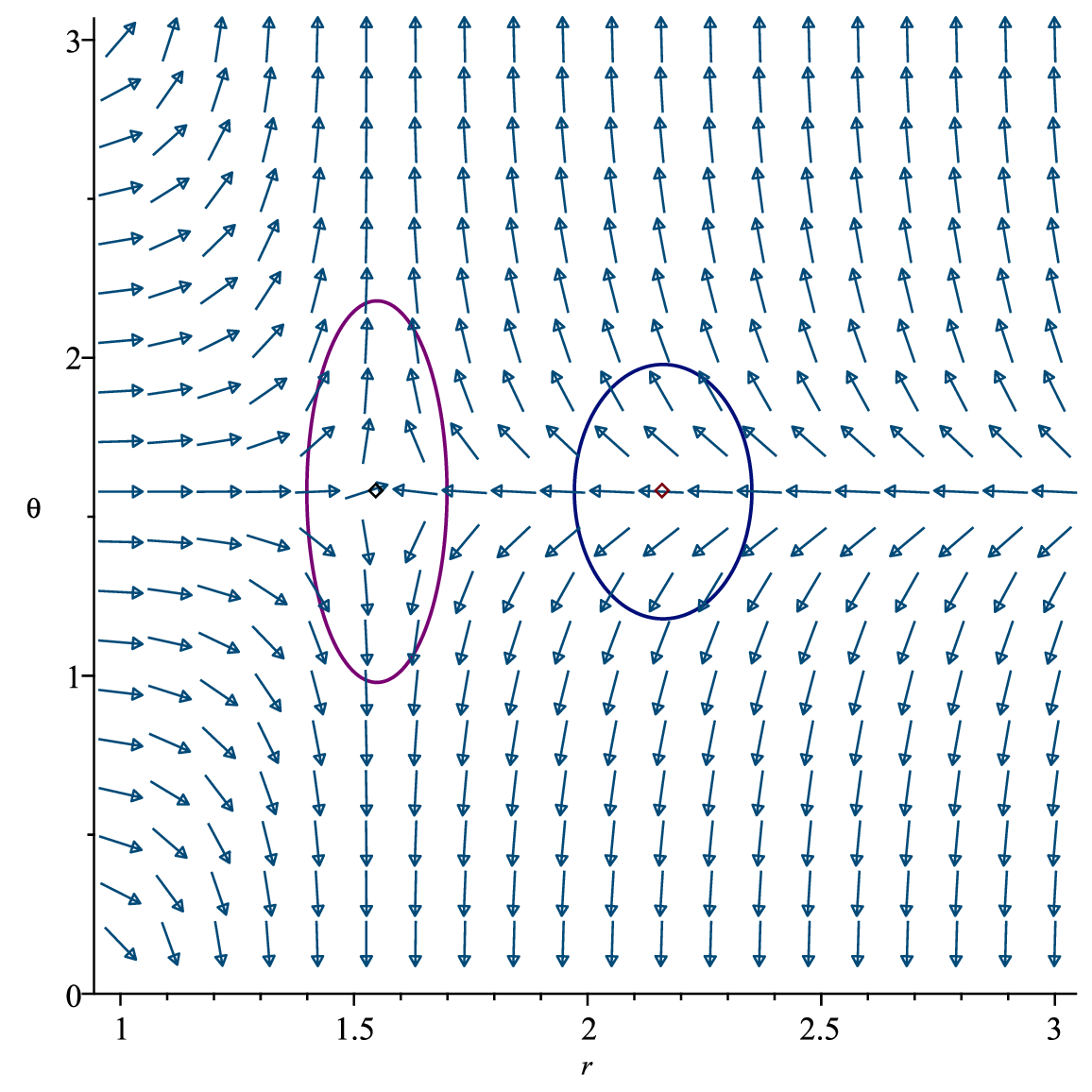}
 \label{3c}}
  \caption{\small{The PSs is located at $(r,\theta)=(2.632334314,1.57)$ for fig 3a with respect to $(q=0.09, z=1, M=13, \overline{\theta}=-2)$, $(r,\theta)=(1.1244,1.57)$ and $(r,\theta)=(2.0599,1.57)$ for fig 3b with respect to $(q=2, z=1, M=4, \overline{\theta}=-2)$ and $(r,\theta)=(1.548991117,1.57)$ with respect $(q=0.12, z=1, M=0.92, \overline{\theta}=-2)$. The contours $C_1$ (purple loop) and $C_2$ (blue loop) are two closed curves, where for fig 3a $C_2$ encircles the PS but $C_1$ does not and for fig 3c is vise versa. For fig 3b $C_1$ and $C_{2}$ encircles the PSs}}
 \label{3}
 \end{center}
 \end{figure}
\begin{figure}[h!]
 \begin{center}
 \subfigure[]{
 \includegraphics[height=4.5cm,width=6cm]{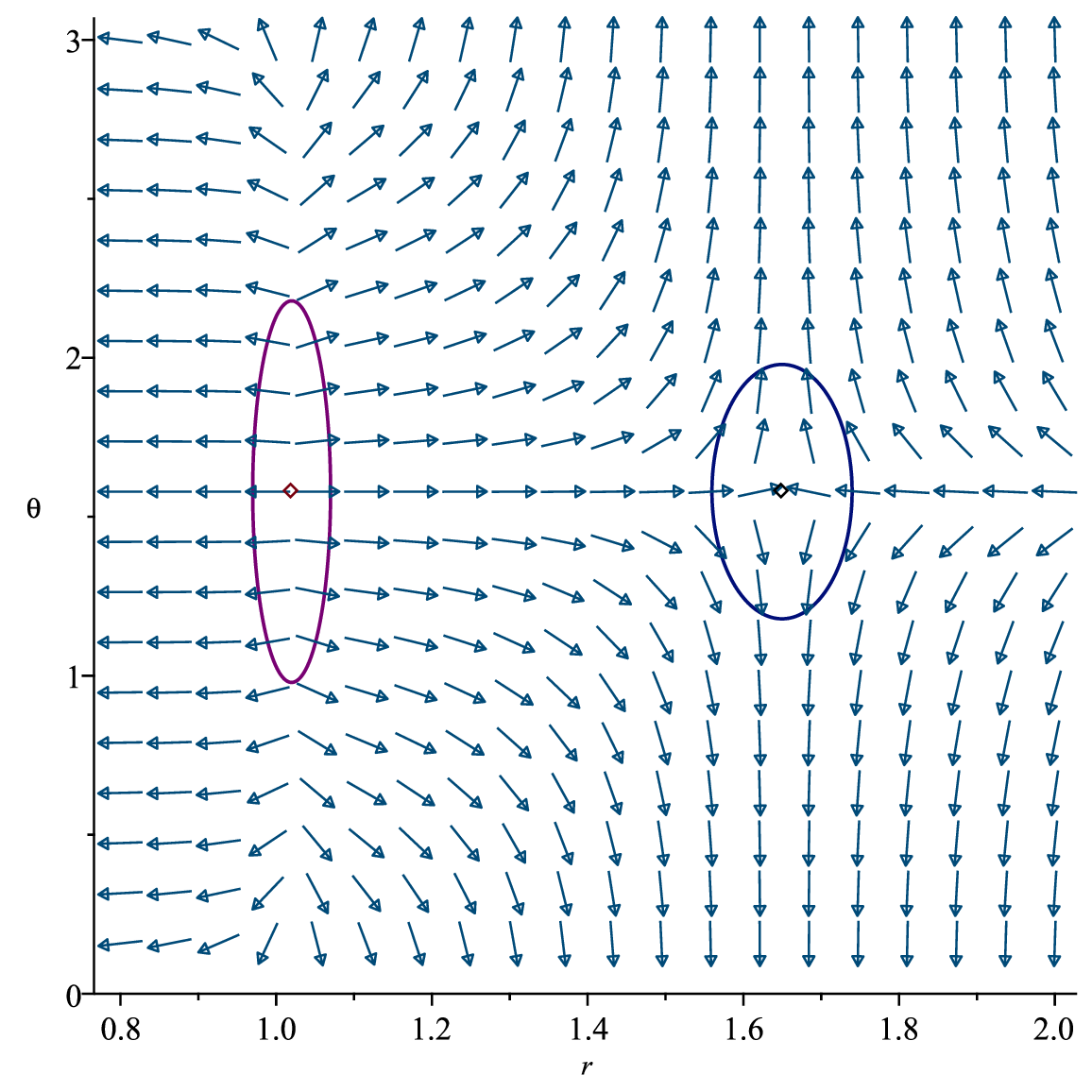}
 \label{4a}}
 \subfigure[]{
 \includegraphics[height=4.5cm,width=6cm]{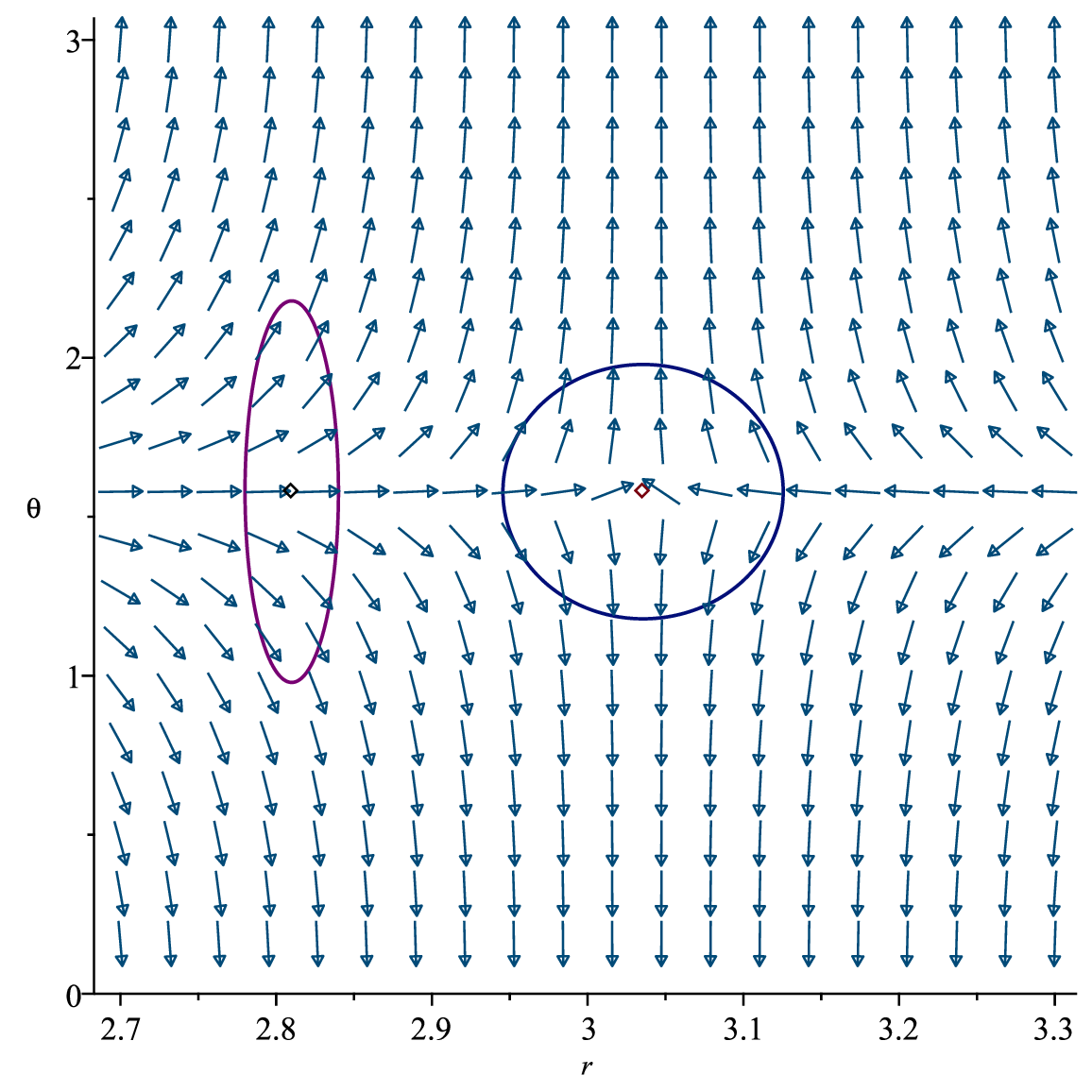}
 \label{4b}}
 \subfigure[]{
 \includegraphics[height=4.5cm,width=6cm]{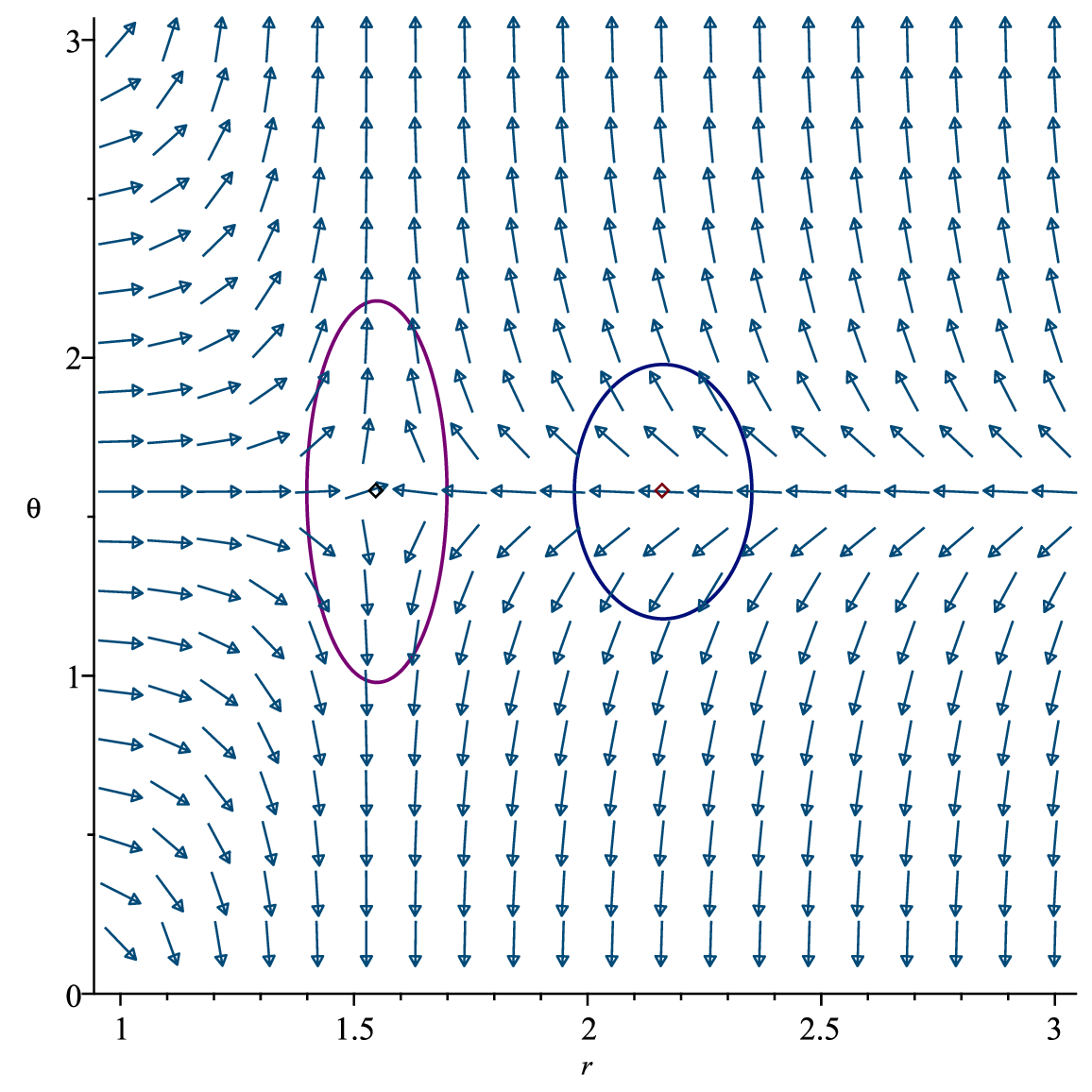}
 \label{4b}}
 \subfigure[]{
 \includegraphics[height=4.5cm,width=6cm]{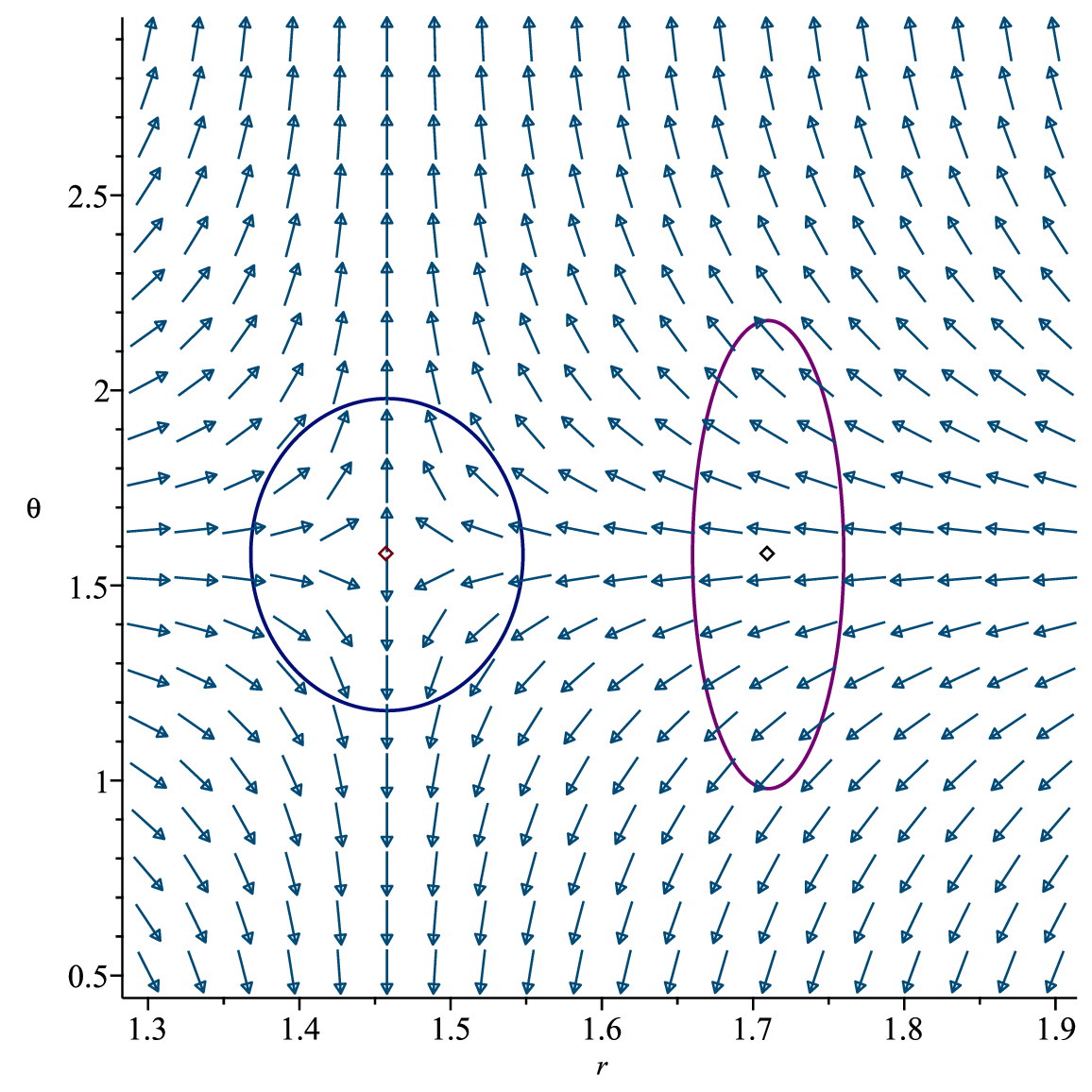}
 \label{4c}}
  \caption{\small{The PSs are located at $(r,\theta)=(1.019869106,1.57)$ and $(r,\theta)=(1.6496,1.57)$ for fig 4a with respect to $(q=0.9, z=1, M=1.33, \overline{\theta}=-3)$, $(r,\theta)=(3.0356074321,1.57)$ for fig 4b with respect to $(q=2, z=1, M=16, \overline{\theta}=-1)$, $(r,\theta)=(1.54899,1.57)$ with respect to $(q=0.12, z=1, M=0.92, \overline{\theta}=-2)$ and $(r,\theta)=(1.457880796,1.57)$ for fig 4d with respect to $(q=0.009, z=1, M=0.85, \overline{\theta}=-1)$. The contours $C_1$ (purple loop) and $C_2$ (blue loop) are two closed curves, where for fig 4b, 4d $C_2$ encircles the PSs but $C_1$ does not and for fig 4c is vise versa. For fig 4a $C_1$ and $C_{2}$ encircles the PSs}}
 \label{4}
 \end{center}
 \end{figure}

\begin{center}
\begin{table}
  \centering
\begin{tabular}{|p{3cm}|p{6.5cm}|p{3cm}|p{3cm}|}
  \hline
  % after \\: \hline or \cline{col1-col2} \cline{col3-col4} ...
  \centering{Hyperscaling violating black hole}  & \centering{Conditions} &\centering{Topological Charge}& Total Topological Charge\\[3mm]
   \hline
  \centering{Case 1:} & $q=0.09, z=1, M=13, \overline{\theta}=-2$ & \centering{-1} & $-1$\\[3mm]
   \hline
  \centering{Case 2:} & $q=2, z=1, M=4, \overline{\theta}=-2$ & \centering{$+1, -1$} & $0$ \\[3mm]
   \hline
   \centering{Case 3:} & $q=0.12, z=1, M=0.92, \overline{\theta}=-2$ & \centering{-1} & $-1$ \\[3mm]
   \hline
   \centering{Case 4:} & $q=0.9, z=1, M=1.33, \overline{\theta}=-3$ & \centering{+1, -1}& $0$ \\[3mm]
   \hline
   \centering{Case 5:} & $q=2, z=1, M=16, \overline{\theta}=-1$ & \centering{-1} & $-1$ \\[3mm]
   \hline
   \centering{Case 6:} & $q=0.12, z=1, M=0.92, \overline{\theta}=-1$ & \centering{-1} & $-1$ \\[3mm]
   \hline
   \centering{Case 7:} & $q=0.009, z=1, M=0.85, \overline{\theta}=-1$ & \centering{-1} & $-1$ \\[3mm]
   \hline
\end{tabular}
\caption{Summary of the results for PSs (z=1 $\&$ $\overline{\theta}<0$)}\label{2}
\end{table}
 \end{center}

\subsection{$1<z<2$}

\begin{figure}[h!]
 \begin{center}
 \subfigure[]{
 \includegraphics[height=4.5cm,width=5cm]{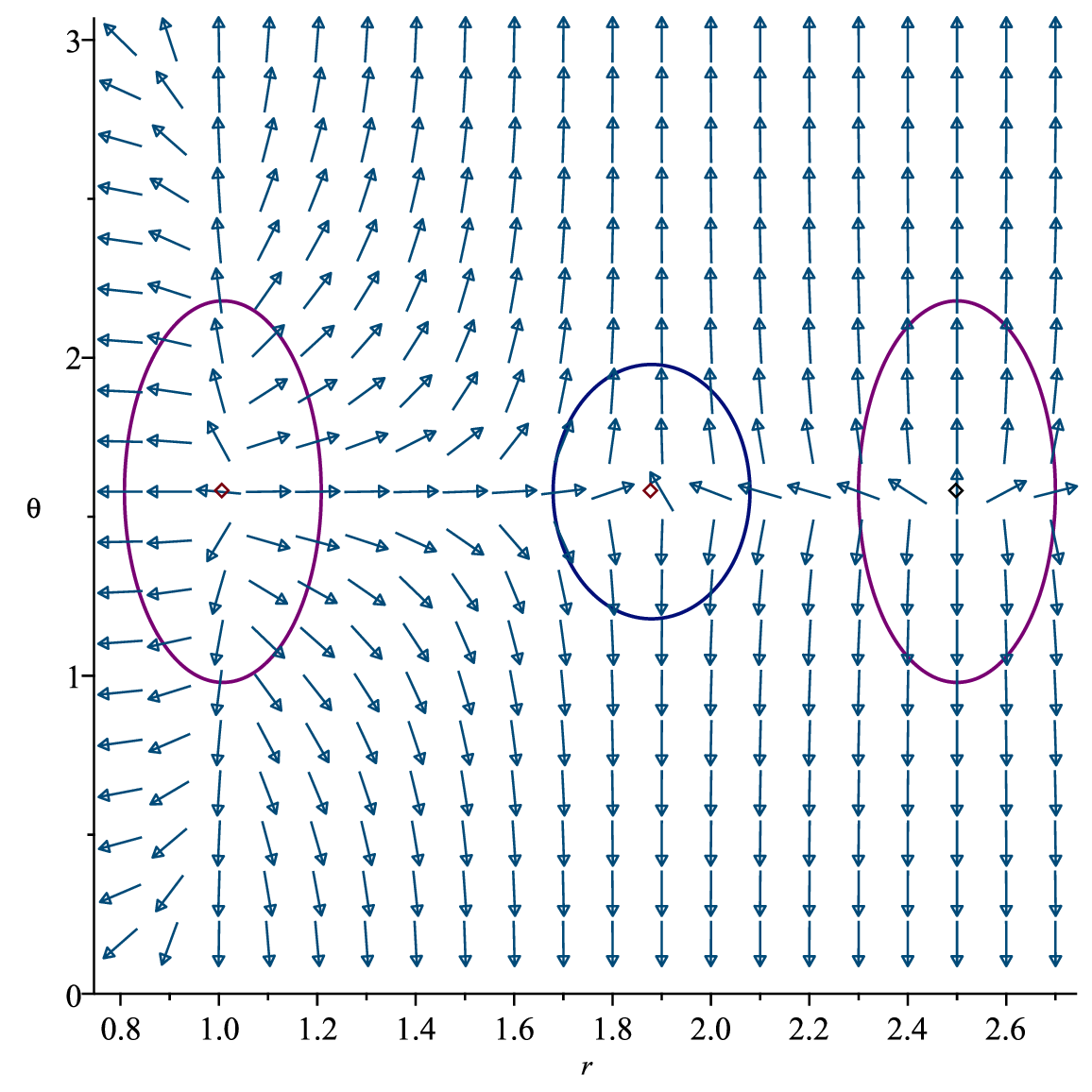}
 \label{5a}}
 \subfigure[]{
 \includegraphics[height=4.5cm,width=5cm]{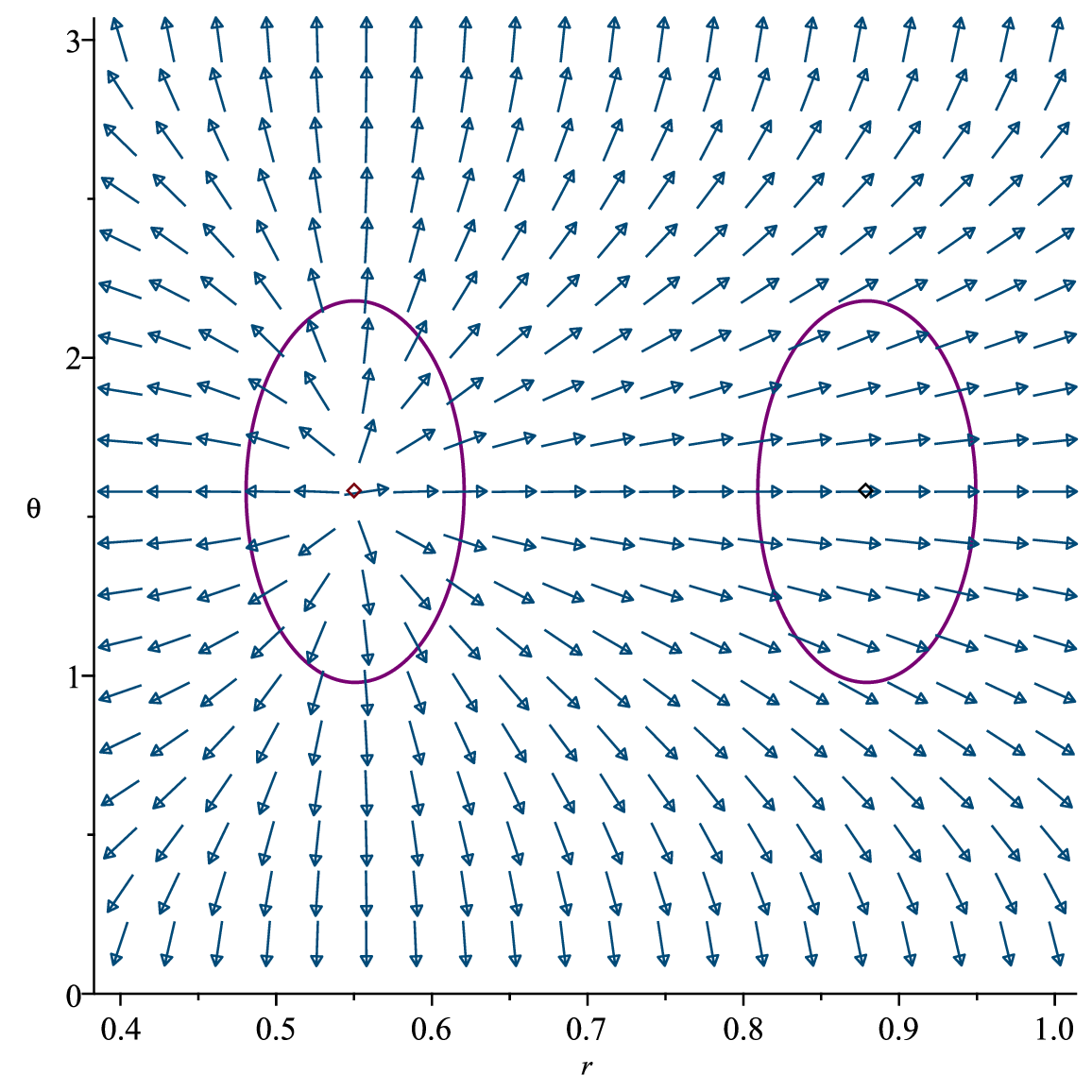}
 \label{5b}}
 \subfigure[]{
 \includegraphics[height=4.5cm,width=5cm]{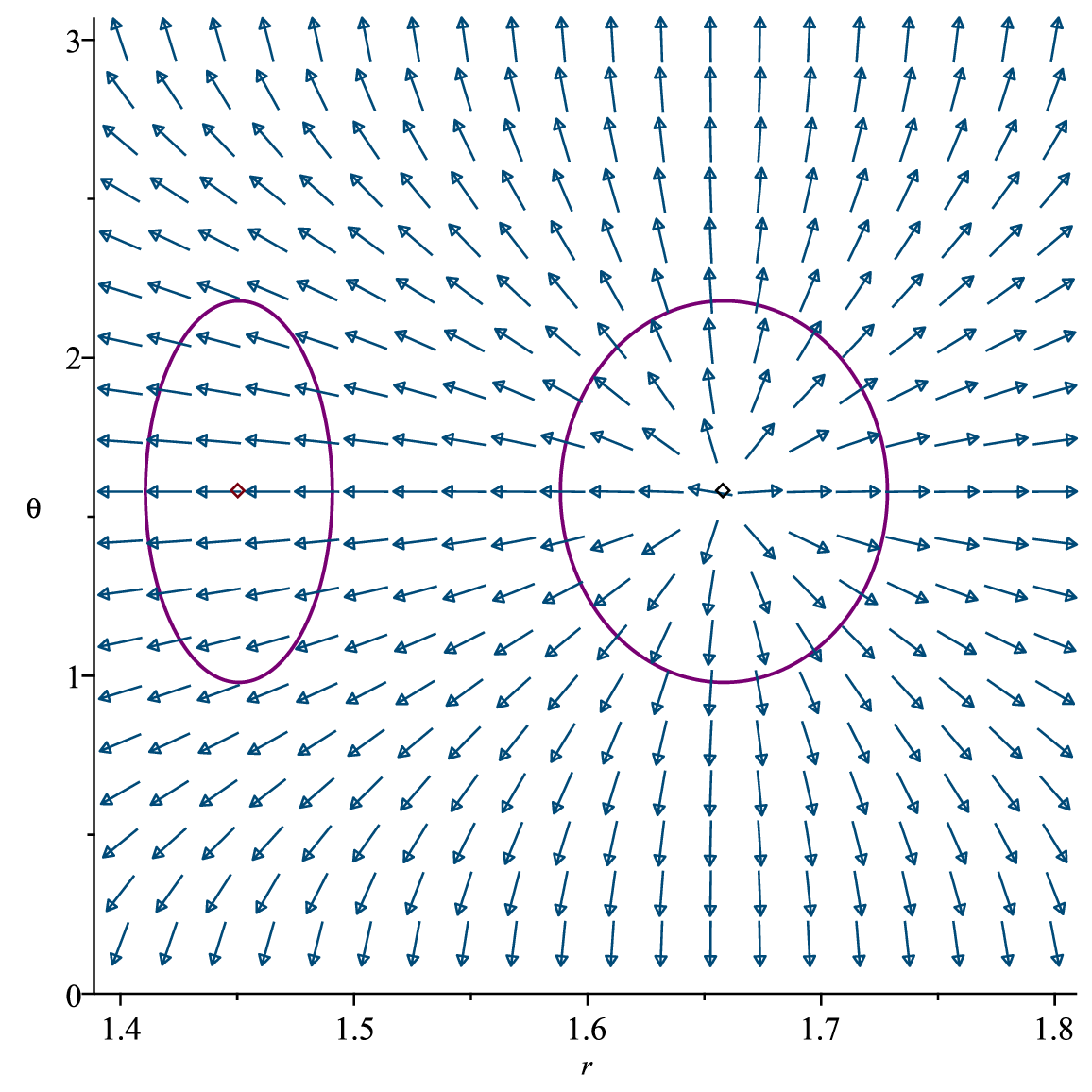}
 \label{5c}}
  \caption{\small{The PSs are located at $(r,\theta)=(2.499938457,1.57)$, $(r,\theta)=(1.879279979,1.57)$, $(r,\theta)=(1.008096108,1.57)$ for fig 5a, $(r,\theta)=(0.5507053204,1.57)$ and $(r,\theta)=(1.658351239,1.57)$ for fig 5b and 5c with respect to $(q=0.9, z=1.1, M=1.6, \overline{\theta}=0)$, $(q=1, z=1.9, M=4, \overline{\theta}=3)$ and $(q=2, z=1.9, M=0.7, \overline{\theta}=3)$, respectively. For fig 5a, contours (purple loop) and (blue loop) encircles the PSs. For fig 5b, the left loop encircles the PS but for fig 5c is vise versa.}}
 \label{5}
 \end{center}
 \end{figure}

 \begin{figure}[h!]
 \begin{center}
 \subfigure[]{
 \includegraphics[height=4.5cm,width=5cm]{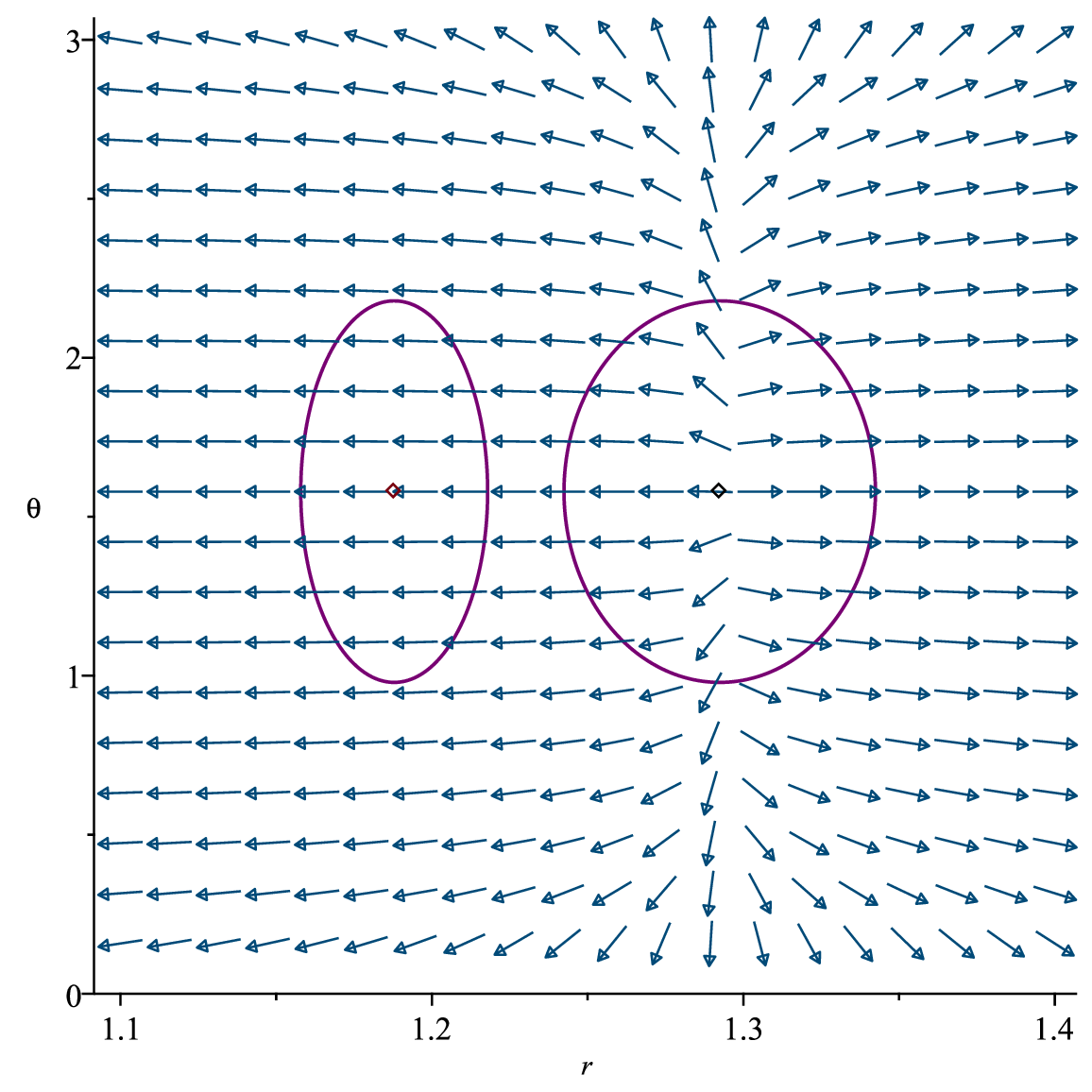}
 \label{6a}}
 \subfigure[]{
 \includegraphics[height=4.5cm,width=5cm]{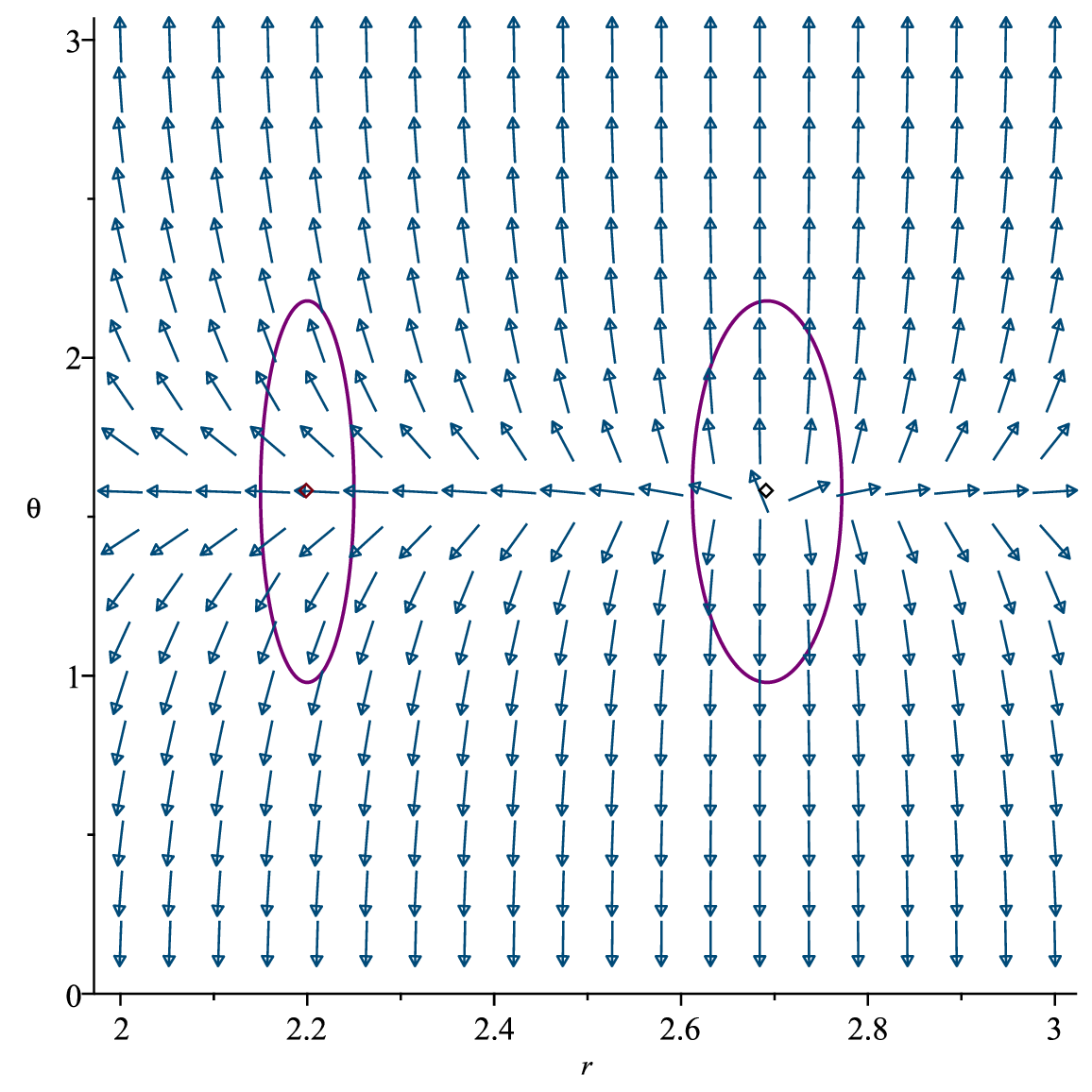}
 \label{6b}}
  \caption{\small{The PSs are located at $(r,\theta)=(1.292413995,1.57)$ for fig 6a and $(r,\theta)=(2.692061862,1.57)$ for fig 6b with respect to $(q=3, z=1.5, M=8, \overline{\theta}=1)$ and $(q=0.9, z=1.9, M=1.1, \overline{\theta}=0)$, respectively. For fig 6a and 6b, the right loop encircles the PSs.}}
 \label{6}
 \end{center}
 \end{figure}
 \begin{center}
 \begin{table}
  \centering
\begin{tabular}{|p{3cm}|p{6.5cm}|p{3cm}|p{3cm}|}
  \hline
  % after \\: \hline or \cline{col1-col2} \cline{col3-col4} ...
  \centering{Hyperscaling violating black hole}  & \centering{Conditions} & \centering{Topological Charge}& Total Topological Charge\\[3mm]
   \hline
  \centering{Case 1:} & $q=0.9, z=1.1, M=1.6, \overline{\theta}=0$ & \centering{+1, -1, +1} & $+1$\\[3mm]
   \hline
  \centering{Case 2:} & $q=1, z=1.9, M=4, \overline{\theta}=3$ & \centering{$+1$} & $+1$ \\[3mm]
   \hline
   \centering{Case 3:} & $q=2, z=1.9, M=0.7, \overline{\theta}=3$ & \centering{+1} & $+1$ \\[3mm]
   \hline
   \centering{Case 4:} & $q=3, z=1.5, M=8, \overline{\theta}=1$ & \centering{+1} & $+1$ \\[3mm]
   \hline
   \centering{Case 5:} & $q=0.9, z=1.9, M=1.1, \overline{\theta}=0$ & \centering{+1} & $+1$ \\[3mm]
   \hline
\end{tabular}
\caption{Summary of the results for PSs ($1<z<2$)}\label{3}
\end{table}
 \end{center}
\newpage
\subsection{$z\geq2$}

\begin{figure}[h!]
 \begin{center}
 \subfigure[]{
 \includegraphics[height=5cm,width=5cm]{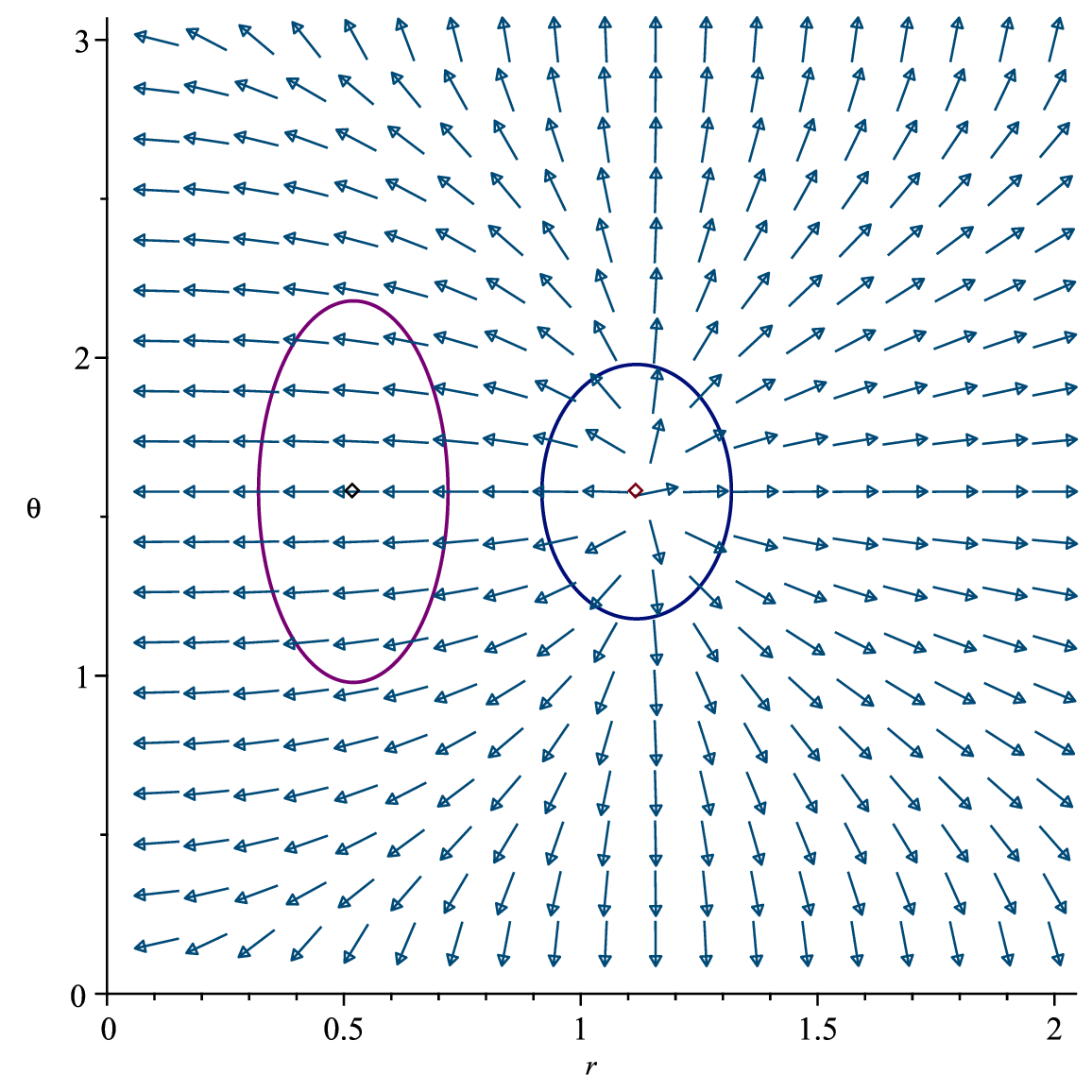}
 \label{7a}}
 \subfigure[]{
 \includegraphics[height=5cm,width=5cm]{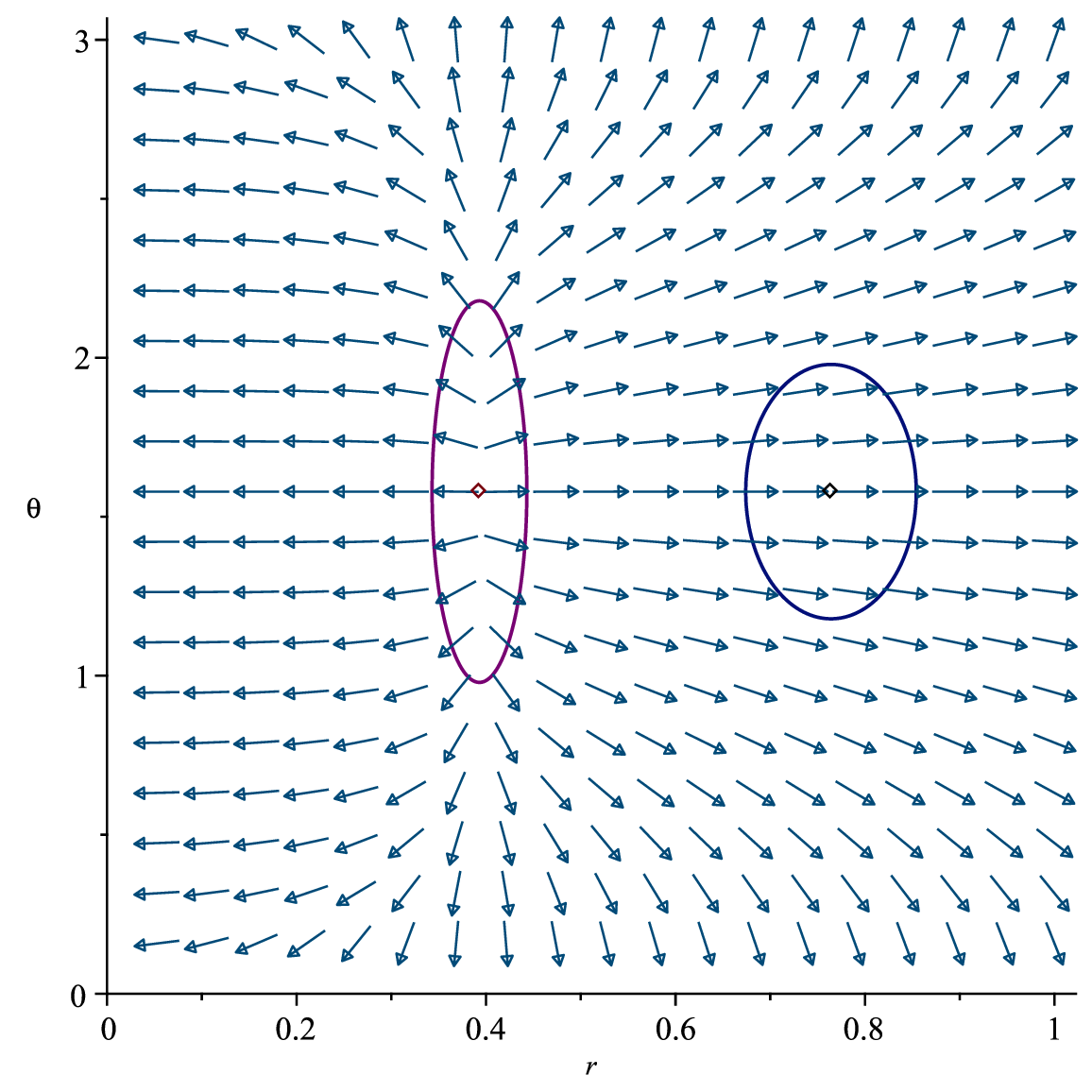}
 \label{7b}}
 \subfigure[]{
 \includegraphics[height=5cm,width=5cm]{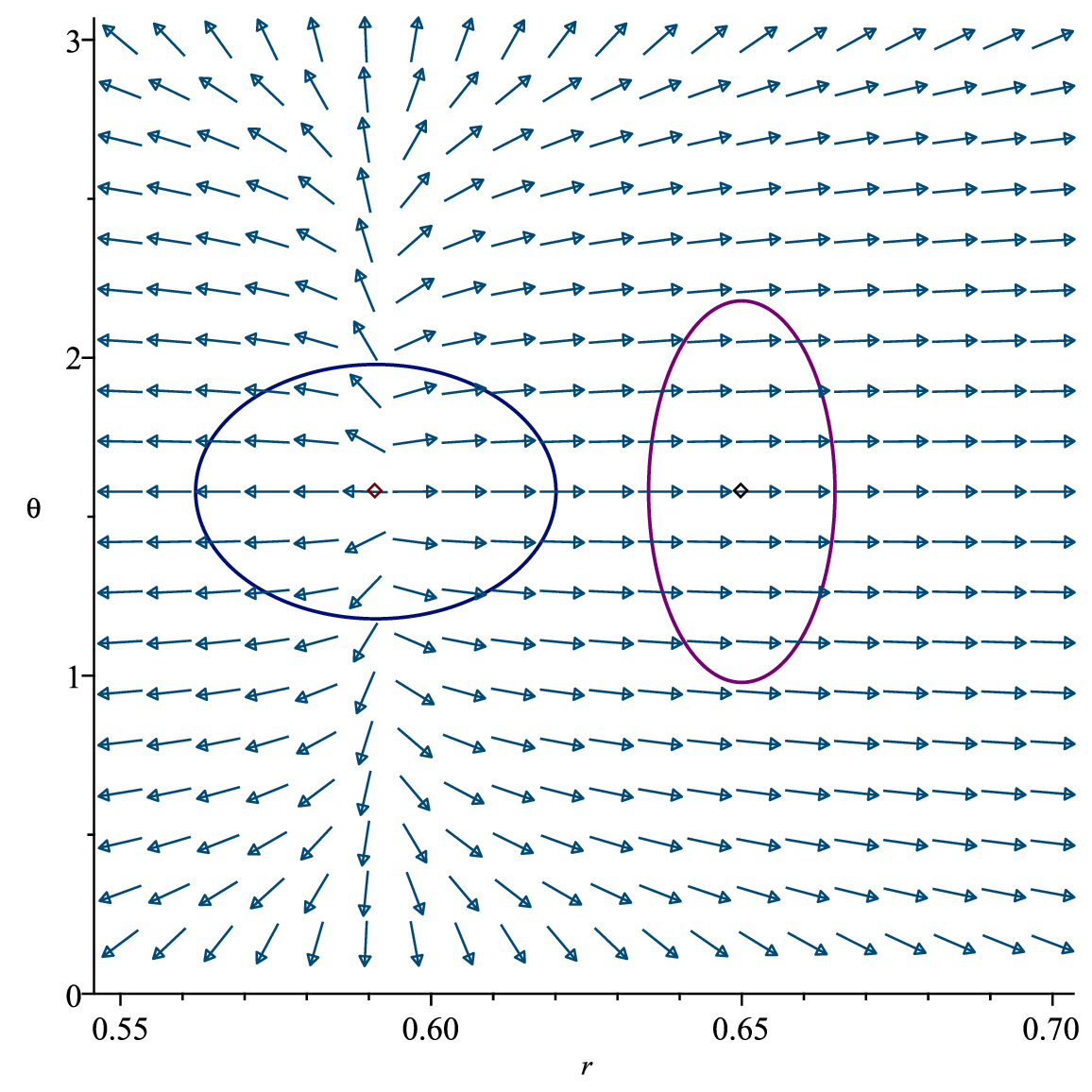}
 \label{7c}}
  \caption{\small{The PSs are located at $(r,\theta)=(1.118,1.57)$, $(r,\theta)=(0.3930,1.57)$ and $(r,\theta)=(0.5911041307,1.57)$ with respect to $(q=2, z=2, M=4, \overline{\theta}=3)$, $(q=0.9, z=2, M=4, \overline{\theta}=3)$ and $(q=0.6, z=3, M=0.05, \overline{\theta}=3)$ for fig 7a, 7b and 7c, respectively. The contours $C_1$ (purple loop) and $C_2$ (blue loop) are two closed curves, where for fig 7a and 7c $C_2$ encircles the PSs but $C_1$ does not and for fig 7b is vise versa.}}
 \label{7}
 \end{center}
 \end{figure}
\begin{center}
 \begin{table}
  \centering
\begin{tabular}{|p{3cm}|p{6cm}|p{3cm}|p{3cm}|}
  \hline
  % after \\: \hline or \cline{col1-col2} \cline{col3-col4} ...
  \centering{Hyperscaling violating black hole}  & \centering{Conditions} &\centering{Topological Charge}& Total Topological Charge\\[3mm]
   \hline
  \centering{Case 1:} & $q=2, z=2, M=4, \overline{\theta}=3$ & \centering{+1} & $+1$\\[3mm]
   \hline
  \centering{Case 2:} & $q=0.9, z=2, M=4, \overline{\theta}=3$ & \centering{$+1$} & $+1$ \\[3mm]
   \hline
   \centering{Case 3:} & $q=0.6, z=3, M=0.05, \overline{\theta}=3$ & \centering{+1} & $+1$ \\[3mm]
   \hline
\end{tabular}
\caption{Summary of the results for PSs ($z\geq 2$)}\label{4}
\end{table}
 \end{center}
 \newpage
\section{Temperature method in hyperscaling violating black holes}
In this section, we are going to apply the temperature method to the topology structure of hyperscaling violating black holes and discuss the results in detail, see calculations in Appendix B. Also we see the calculations $(T, \Phi, \phi^r, \phi^{\theta}, q_{cp_1}^2, r_{cp_1}^2, q_{cp_2}^2, r_{cp_2}^2, T_{cp_1}, T_{cp_2})$ in Appendix B. Here we will only discuss the outcomes of the work based on such Appendix B. In here one can say that, the second critical point is entirely in agreement with \cite{3}. Therefore, in the topology method, we will have one additional critical point, and then we will obtain the topological charge corresponding to each of the critical points. In order to determine the topological charges of the critical points, we analyze them in the state where $k=1$. However, it is important to note that if $k=0,-1$, based on equations \eqref{1}, \eqref{93}, \eqref{94}, \eqref{96}, \eqref{97} and Table 1 in \cite{3,6}, no critical point exists and therefore no topological charge can be determined.\\
Furthermore, based on the relationship between \eqref{1} and \eqref{96}, it is evident that physical critical points can only exist when $k=1$ and $1\leq z<2$. Furthermore, through the analysis and examination of the aforementioned relationships, we arranged following table,
\begin{eqnarray}\label{21}
1\leq z<2 \rightarrow \left\{
               \begin{array}{ll}
                 1\leq z<\frac{1}{2}(\overline{\theta}-d+2+\sqrt{(d-\overline{\theta})^2+4}) , & \hbox{Two Critical Point ;} \\
                 \frac{1}{2}(\overline{\theta}-d+2+\sqrt{(d-\overline{\theta})^2+4})\leq z<2, & \hbox{One  Critical Point .}
               \end{array}
             \right.
\end{eqnarray}
In this case, there can be different topological classes. By using following formula we obtain the normalized vector field,
\begin{equation}\label{22}
\begin{split}
n = \left(\frac{\phi^r}{\|\phi\|},\frac{\phi^{\theta}}{\|\phi\|}\right).
\end{split}
\end{equation}
In order to examine the topological charge of each area, we will analyze them in two separate parts.
\subsection{Case I: $1\leq z<\frac{1}{2}(\overline{\theta}-d+2+\sqrt{(d-\overline{\theta})^2+4})$}
\begin{figure}[h!]
 \begin{center}
 \includegraphics[height=5.5cm,width=5.5cm]{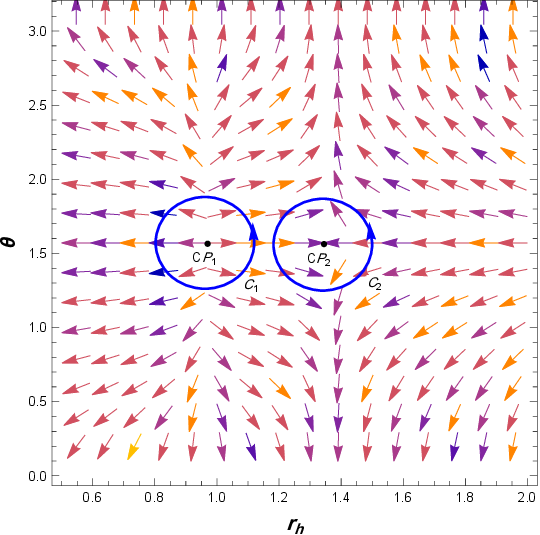}
 \caption{\small{The  arrows represent the vector field $n$ on  the $r_h-\theta$ plane for the HSV  black hole with the $q=1, z=1, \overline{\theta}=-1, d=4.$}}
 \label{fig5.1}
 \end{center}
 \end{figure}

Since we have two critical points within this area, we can have varying topological charges. By using  relations \eqref{90}, \eqref{91},  \eqref{22} and examining  $(z=1, \overline{\theta}=-1, d=4)$ in Figure \eqref{fig5.1}, one can determine the topological charges of each critical point.
Therefore, based on figure \eqref{fig5.1}, we can obtain that $Q_{CP_1}=+1$ and $Q_{CP_2}=-1$ . Additionally, we can conclude that $Q_{total}=Q_{CP_1} + Q_{CP_2}=0$.
\begin{figure}[h!]
 \begin{center}
 \subfigure[]{
 \includegraphics[height=5cm,width=6cm]{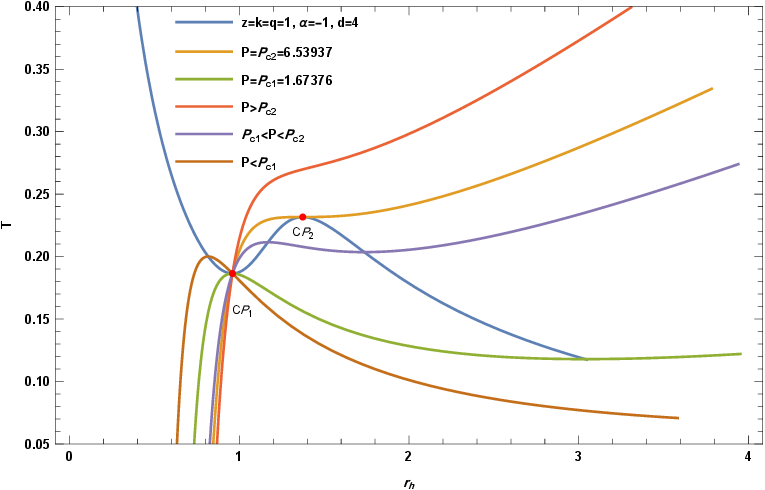}
 \label{fig5.2a}}
 \subfigure[]{
 \includegraphics[height=5cm,width=6cm]{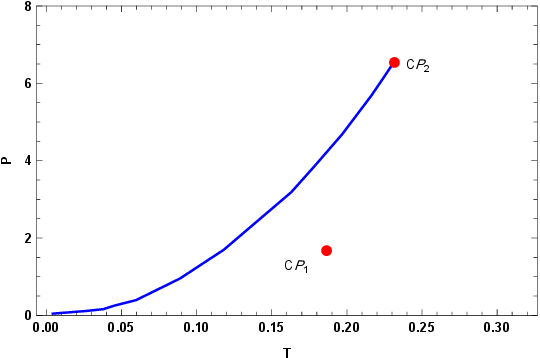}
 \label{fig5.2b}}
 \caption{\small{(a) Isobaric curves and spinodal curve (blue line)
for the Hyperscaling violating black hole. (b) Phase diagram showing the first-order phase transitions near the conventional critical point $CP_2$ . We have set $q = 1$, $z=1$, $\overline{\theta}=-1$ and $d=4$.}}
 \label{fig5.2}
 \end{center}
 \end{figure}
As shown in figure \eqref{fig5.2a}, the novel and conventional critical points show the local minimum and local maximum of the spinodal curve, respectively, which is in agreement with \cite{16,26}. Also, according to figure \eqref{fig5.2b}, it is observed that the first-order phase transition takes place in proximity to the conventional critical point $(CP_2)$, whereas the novel critical point $(CP_1)$ does not have any impact.
\newpage
\subsection{Case II: $\frac{1}{2}(\overline{\theta}-d+2+\sqrt{(d-\overline{\theta})^2+4})\leq z<2$}
\begin{figure}[h!]
 \begin{center}
 \includegraphics[height=5.5cm,width=5.5cm]{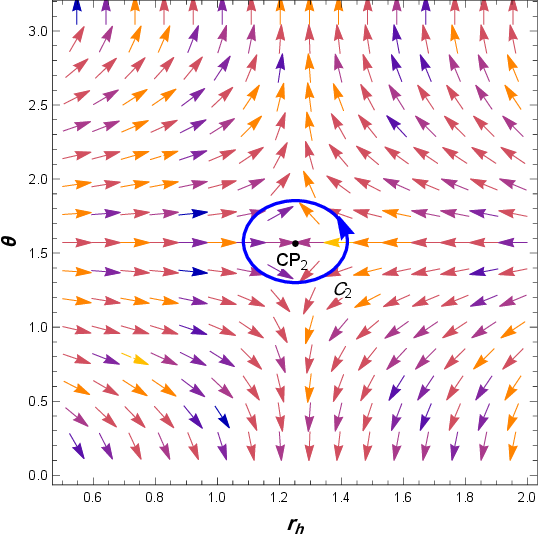}
 \caption{\small{The  arrows represent the vector field $n$ on  the $r_h-\theta$ plane for the HSV  black hole with the $q=1, z=1.5, \overline{\theta}=1, d=4.$}}
 \label{fig5.3}
 \end{center}
 \end{figure}

In this section, by using the relations \eqref{90}, \eqref{91}, and \eqref{22} and considering $(z=1.5, \overline{\theta}=1, d=4)$, with the assistance of figure \eqref{fig5.3}, we find the topological charge related to the critical point.
In this area, there exists a critical point with a topological charge of $Q_{CP_2}=-1$. Therefore, we have a total topological charge of $Q_{total}= Q_{CP_2}=-1$.
According to figure \eqref{fig5.3}, in this area (Case II), there exists a critical point with a topological charge of $Q_{CP_2}=-1$. Therefore, we have a total topological charge of $Q_{total}= Q_{CP_2}=-1$.
 \begin{figure}[h!]
 \begin{center}
 \subfigure[]{
 \includegraphics[height=5cm,width=6cm]{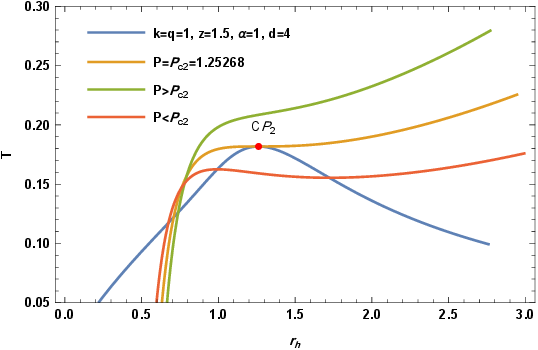}
 \label{fig5.4a}}
 \subfigure[]{
 \includegraphics[height=5cm,width=6cm]{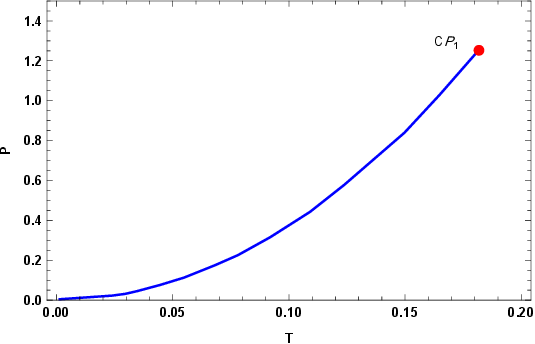}
 \label{fig5.4b}}
 \caption{\small{(a) Isobaric curves and spinodal curve (blue line)
for the Hyperscaling violating black hole. (b) Phase diagram showing the first-order phase transitions near the conventional critical point $CP_2$ . We have set $q = 1$, $z=1.5$, $\overline{\theta}=1$ and $d=4$.}}
 \label{fig5.4}
 \end{center}
 \end{figure}
As shown in figure \eqref{fig5.4a}, the conventional critical point $(CP_2)$ is the maximum of the spinodal curve. Additionally, based on figure \eqref{fig5.4b}, it is evident that a first-order phase transition will occur in this area (Case II).\\
Therefore, we can conclude that the HSV black hole only has critical points for $k = 1$ and at interval $1\leq z<2$ we can define a topological charge.
Also, we found that this region is divided into two different topological classes, but they have  same phase structure.
The proposal in \cite{16} suggests that topological change can serve as a prognostic indicator for changes in phase structures. However, our findings contradict this claim while being consistent with \cite{26}.
 \begin{center}
\begin{table}
  \centering
\begin{tabular}{|p{5cm}|p{5cm}|p{3cm}|p{3cm}|}
  \hline
  % after \\: \hline or \cline{col1-col2} \cline{col3-col4} ...
  \centering{Hyperscaling violating black hole\\ $k=1$ and $1\leq z<2$ }  & \centering{Conditions} &\centering{Topological Charge}& Total Topological Charge\\[3mm]
   \hline
  \centering{Case I:} & $1\leq z<\frac{1}{2}(\overline{\theta}-d+2+\sqrt{(d-\overline{\theta})^2+4})$ & \centering{$Q_{t|CP_1}=1$,\hspace{3cm} $Q_{t|CP_2}=-1$} & $0$\\[3mm]
   \hline
  \centering{Case II:} & $\frac{1}{2}(\overline{\theta}-d+2+\sqrt{(d-\overline{\theta})^2+4})\leq z<2$ & \centering{$Q_{t|CP_2}=-1$}& $-1$ \\[3mm]
   \hline
\end{tabular}
\caption{Summary of the results for Hyperscaling violating black hole.}\label{20}
\end{table}
 \end{center}

\section{Generalized Helmholtz free energy method}
We use different quantities to describe the thermodynamic properties of black holes. For example, we can use mass and temperature as two variables to express the generalized free energy. Since mass and energy are related in black holes, we can write our generalized free energy function as a standard thermodynamic function like following,
\begin{equation}\label{23}
\mathcal{F}=M-\frac{S}{\tau}.
\end{equation}
The corresponding equation shows that $\tau$ and T (the reciprocal of $\tau$) are the Euclidean time period and the ensemble's temperature, respectively. The generalized free energy is only on-shell when $\tau= \tau_{H} =\frac{1}{T_{H}}$. We can construct a vector $\phi$ as folows,
\begin{equation}\label{24}
\phi=(\phi^r,\phi^\theta)=\big(\frac{\partial\mathcal{F}}{\partial r_{H}},-\cot\theta\csc\theta\big).
\end{equation}
The vector points outward when $\theta= 0, \pi$ and $\phi^{\theta}$ diverges. The range of $r_{H}$ is from $0$ to $\infty$, and the range of $\theta$ is from $0$ to $\pi$. Here, we use the Helmholtz free energy method on the above mentioned black hole and compare the obtained results with other existing studies. To examine the thermodynamic topology from the mentioned method, we need to introduce quantities such as mass, free energy, entropy, and the radius of the AdS, which are calculated as follows. We will present the obtained results by the above method for various free parameters in several figures with respect to $(z=1, 1.3, 1.9)$.
\begin{equation}\label{25}
M =\ell^{-z -1} \left(1+\frac{9 \ell^{2}}{r_H^{2} \left(2-\overline{\theta} +z \right)^{2}}+\frac{q^{2}}{r_H^{6-2 \overline{\theta} +2 z}}\right) r_H^{z -\overline{\theta} +4} \left(4-\overline{\theta} \right)
\end{equation}
the radius of the AdS which is given by $$\ell^{2}=\frac{\left(z -\overline{\theta} +4\right) \left(3-\overline{\theta} +z \right)}{{\mathrm e}^{\frac{\overline{\theta}}{2 \sqrt{\left(z -1-\frac{\overline{\theta}}{4}\right) \left(8-2 \overline{\theta} \right)}}} P}$$. Also the entropy is as $S =r_H^{4-\overline{\theta}}$. Also, we will have
\begin{equation}\label{288}
\begin{split}
&\mathcal{F} =\left(\sqrt{\frac{\left(z -\overline{\theta} +4\right) \left(3-\overline{\theta} +z \right)}{{\mathrm e}^{\frac{\overline{\theta}}{2 \sqrt{\left(z -1-\frac{\overline{\theta}}{4}\right) \left(8-2 \overline{\theta} \right)}}} P}}\right)^{-z -1}\\
&\times \left(1+\frac{9 \left(z -\overline{\theta} +4\right) \left(3-\overline{\theta} +z \right)}{{\mathrm e}^{\frac{\overline{\theta}}{2 \sqrt{\left(z -1-\frac{\overline{\theta}}{4}\right) \left(8-2 \overline{\theta} \right)}}} P \,r_H^{2} \left(2-\overline{\theta} +z \right)^{2}}+\frac{q^{2}}{r_H^{6-2 \overline{\theta} +2 z}}\right)
\times r_H^{z -\overline{\theta} +4} \left(4-\overline{\theta} \right)-\frac{r_H^{4-\overline{\theta}}}{\tau}
\end{split}
\end{equation}
\newpage
\subsection{z=1}

\begin{figure}[h!]
 \begin{center}
 \subfigure[]{
 \includegraphics[height=4cm,width=5.5cm]{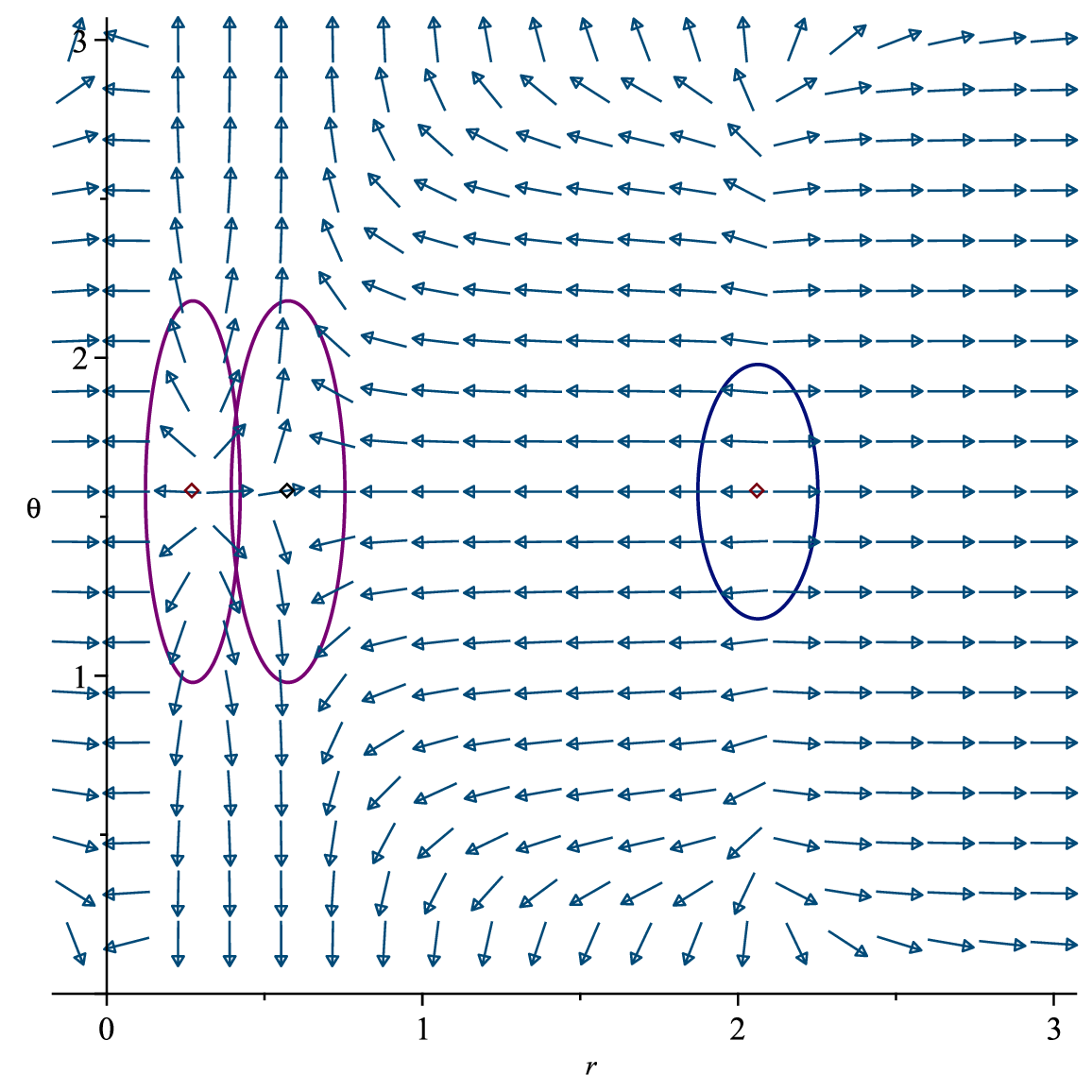}
 \label{12a}}
 \subfigure[]{
 \includegraphics[height=4cm,width=5.5cm]{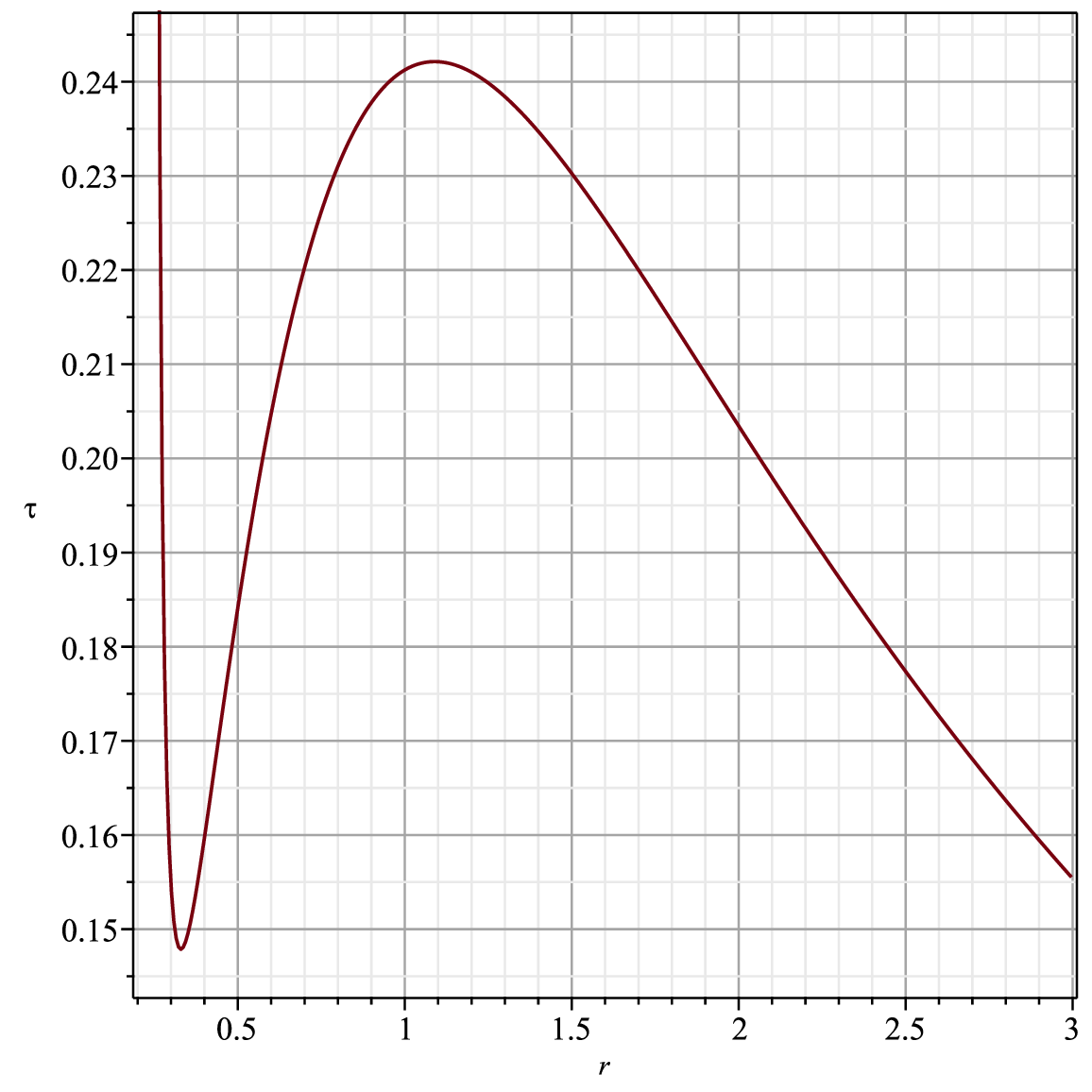}
 \label{12b}}
  \caption{\small{Figure 12a shows the vector field $n$ on a part of the $(r_H-\theta)$ plane with $(q=0.005, z=1, \tau=0.2, p=13, \overline{\theta}=-1)$. The ZPs are located at $(r_H-\theta)=(0.2729036582,1.57), (0.5746163406,1.57), (2.062792783,1.57)$ inside the circular loop. The contours (blue loop) and (purple loop) are three closed curves, where encircles the ZPs. Figure 12b displays the plot of the curve given by equation $\tau$.}}
 \label{12}
 \end{center}
 \end{figure}

\subsection{z=1.3}

\begin{figure}[h!]
 \begin{center}
 \subfigure[]{
 \includegraphics[height=4cm,width=5cm]{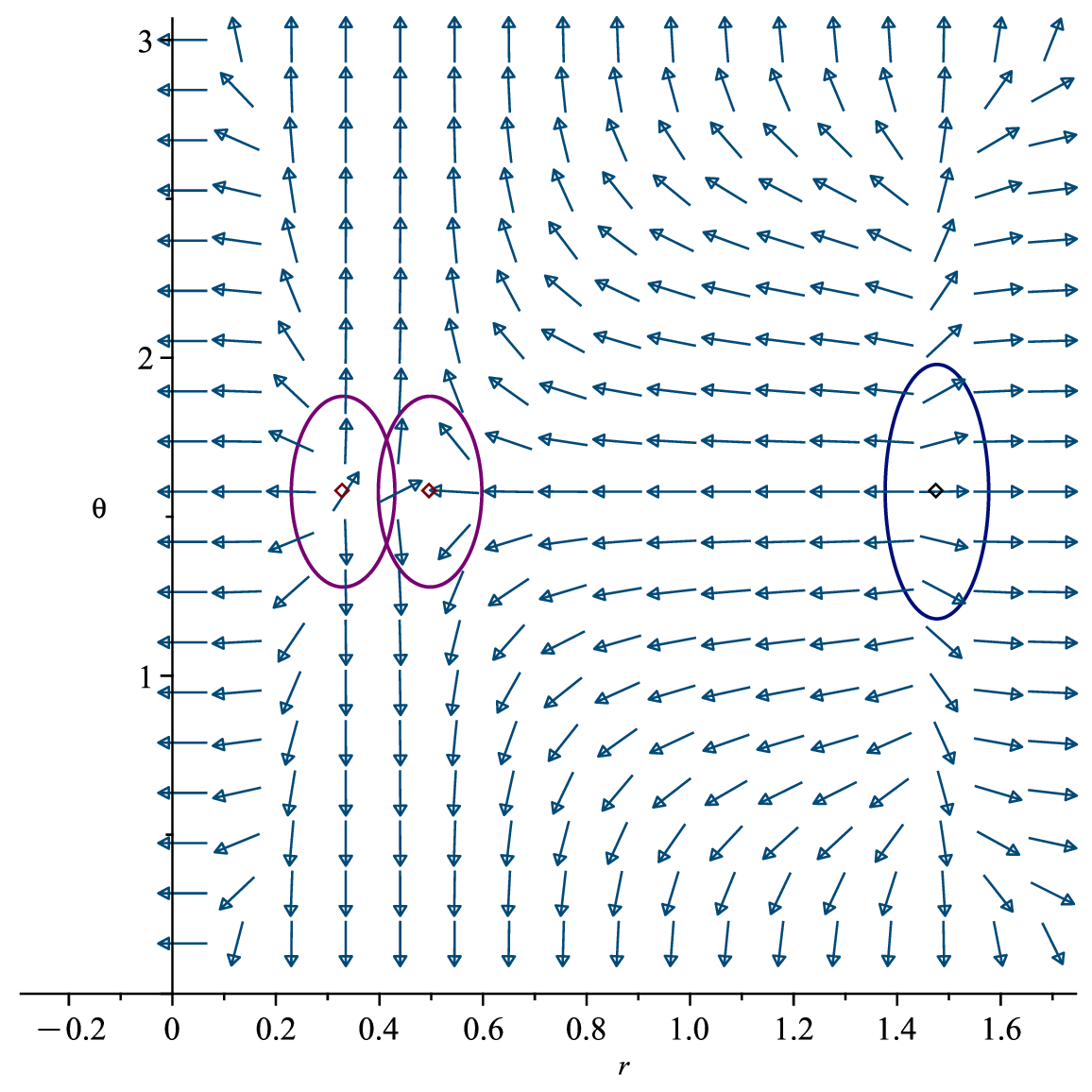}
 \label{13a}}
 \subfigure[]{
 \includegraphics[height=4cm,width=5cm]{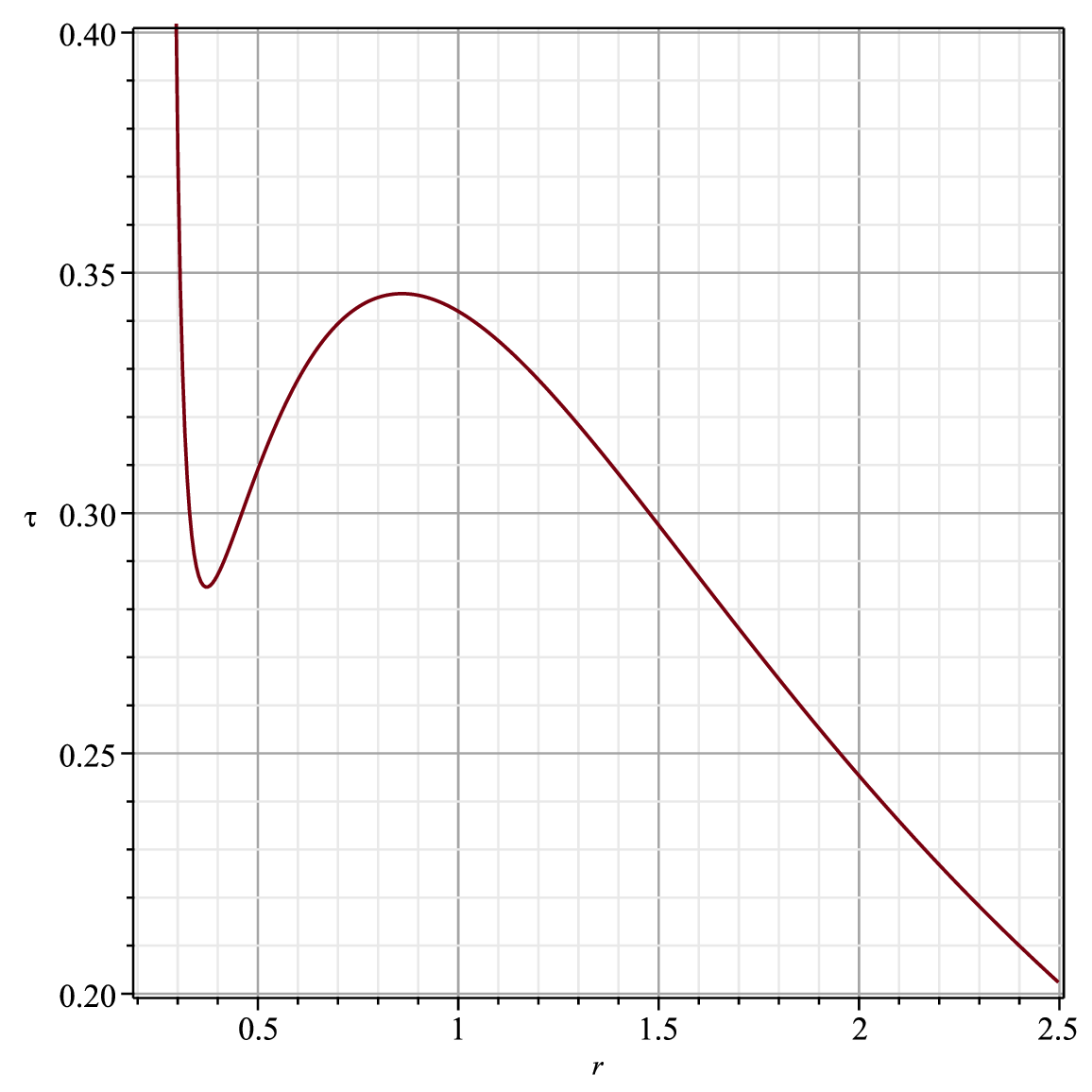}
 \label{13b}}
  \caption{\small{Figure 13a shows the vector field $n$ on a part of the $(r_H-\theta)$ plane with $(q=0.005, z=1.3, \tau=0.3, P=10, \overline{\theta}=-1)$. The ZPs are located at $(r_H-\theta)=(0.3297766982,1.57), (0.497766982, 1.57) ,(1.47682725,1.57)$ inside the circular loop. Figure 13b displays the plot of the curve given by equation $\tau$.}}
 \label{13}
 \end{center}
 \end{figure}

\subsection{z=1.9}

\begin{figure}[h!]
 \begin{center}
 \subfigure[]{
 \includegraphics[height=4cm,width=5cm]{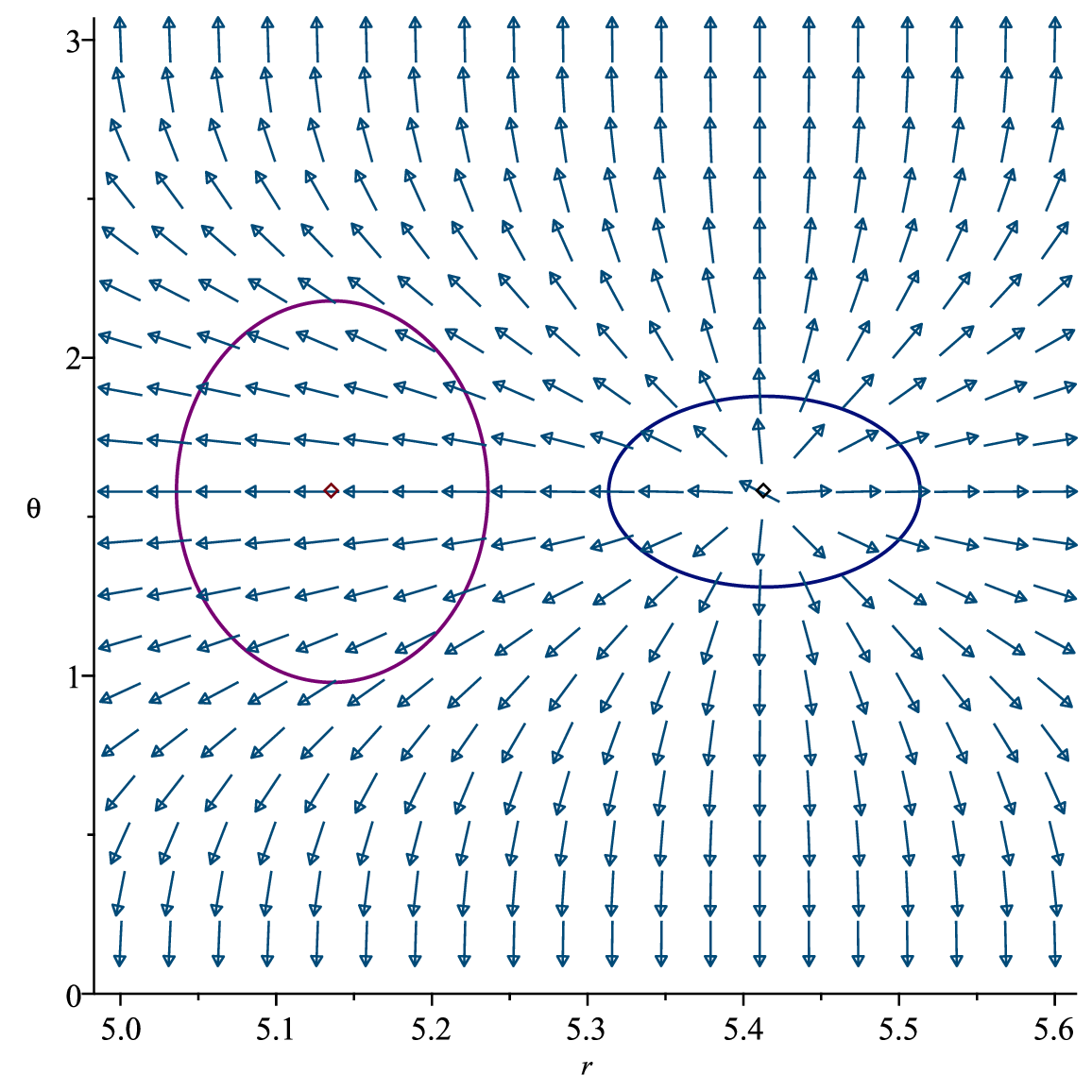}
 \label{14a}}
 \subfigure[]{
 \includegraphics[height=4cm,width=5cm]{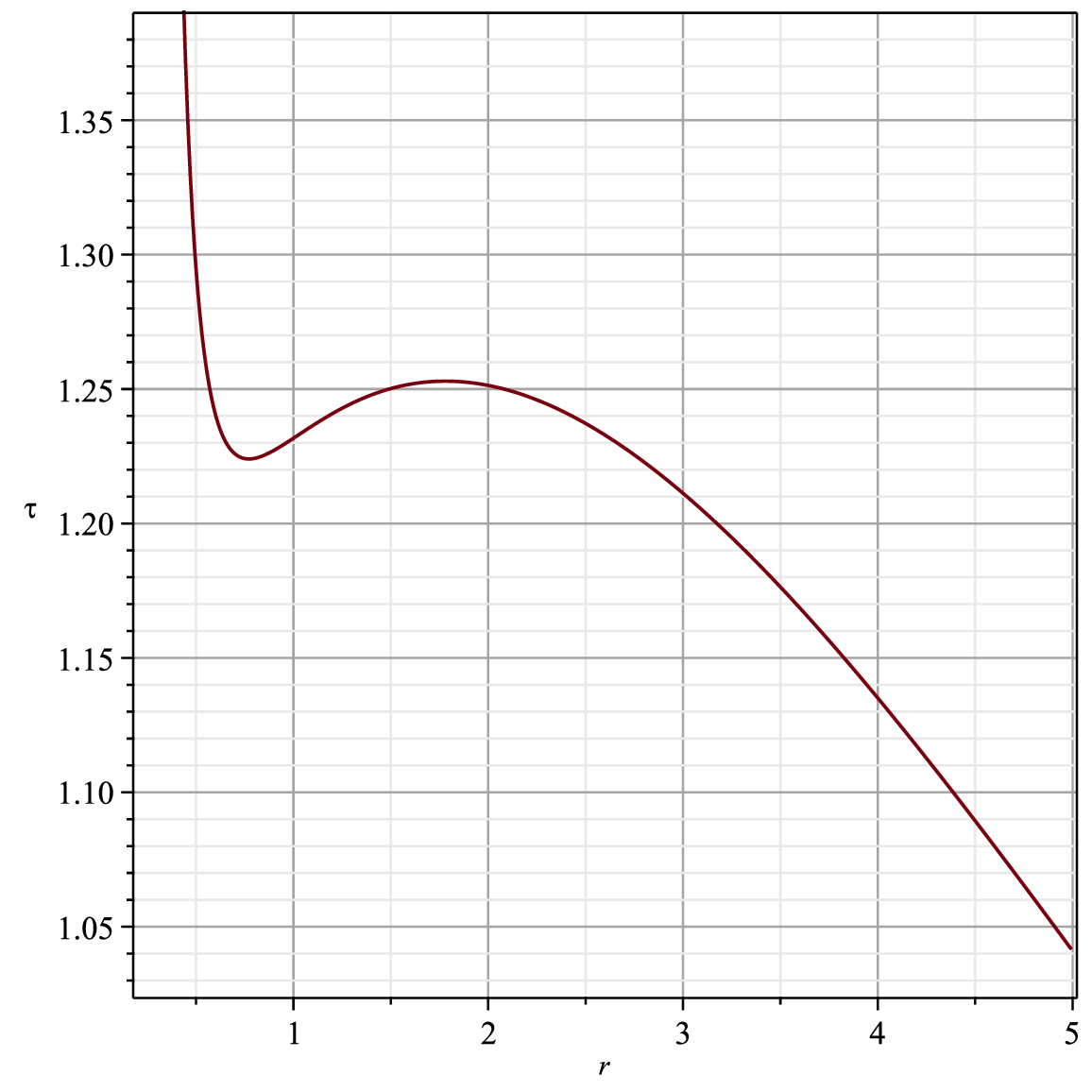}
 \label{14b}}
  \caption{\small{Figure 14a shows the vector field $n$ on a part of the $(r_H-\theta)$ plane with $(q=1, z=1.9, \tau=1, p=0.1, \overline{\theta}=2)$. The ZP is located at $(r_H-\theta)=(5.413521655,1.57)$ inside the circular loop. The contours $C_1$ (blue loop) and $C_2$ (purple loop) are two closed curves, where $C_1$ encircles the ZP but $C_2$ does not. Figure 14b displays the plot of the curve given by equation $\tau$.}}
 \label{14}
 \end{center}
 \end{figure}

Similar results were obtained in connection with the changes of free parameters ($q, \overline{\theta}, p, \tau$) for ($z=1, 1.3, 1.9$), and we added the related explanations and plots to Appendix B. Figures (12-14) show the $(r-\theta)$ plane in the left plot and the variation of $\tau$ in terms of $r$ in the right plots for ($z=1, 1.3, 1.9$), respectively. Figures (15-21) in Appendix C show the $(r-\theta)$ plane in the left plot and the variation of $\tau$ in terms of $r$ in the right plots with respect to different free parameters such as ($q, \tau, a, z, P$). We have analyzed these figures for hyperscaling violating black holes. The analysis is divided into two parts. The first part displays the normalized field lines. We encounter in Figures (12a, 13a) three zero points (ZPs) and in Figure (14a), one ZP for ($z=1, 1.3, 1.9$), respectively, and their locations are determined in the Figures. Figure (12a) has three ZPs which represent the three topological numbers ($\omega=+1, \omega=-1, \omega=+1$) with a total topological number $W=1$. Also, Figure (13a) shows three ZPs with topological numbers ($\omega=+1, \omega=-1, \omega=+1$) and a total topological number $W=+1$. Figure (14a) only has one ZP, which represents the topological number ($\omega=+1$) and $W=+1$ for this black hole. The winding number is a measure of how many times the field lines of the black hole wrap around a contour in the $(r-\theta)$ plane. The interesting point is that similar results are obtained for other free parameters corresponding to ($z=1, 1.3, 1.9$) for the mentioned black hole, the details of which can be seen in Appendix C. That is, Figures (15a) and (16a) behave similarly to (12a), Figures (17a) and (18a) behave similarly to (13a), and Figures (19a), (20a) and (21a) behave similarly to Figure (14a). In Figures 12b-21b, we have plotted the curve corresponding to the equation of $\tau$ for different values of the free parameters.
Without loss of generality, we have studied the topological properties of hyperscaling violating black holes in this paper by choosing different free parameters in each step. The results of this study on a black hole show that it only features positive total topological numbers ($W=+1$). The stability of the black hole is discussed by examining the winding numbers and the specific heat capacity. The positive value of the winding numbers implies that the on-shell black hole is thermodynamically stable, which can be confirmed by calculating the specific heat capacity.
In Figures (12b-21b), we have plotted the curve corresponding to equations of $\tau$ for various values of the free parameters for the mentioned black holes. In these Figures, there are some regions where $\tau$ changes in different ways, such as increasing, decreasing, or oscillating. We summarize the results of the discussion in Table (6).
\begin{center}
\begin{table}
  \centering
\begin{tabular}{|p{4cm}|p{4cm}||p{4cm}|p{4cm}|}
  \hline
  % after \\: \hline or \cline{col1-col2} \cline{col3-col4} ...
  \centering{HSV black holes}  & \hspace{2cm} Total Topological Numbers & Generation Point & Annihilation Point\\[3mm]
   \hline
  \centering{z=1} & \hspace{1cm} $W=+1$ & \hspace{2cm}0 or 1 & \hspace{2cm}0 or 1  \\[3mm]
   \hline
  \centering{z=1.3} & \hspace{1cm} $W=+1$ & \hspace{2cm}0 or 1 & \hspace{2cm}0 or 1 \\[3mm]
   \hline
   \centering{z=1.9}  & \hspace{1cm} $W=+1$ & \hspace{2cm}0 or 1 & \hspace{2cm}0 or 1  \\[3mm]
  \hline
\end{tabular}
\caption{Summary of the results.}\label{6}
\end{table}
 \end{center}
\section{Conclusions}
It was shown that a standard ring of light can be imagined outside the event horizon for stationary rotating four-dimensional black holes with axial symmetry using the topological method. Then, this idea was extended to non-rotating black holes using Duan's mapping. This means that the topological current is non-zero only at the zero point of the vector field that determines the location of the photon sphere. Therefore, a topological charge can be attributed to each photon sphere, and the total topological charge is always equal to a negative one.
Based on this concept, in this paper, we investigated the topological charge and the conditions of existence of the photon sphere (PS) for a HSV black hole with various values of the parameters of this model. Then, after carrying out a detailed analysis, we "proposed" a new topological class for naked singularities viz Q=+1 with respect to $z\geq1$ in addition to the conventional topological classes (Q=-1) for mentioned black hole and (Q=0) for the naked singularities. We also determined that $z\geq2$, it either shows a naked singularity form with total topological charge $+1$ or has no solution. Therefore, we have the black hole solution only in $1\leq z<2$. Then, we used two different methods, i.e., the temperature (Duan's topological current $\Phi$-mapping theory) and the generalized Helmholtz free energy, to study the topological classes of our black hole. By considering this black hole, we discussed the critical and zero points (topological charges and topological numbers) for different parameters of the hyperscaling violating black holes, such as ($z, \overline{\theta}$) and other free parameters, and studied their thermodynamic topology. We observed that for a given value of the parameters $z$, $\overline{\theta}$, and other free parameters, there were two total topological charges $(Q_{total}=-1, 0)$ for the temperature method and two total topological numbers $(W=+1)$ for the generalized Helmholtz free energy method. Also, we summarized the results for each study as photon sphere, temperature, and generalized Helmholtz free energy in some Figures and tables. Finally, we compared our results with other related studies in the literature. There is some interesting research in this area for the future. It may be interesting to employ the above-mentioned black hole and investigate the hidden conformal symmetry on the black hole photon sphere.  Also, in the future, we can study the photon sphere and quasinormal modes in AdS/CFT in the corresponding black hole and show how we can connect the critical points with confinement and non-confinement in the QCD system.

\section{Appendix A: Some points about photon sphere}
It be used two variables, $\overline{\theta}$ and z, to describe the shape and size of the black hole. $\overline{\theta}$ is the polar angle, which measures how far the point is from the north pole of the black hole. z is a dimensionless parameter that depends on the mass and charge of the black hole. The mass and charge are two important quantities that characterize a black hole. The mass measures how much matter and energy are contained in the black hole, and the charge measures how much electric charge is carried by the black hole.
So claims that in some regions of the parameter space, the behavior of the black hole with respect to $\overline{\theta}$ and z variations is almost the same with one exception. This means that for each specific value of z and $\overline{\theta}$, there is the fixed ratio M to Q, where M is the mass and Q is the charge of the black hole. If this ratio is smaller than a certain value, then the black hole has one root, which means that it has one event horizon. The event horizon is the boundary of the black hole, beyond which nothing can escape. If this ratio is larger than that value, then the black hole has no root, which means that it has no event horizon and is not a black hole anymore. So we gave an example of this ratio for several different values of z in some tables. The table shows that as z increases, the ratio decreases. This means that for larger values of z, the black hole can have a smaller mass relative to its charge and still be a black hole. We also say that this ratio is usually considered to be a criterion for the stability of the black hole, but in this region and in this range, even if this ratio is smaller than one, the system has the same topological behavior as usual.

 \begin{center}
\begin{table}
  \centering
\begin{tabular}{|p{4cm}|p{4cm}||p{4cm}||p{4cm}|}
  \hline
  \hspace{2cm} $\theta$  & \hspace{1cm} M/Q  & \hspace{1.5cm} Root & \hspace{1cm} Kind\\[3mm]
   \hline
  \hspace{2cm} 0 & \hspace{1cm} $<3$ & \hspace{2cm} 1 & \hspace{1cm} divergent \\[3mm]
   \hline
  \hspace{2cm} -1 & \hspace{1cm} $<2.5$ & \hspace{2cm} 1 & \hspace{1cm} divergent \\[3mm]
  \hline
\end{tabular}
\caption{Summary of the results of $z=1.2$}\label{8}
\end{table}
 \end{center}

\begin{center}
\begin{table}
  \centering
\begin{tabular}{|p{4cm}|p{4cm}||p{4cm}||p{4cm}|}
  \hline
  \hspace{2cm} $\theta$  & \hspace{1cm} M/Q  & \hspace{1.5cm} Root & \hspace{1cm} Kind\\[3mm]
   \hline
  \hspace{2cm} 1 & \hspace{1cm} $<3.5$ & \hspace{2cm} 1 & \hspace{1cm} divergent \\[3mm]
   \hline
  \hspace{2cm} -1 & \hspace{1cm} $<2.5$ & \hspace{2cm} 1 & \hspace{1cm} divergent \\[3mm]
  \hline
\end{tabular}
\caption{Summary of the results of $z=1.3$}\label{9}
\end{table}
 \end{center}

\begin{center}
\begin{table}
  \centering
\begin{tabular}{|p{4cm}|p{4cm}||p{4cm}||p{4cm}|}
  \hline
  \hspace{2cm} $\theta$  & \hspace{1cm} M/Q  & \hspace{1.5cm} Root & \hspace{1cm} Kind\\[3mm]
   \hline
  \hspace{2cm} 2 & \hspace{1cm} $<4.5$ & \hspace{2cm} 1 & \hspace{1cm} divergent \\[3mm]
   \hline
  \hspace{2cm} 1 & \hspace{1cm} $<3$ & \hspace{2cm} 1 & \hspace{1cm} divergent \\[3mm]
  \hline
\end{tabular}
\caption{Summary of the results of $z=1.5$}\label{10}
\end{table}
 \end{center}

\begin{center}
\begin{table}
  \centering
\begin{tabular}{|p{4cm}|p{4cm}||p{4cm}||p{4cm}|}
  \hline
  \hspace{2cm} $\theta$  & \hspace{1cm} M/Q  & \hspace{1.5cm} Root & \hspace{1cm} Kind\\[3mm]
   \hline
  \hspace{2cm} 3 & \hspace{1cm} $<7$ & \hspace{2cm} 1 & \hspace{1cm} divergent \\[3mm]
   \hline
  \hspace{2cm} 2 & \hspace{1cm} $<4$ & \hspace{2cm} 1 & \hspace{1cm} divergent \\[3mm]
  \hline
\end{tabular}
\caption{Summary of the results of $z=1.9$}\label{11}
\end{table}
 \end{center}

 \begin{center}
\begin{table}
  \centering
\begin{tabular}{|p{4cm}|p{4cm}||p{4cm}||p{4cm}|}
  \hline
  \hspace{2cm} $\theta$  & \hspace{1cm} M/Q  & \hspace{1.5cm} Root & \hspace{1cm} Kind\\[3mm]
   \hline
  \hspace{2cm} 3 & \hspace{1cm} $<6$ & \hspace{2cm} 1 & \hspace{1cm} divergent \\[3mm]
   \hline
  \hspace{2cm} 2 & \hspace{1cm} $<4$ & \hspace{2cm} 1 & \hspace{1cm} divergent \\[3mm]
  \hline
  \hspace{2cm} 1 & \hspace{1cm} $<3$ & \hspace{2cm} 1 & \hspace{1cm} divergent \\[3mm]
  \hline
  \hspace{2cm} 0 & \hspace{1cm} $<2.5$ & \hspace{2cm} 1 & \hspace{1cm} divergent \\[3mm]
  \hline
  \hspace{2cm} -1 & \hspace{1cm} $<2.3$ & \hspace{2cm} 1 & \hspace{1cm} divergent \\[3mm]
  \hline
\end{tabular}
\caption{Summary of the results of $z=2$}\label{12}
\end{table}
 \end{center}

  \begin{center}
\begin{table}
  \centering
\begin{tabular}{|p{4cm}|p{4cm}||p{4cm}||p{4cm}|}
  \hline
  \hspace{2cm} $\theta$  & \hspace{1cm} M/Q  & \hspace{1.5cm} Root & \hspace{1cm} Kind\\[3mm]
   \hline
  \hspace{2cm} 3 & \hspace{1cm} $<4$ & \hspace{2cm} 1 & \hspace{1cm} divergent \\[3mm]
   \hline
  \hspace{2cm} 2 & \hspace{1cm} $<3$ & \hspace{2cm} 1 & \hspace{1cm} divergent \\[3mm]
  \hline
  \hspace{2cm} 1 & \hspace{1cm} $<2.8$ & \hspace{2cm} 1 & \hspace{1cm} divergent \\[3mm]
  \hline
  \hspace{2cm} 0 & \hspace{1cm} $<2.6$ & \hspace{2cm} 1 & \hspace{1cm} divergent \\[3mm]
  \hline
  \hspace{2cm} -2 & \hspace{1cm} $<2.2$ & \hspace{2cm} 1 & \hspace{1cm} divergent \\[3mm]
  \hline
\end{tabular}
\caption{Summary of the results of $z=3$}\label{13}
\end{table}
 \end{center}
\section{Appendix B: Calculation for the temperature method}
At first, we rewrite the temperature using equations \eqref{3} and \eqref{5} considering $16\pi G=r_F=\phi_0=1$,
\begin{equation}\label{87}
\begin{split}
&T(r_h,P,Q)=\frac{r_h^z }{4 \pi } \bigg(\frac{(-\overline{\theta} +d+z-1) (-\overline{\theta} +d+z)}{P e^{\frac{2 \overline{\theta} }{d \sqrt{2 (d-\overline{\theta} ) (-\frac{\overline{\theta} }{d}+z-1)}}}}\bigg)^{\frac{1}{2} (-z-1)} \\
&\times \bigg(-\overline{\theta} +\frac{(d-1)^2 k (-\overline{\theta} +d+z-1) (-\overline{\theta} +d+z)}{P r_h^2 e^{\frac{2 \overline{\theta} }{d \sqrt{2 (d-\overline{\theta} ) (-\frac{\overline{\theta} }{d}+z-1)}}} (-\overline{\theta} +d+z-2)}\\&-q^2 (-\overline{\theta} +d+z-2) r_h^{-2 (-\overline{\theta} +d+z-1)}+d+z\bigg)
\end{split}
\end{equation}
By setting $\left(\frac{\partial T}{\partial r_h}\right)_{P,Q}=0$, it becomes feasible to eliminate a parameter from equation \eqref{87} for temperature. So, we have,
\begin{equation}\label{88}
\begin{split}
&T(r_h,Q)=\bigg(-d^2 q^2 r_h^{2 \overline{\theta} +2}+2 d q^2 r_h^{2 \overline{\theta} +2} (\overline{\theta} -z+2)\\&+(\overline{\theta} -z) r_h^{2 (d+z)}-d r_h^{2 (d+z)}-q^2 r_h^{2 \overline{\theta} +2} (\overline{\theta} -z+2)^2\bigg) \\
&\times\bigg(-\frac{r_h^{-2 (d+z-1)} (-\overline{\theta} +d+z-2)}{(d-1)^2 k (z-2)}\bigg)^{\frac{1}{2} (-z-1)} \frac{r_h^{-2 d-z}}{2 \pi  (z-2)}\\
& \times\bigg(q^2 r_h^{2 \overline{\theta} +2} (2 \overline{\theta} ^2+6 \overline{\theta} +2 d^2+d (-4 \overline{\theta} +3 z-6)\\&+z^2-3 \overline{\theta}  z-4 z+4)+z r_h^{2 (d+z)} (-\overline{\theta} +d+z)\bigg)^{\frac{1}{2} (-z-1)}
\end{split}
\end{equation}
Now, we define the new thermodynamic function  as follows,
\begin{equation}\label{89}
\begin{split}
&\Phi=\frac{1}{\sin \theta}T(r_h,Q)
\end{split}
\end{equation}
We can obtain the components of the vector field by incorporating relations \eqref{88} and \eqref{89} into equation \eqref{10},
\begin{equation}\label{90}
\begin{split}
&\phi ^r= -\frac{r^{-4 d-3 z+1} (-\overline{\theta} +d+z-2) (-\frac{r^{-2 (d+z-1)} (-\overline{\theta} +d+z-2)}{(d-1)^2 k (z-2)})^{\frac{1}{2} (-z-3)}}{2 \pi  (d-1)^2 k (z-2)^2}\\
&\times\bigg(q^2 r^{2 \overline{\theta} +2} (2 \overline{\theta} ^2+6 \overline{\theta} +2 d^2+d (-4 \overline{\theta} +3 z-6)+z^2-3 \overline{\theta}  z-4 z+4)+z r^{2 (d+z)} (-\overline{\theta} +d+z)\bigg)^{\frac{1}{2} (-z-3)}\\
&\times\bigg(d^2 q^2 (z-1) r^{2 \overline{\theta} +2}+(z-\overline{\theta} ) r^{2 (d+z)}+q^2 r^{2 \overline{\theta} +2} \bigg[-\overline{\theta}  (\overline{\theta} +2)+z^3-2 (\overline{\theta} +2) z^2+(\overline{\theta} ^2+5 \overline{\theta} +4) z\bigg]\\
&+d \bigg[r^{2 (d+z)}-q^2 r^{2 \overline{\theta} +2} bigg(-2 (\overline{\theta} +1)-2 z^2+(2 \overline{\theta} +5) z\bigg)\bigg]\bigg)\csc (\theta )
\end{split}
\end{equation}
\begin{equation}\label{91}
\begin{split}
&\phi ^{\theta}= -\frac{r^{-2 d-z}}{2 \pi  (z-2)} \cot (\theta ) \csc (\theta )
\times\bigg(-d^2 q^2 r^{2 \overline{\theta} +2}+2 d q^2 r^{2 \overline{\theta} +2} (\overline{\theta} -z+2)+(\overline{\theta} -z) r^{2 (d+z)}\\&-d r^{2 (d+z)}-q^2 r^{2 \overline{\theta} +2} (\overline{\theta} -z+2)^2\bigg)
\times\bigg(-\frac{r^{-2 (d+z-1)} (-\overline{\theta} +d+z-2)}{(d-1)^2 k (z-2)}\bigg)^{\frac{1}{2} (-z-1)}\\
&\times\bigg(q^2 r^{2 \overline{\theta} +2} (2 \overline{\theta} ^2+6 \overline{\theta} +2 d^2+d (-4 \overline{\theta} +3 z-6)+z^2-3 \overline{\theta}  z-4 z+4)+z r^{2 (d+z)} (-\overline{\theta} +d+z)\bigg)^{\frac{1}{2} (-z-1)}
\end{split}
\end{equation}
Now, by using equation \eqref{88} with$\phi^r=0$, we can obtain two critical points as follows,
\begin{equation}\label{92}
\begin{split}
&q_{cp_1}^2=-(-\overline{\theta} +d+z) r^{-2 \overline{\theta} +2 d+2 z-2}\bigg/\bigg(-\overline{\theta} ^2-2 \overline{\theta} +d^2 z-d^2+2 \overline{\theta}  d\\&+2 d z^2-2 \overline{\theta}  d z-5 d z+2 d+z^3-2 \overline{\theta}  z^2-4 z^2+\overline{\theta} ^2 z+5 \overline{\theta}  z+4 z\bigg)
\end{split}
\end{equation}
\begin{equation}\label{93}
\begin{split}
r_{cp_1}^2=-\frac{(d-1)^2 k \ell^2 \left(\overline{\theta} +d (z-1)+z^2-(\overline{\theta} +2) z\right)}{(z+1) (-\overline{\theta} +d+z-2) (-\overline{\theta} +d+z-1) (-\overline{\theta} +d+z)}
\end{split}
\end{equation}
\begin{equation}\label{94}
\begin{split}
&T_{cp_1}=-\frac{r^z (-\overline{\theta} +d+z-1) (-\overline{\theta} +d+z)}{2 \pi (\overline{\theta} +d (z-1)+z^2-(\overline{\theta} +2) z)} \\
&\times\bigg(-\frac{r^2 (z+1)}{(d-1)^2 k \left(\overline{\theta} +d (z-1)+z^2-(\overline{\theta} +2) z\right)}\bigg)^{\frac{1}{2} (-z-1)}\\
&\times \bigg(-\overline{\theta}  (\overline{\theta} ^2+3 \overline{\theta} +2)+d^3+3 d^2 (-\overline{\theta} +z-1)+z^3-3 (\overline{\theta} +1) z^2 \\
&+d \left(3 \overline{\theta} ^2+6 \overline{\theta} +3 z^2-6 (\overline{\theta} +1) z+2\right)+\left(3 \overline{\theta} ^2+6 \overline{\theta} +2\right) z\bigg)^{\frac{1}{2} (-z-1)}
\end{split}
\end{equation}
Furthermore, for the second critical point, we have,
\begin{equation}\label{95}
\begin{split}
q_{cp_2}^2=\frac{z (-\overline{\theta} +d+z) r^{-2 \overline{\theta} +2 d+2 z-2}}{(-2 \overline{\theta} +2 d+z-2) (-\overline{\theta} +d+z-2)^2}
\end{split}
\end{equation}
\begin{equation}\label{96}
\begin{split}
r_{cp_2}^2=\frac{(d-1)^2 k \ell^2 (2-z)}{z (-\overline{\theta} +d+z-1) (-\overline{\theta} +d+z)}
\end{split}
\end{equation}
\begin{equation}\label{97}
\begin{split}
T_{cp_2}=-\frac{r^z (-\overline{\theta} +d+z-1) (-\overline{\theta} +d+z) \left(-\frac{r^2 z \left(\overline{\theta} ^2+\overline{\theta} +d^2+d (-2 \overline{\theta} +2 z-1)+z^2-2 \overline{\theta}  z-z\right)}{(d-1)^2 k (z-2)}\right)^{\frac{1}{2} (-z-1)}}{\pi  (z-2) (-2 (\overline{\theta} +1)+2 d+z)}
\end{split}
\end{equation}

\section{Appendix C: Generalized Helmholtz free energy method plots}
As mentioned in the main text, this section presents the plots of the free energy method for different values of z (1, 1.3, and 1.9) and various free parameters. The results are discussed in detail below.

\subsection{z=1}

\begin{figure}[h!]
 \begin{center}
 \subfigure[]{
 \includegraphics[height=5cm,width=5.5cm]{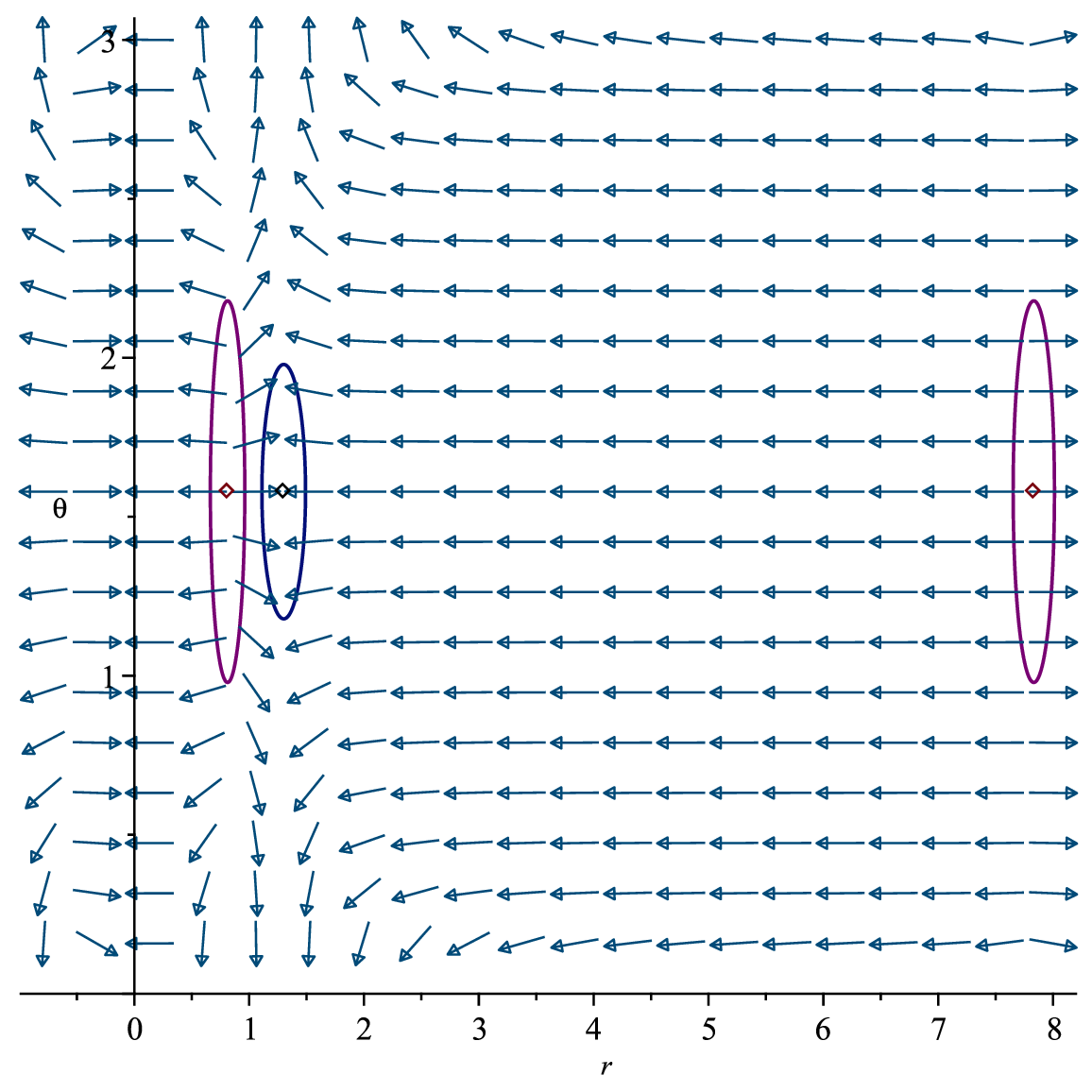}
 \label{15a}}
 \subfigure[]{
 \includegraphics[height=5cm,width=5.5cm]{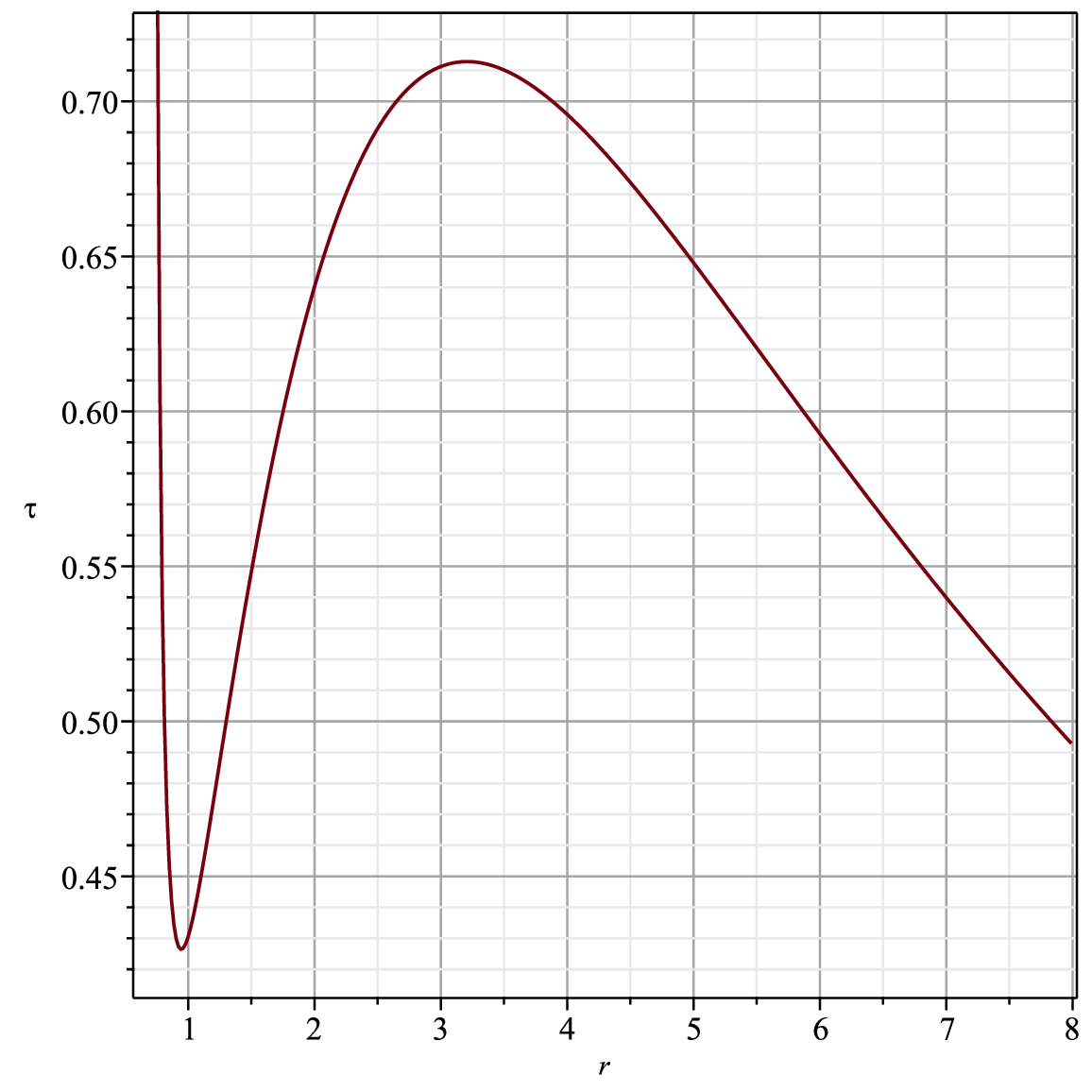}
 \label{15b}}
  \caption{\small{Figure 15a shows the vector field $n$ on a part of the $(r_H-\theta)$ plane with $(q=1, z=1, \tau=0.4, P=1.5, \overline{\theta}=-1)$. The blue arrows indicate the direction of $n$. The ZPs are located at $(r,\theta)=(0.8121954872,1.57), (1.301269606,1.57), (7.832599450,1.57)$ inside the circular loop, where encircles the ZPs. Figure 15b displays the plot of the curve given by equation $\tau$.}}
 \label{15}
 \end{center}
 \end{figure}

\begin{figure}[h!]
 \begin{center}
 \subfigure[]{
 \includegraphics[height=5cm,width=5.5cm]{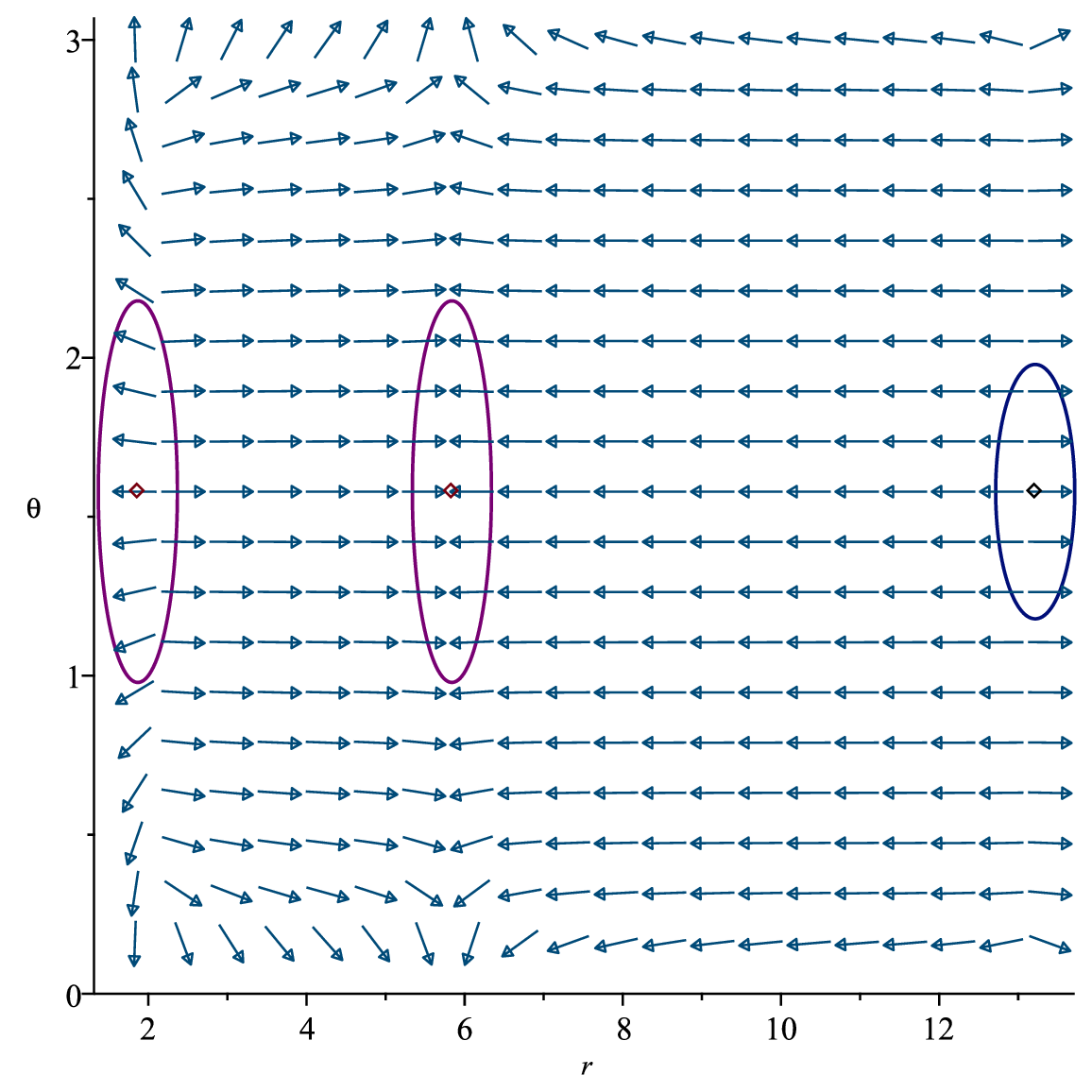}
 \label{16a}}
 \subfigure[]{
 \includegraphics[height=5cm,width=5.5cm]{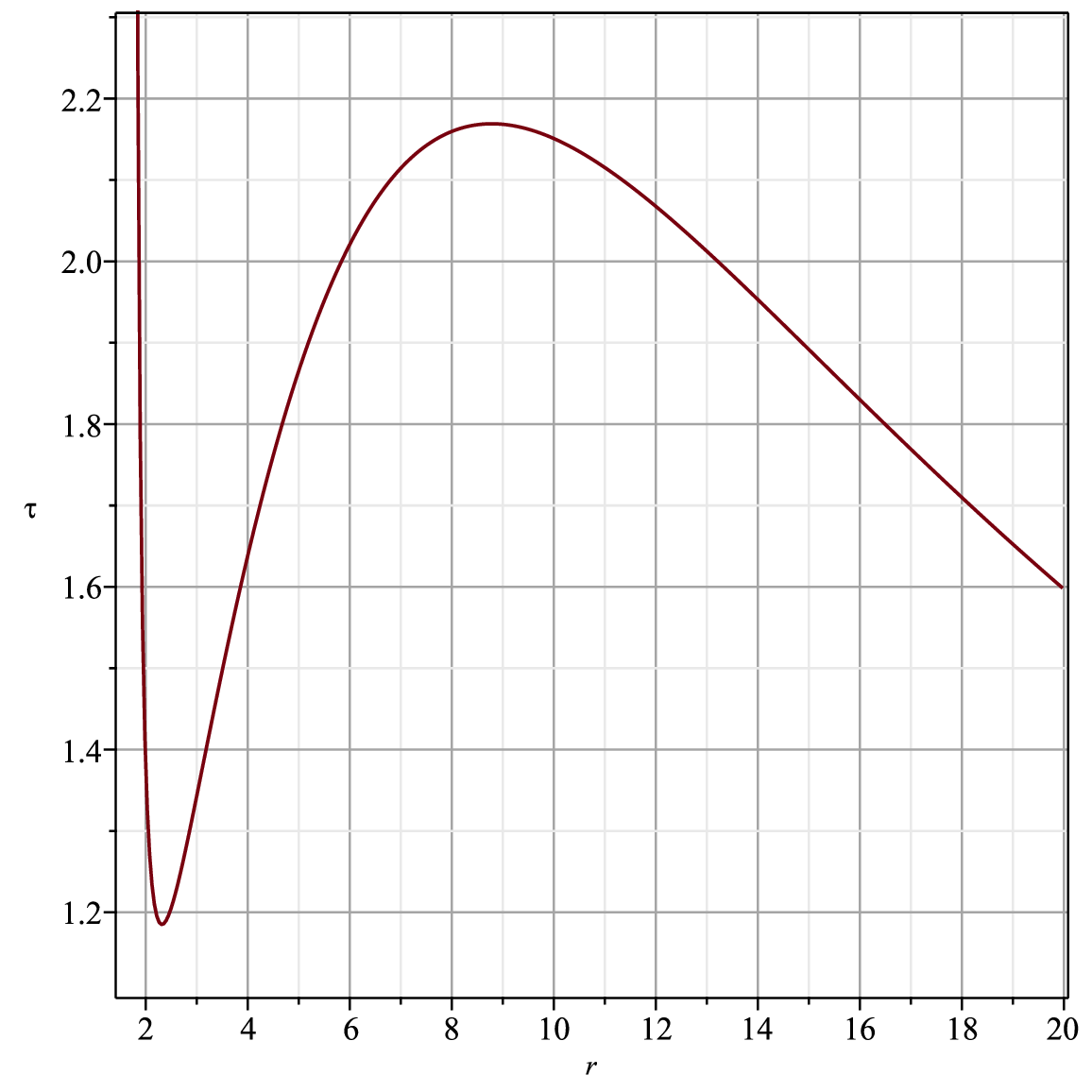}
 \label{16b}}
  \caption{\small{Figure 16a shows the vector field $n$ on a part of the $(r_H-\theta)$ plane with $(q=100, z=1, \tau=2, p=0.2, \overline{\theta}=-1)$. The blue arrows indicate the direction of $n$. The ZPs are located at $(r_H-\theta)=(1.867437105,1.57), (5.839002955,1.57), (13.21508885,1.57)$ inside the circular loop where encircles the ZPs. Figure 16b displays the plot of the curve given by equation $\tau$.}}
 \label{16}
 \end{center}
 \end{figure}

\newpage
\subsection{z=1.3}

\begin{figure}[h!]
 \begin{center}
 \subfigure[]{
 \includegraphics[height=4.3cm,width=5cm]{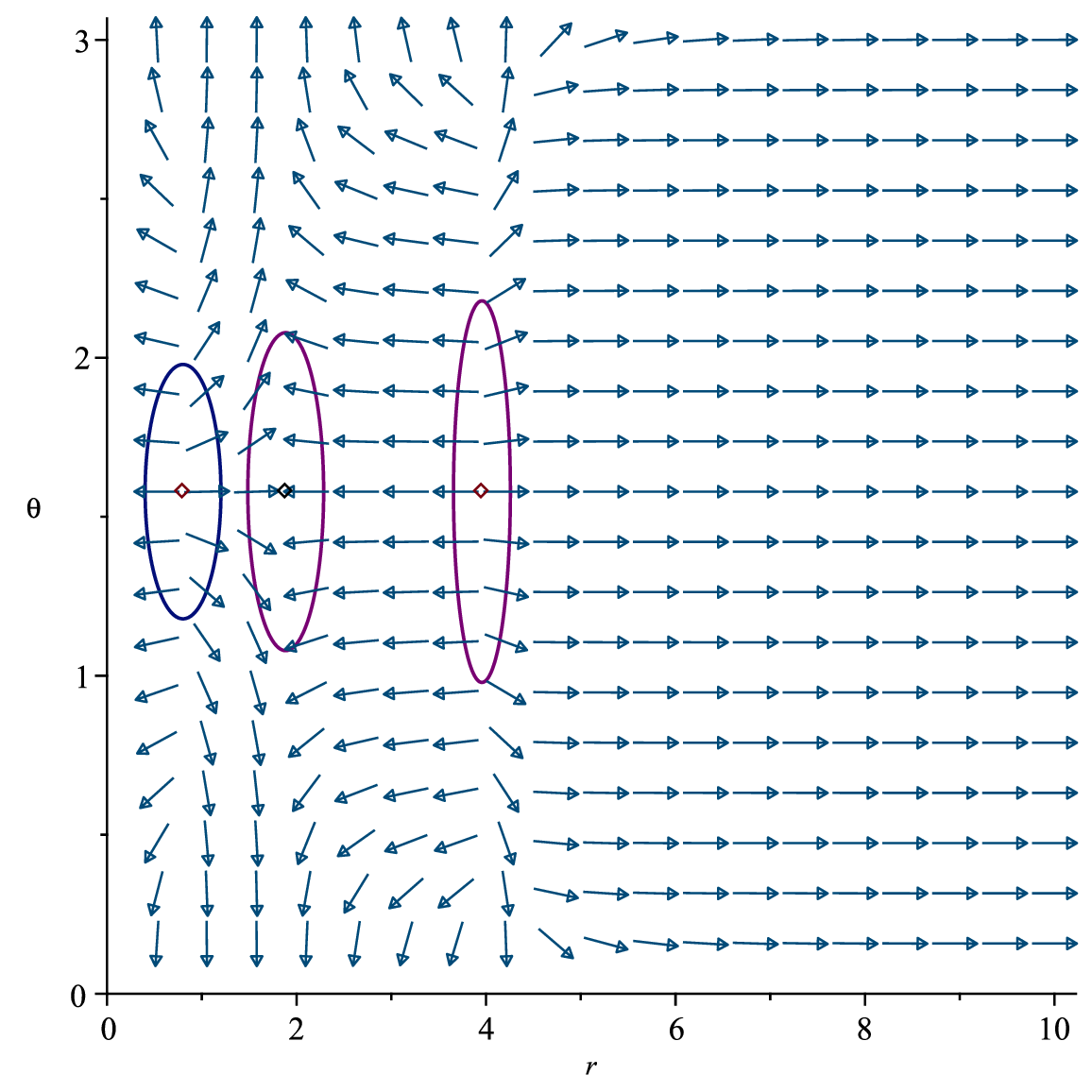}
 \label{17a}}
 \subfigure[]{
 \includegraphics[height=4.3cm,width=5cm]{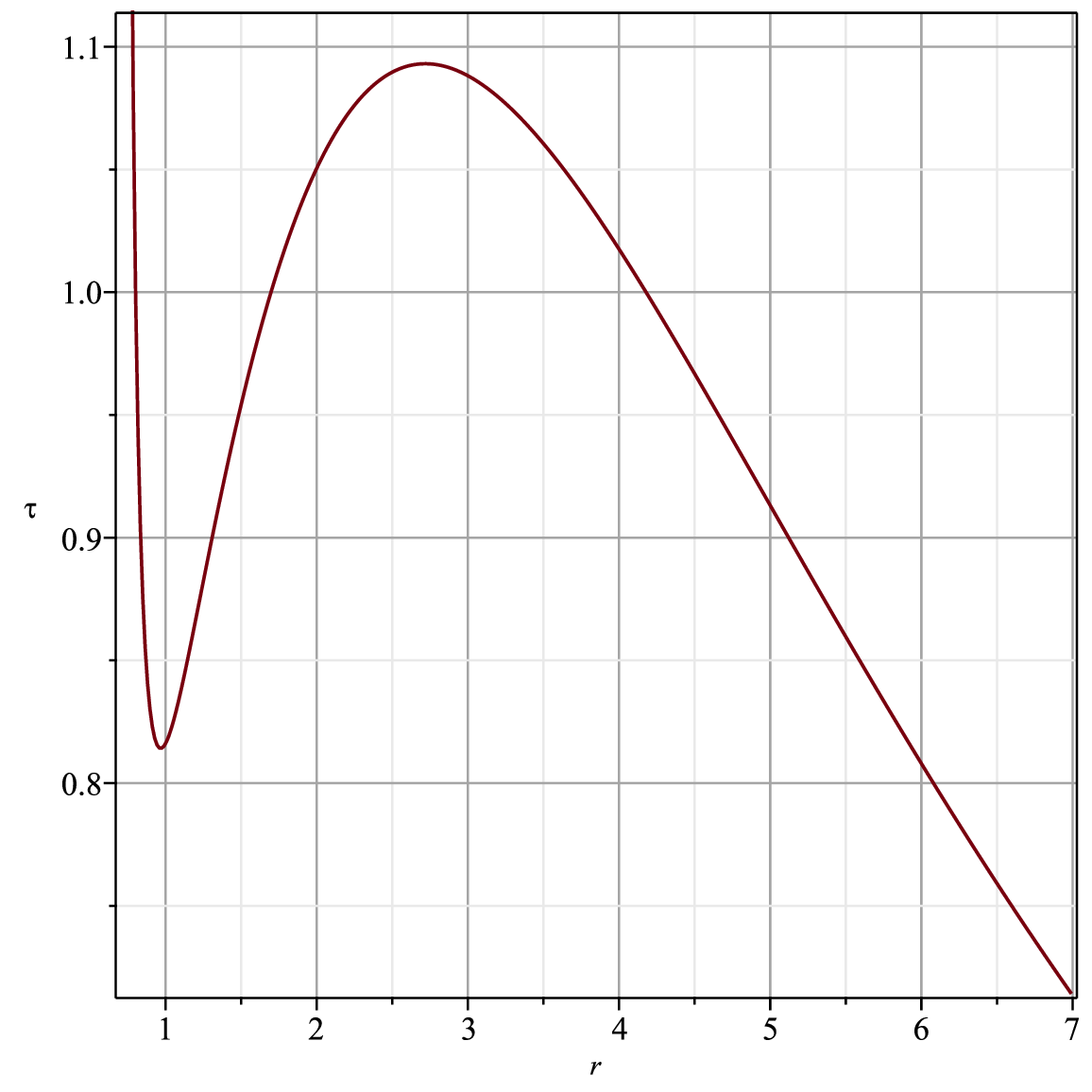}
 \label{17b}}
  \caption{\small{Figure 17a shows the vector field $n$ on a part of the $(r_H-\theta)$ plane with $(q=1, z=1.3, \tau=1, p=1, \overline{\theta}=-1)$. The blue arrows indicate the direction of $n$. The ZPs are located at $(r_H-\theta)=(0.8020587019,1.57), (1.886, 1.57), (3.957,1.57)$ inside the circular loop. The contours (blue loop) and (purple loop) are two closed curves, where encircles the ZPs. Figure 17b displays the plot of the curve given by equation $\tau$.}}
 \label{17}
 \end{center}
 \end{figure}

\begin{figure}[h!]
 \begin{center}
 \subfigure[]{
 \includegraphics[height=4.3cm,width=5cm]{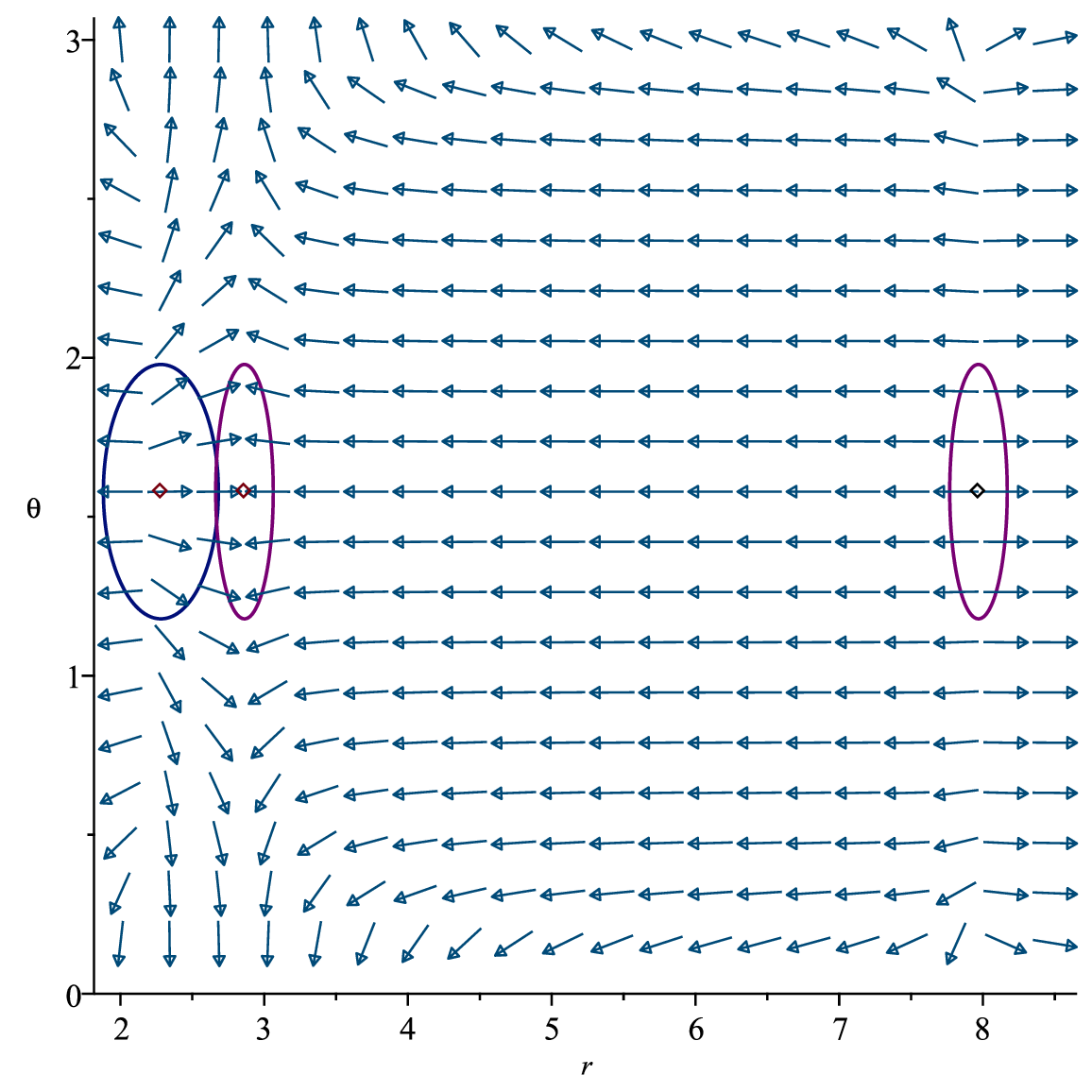}
 \label{18a}}
 \subfigure[]{
 \includegraphics[height=4.3cm,width=5cm]{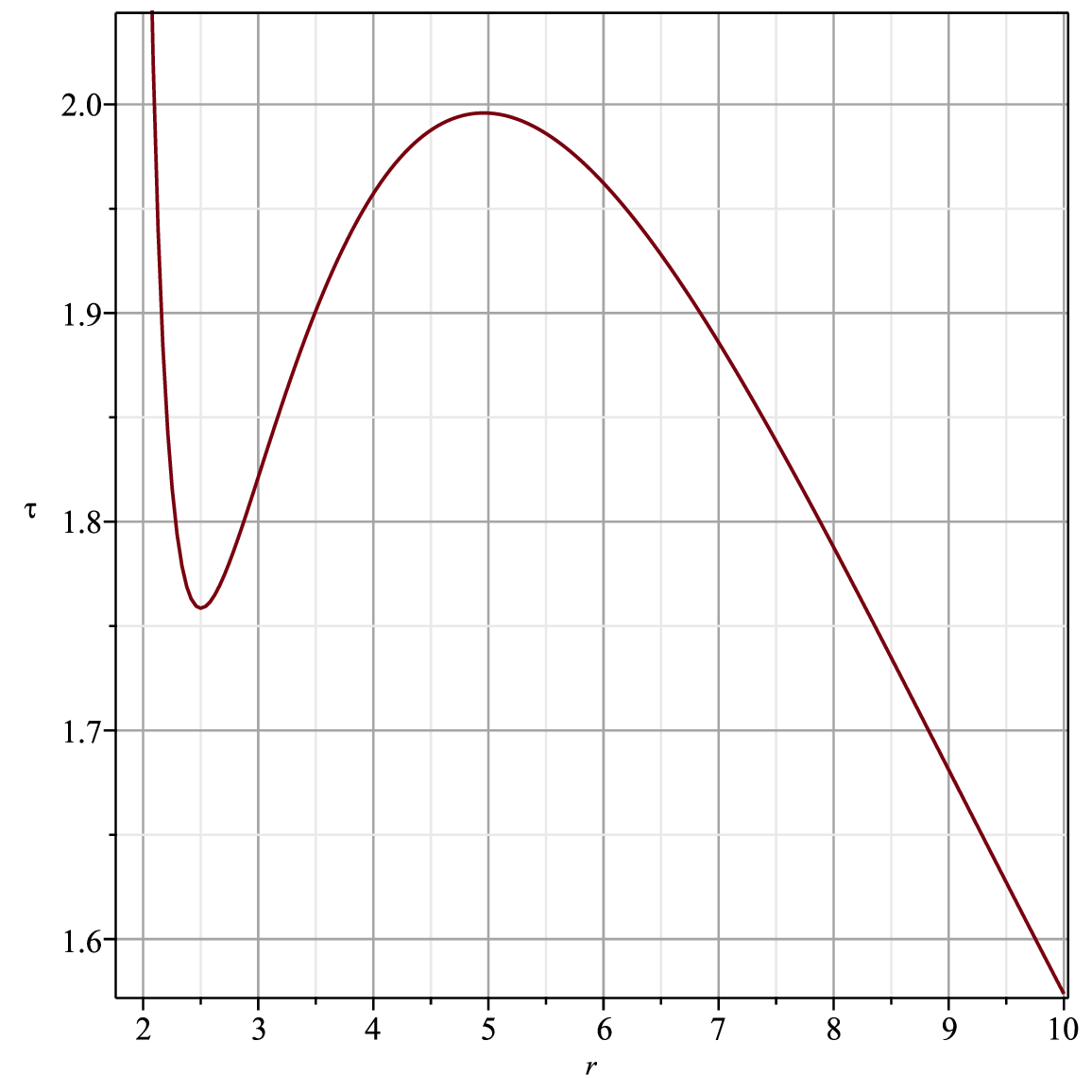}
 \label{18b}}
  \caption{\small{Figure 18a shows the vector field $n$ on a part of the $(r_H-\theta)$ plane with $(q=100, z=1.3, \tau=1.8, p=0.3, \overline{\theta}=-1)$. The blue arrows indicate the direction of $n$. The ZPs are located at $(r_H-\theta)=(2.281369836,1.57), (2.863, 1.57), (7.969,1.57)$ inside the circular loop. The contours (blue loop) and (purple loop) are two closed curves, where encircles the ZPs. Figure 18b displays the plot of the curve given by equation $\tau$.}}
 \label{18}
 \end{center}
 \end{figure}
 \newpage
\subsection{z=1.9}

\begin{figure}[h!]
 \begin{center}
 \subfigure[]{
 \includegraphics[height=4.8cm,width=5.5cm]{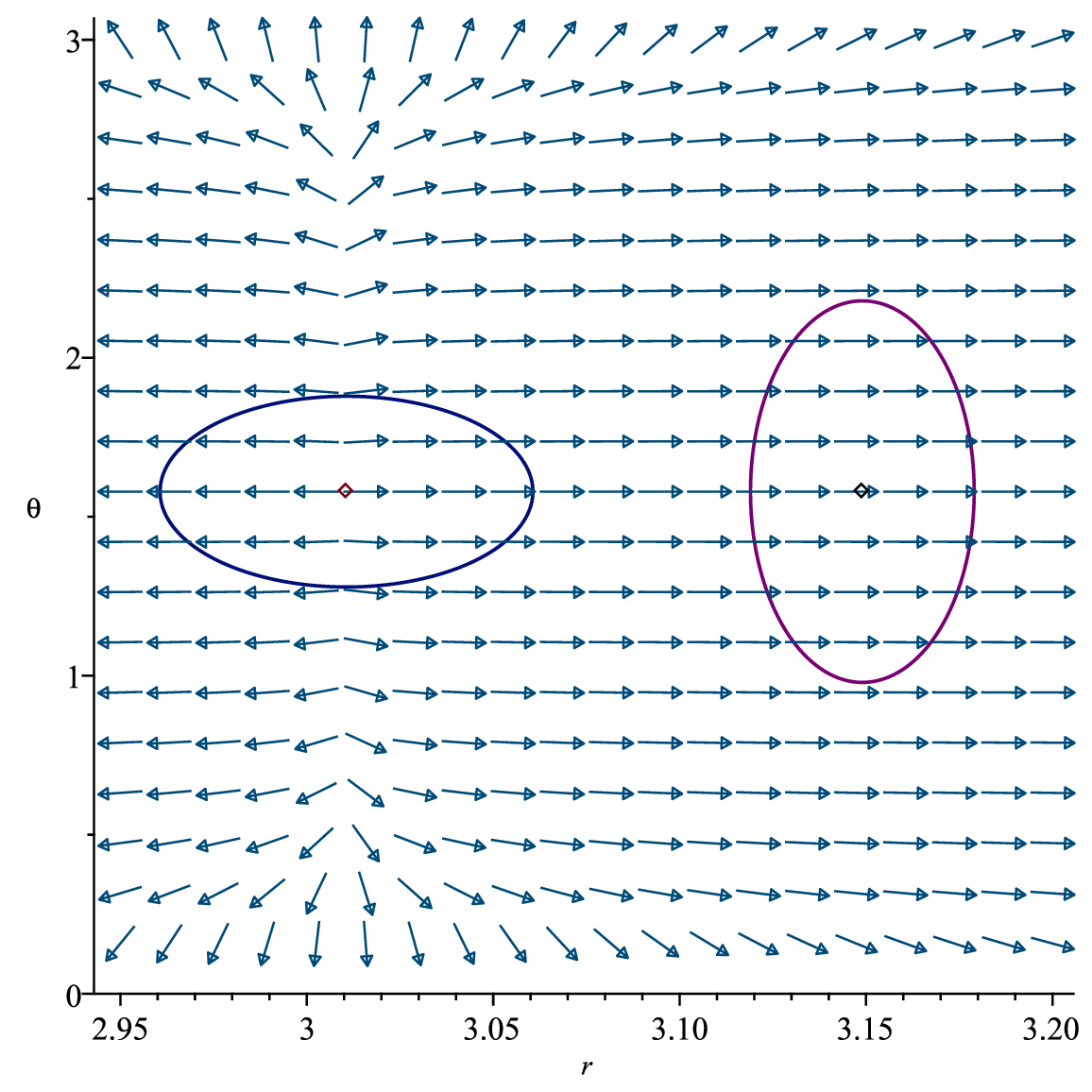}
 \label{19a}}
 \subfigure[]{
 \includegraphics[height=4.8cm,width=5.5cm]{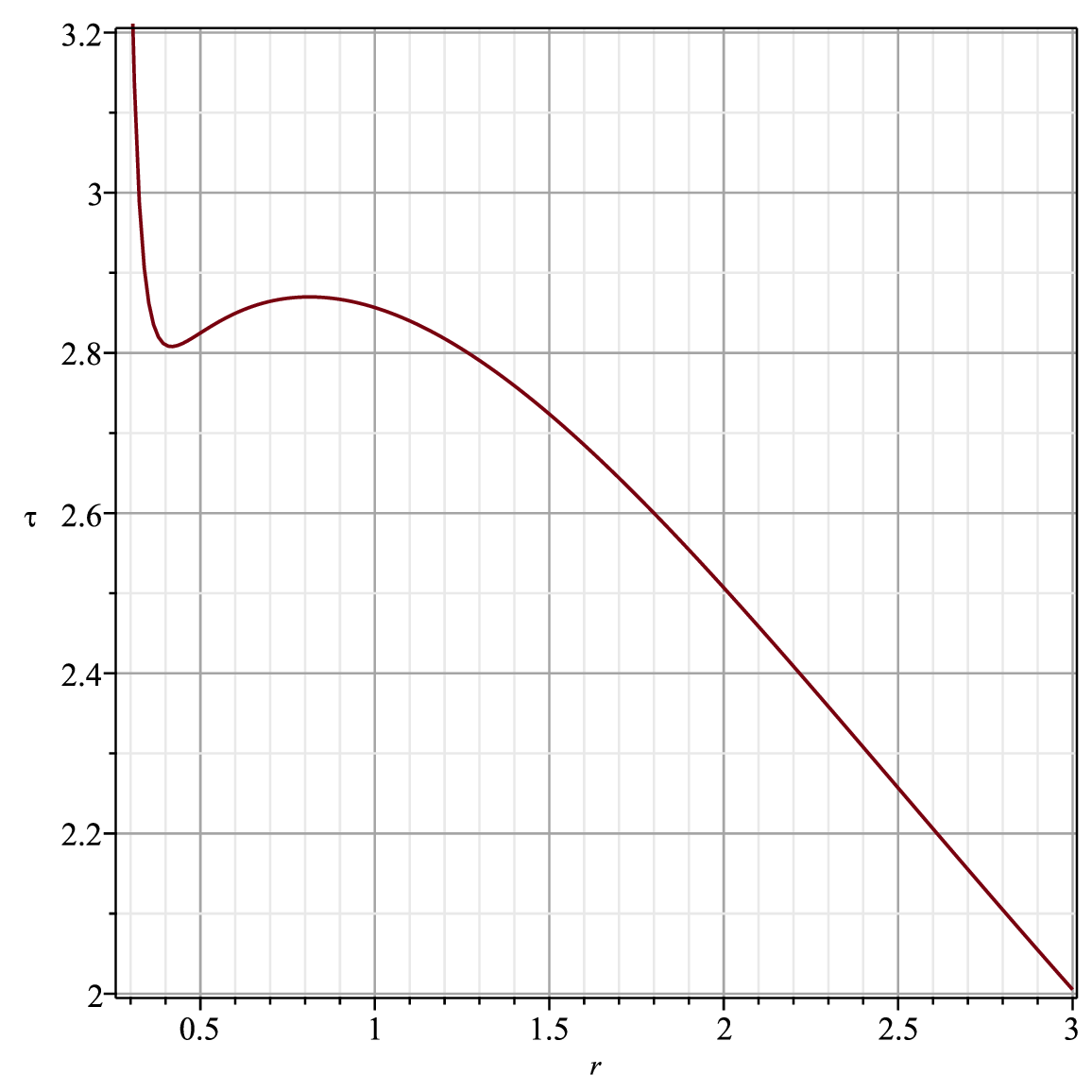}
 \label{19b}}
  \caption{\small{Figure 19a shows the vector field $n$ on a part of the $(r_H-\theta)$ plane with $(q=0.005, z=1.9, \tau=2, p=1, \overline{\theta}=-1)$. The blue arrows indicate the direction of $n$. The ZP is located at $(r_H-\theta)=(3.010641433,1.57)$ inside the circular loop ($C_1$). The contours $C_1$ (blue loop) and $C_2$ (purple loop) are two closed curves, where $C_1$ encircles the ZP but $C_2$ does not. Figure 19b displays the plot of the curve given by equation $\tau$.}}
 \label{19}
 \end{center}
 \end{figure}
\begin{figure}[h!]
 \begin{center}
 \subfigure[]{
 \includegraphics[height=4.8cm,width=5.5cm]{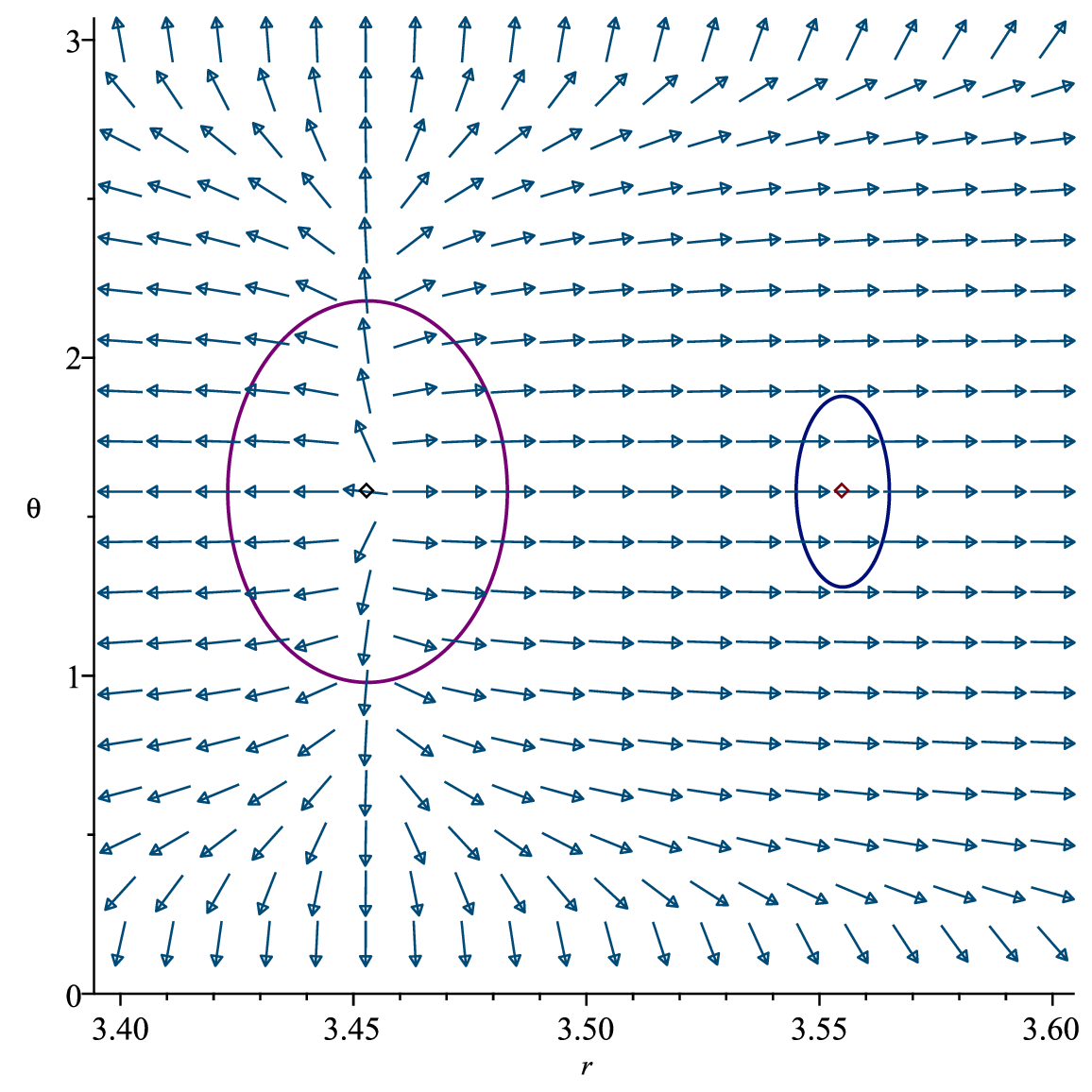}
 \label{20a}}
 \subfigure[]{
 \includegraphics[height=4.8cm,width=5.5cm]{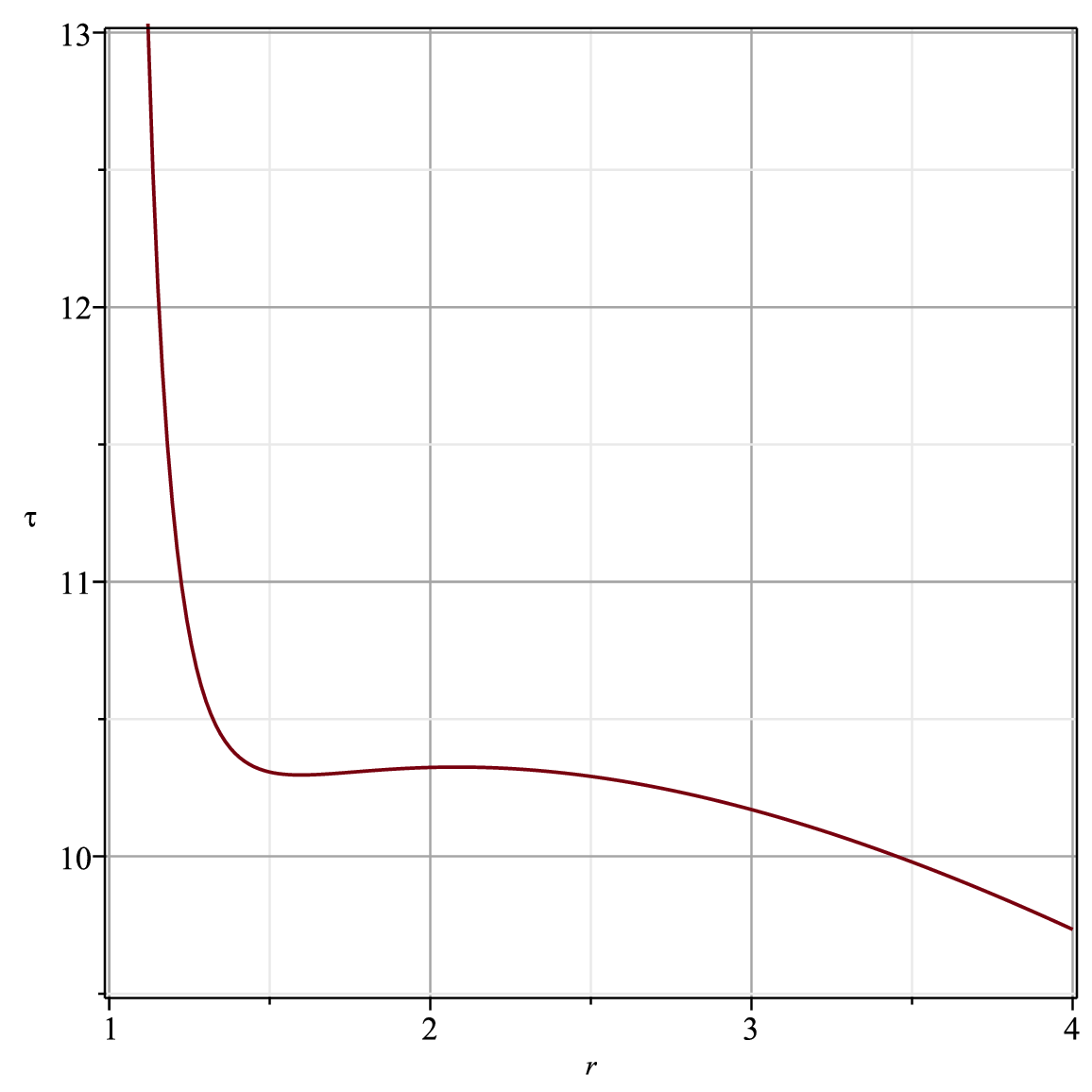}
 \label{20b}}
  \caption{\small{Figure 20a shows the vector field $n$ on a part of the $(r_H-\theta)$ plane with $(q=10, z=1.9, \tau=10, p=0.16, \overline{\theta}=-2)$. The blue arrows indicate the direction of $n$. The ZP is located at $(r_H-\theta)=(3.453034309,1.57)$ inside the circular loop. The contours $C_1$ (blue loop) and $C_2$ (purple loop) are two closed curves, where $C_2$ encircles the ZP but $C_1$ does not. Figure 20b displays the plot of the curve given by equation $\tau$.}}
 \label{20}
 \end{center}
 \end{figure}
\begin{figure}[h!]
 \begin{center}
 \subfigure[]{
 \includegraphics[height=5cm,width=5.5cm]{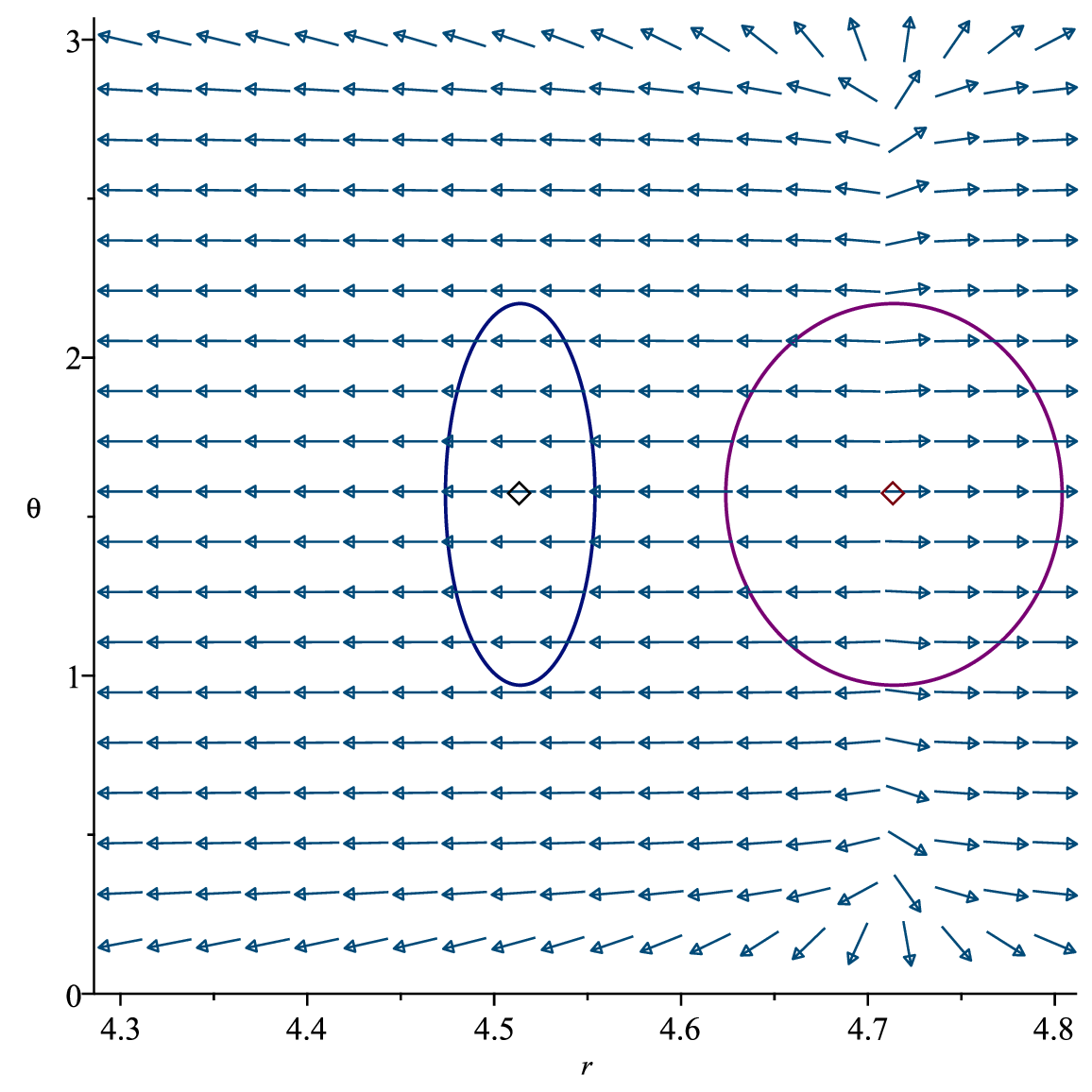}
 \label{21a}}
 \subfigure[]{
 \includegraphics[height=5cm,width=5.5cm]{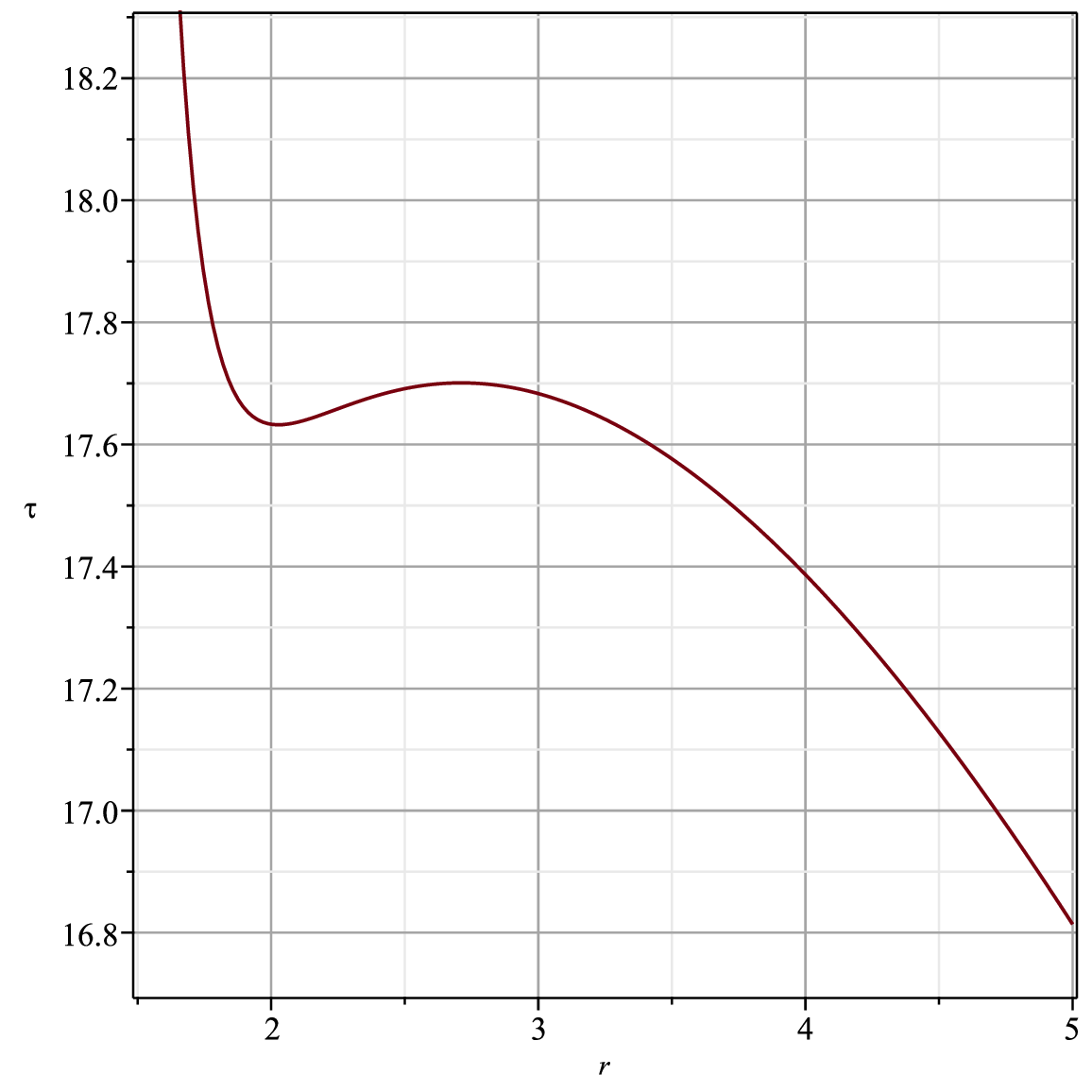}
 \label{21b}}
  \caption{\small{Figure 21a shows the vector field $n$ on a part of the $(r_H-\theta)$ plane with $(q=100, z=1.9, \tau=17, p=0.1, \overline{\theta}=-3)$. The blue arrows indicate the direction of $n$. The ZP is located at $(r_H-\theta)=(4.713933532,1.57)$ inside the circular loop. The contours $C_1$ (blue loop) and $C_2$ (purple loop) are two closed curves, where $C_2$ encircles the ZP but $C_1
  $ does not. Figure 21b displays the plot of the curve given by equation $\tau$.}}
 \label{21}
 \end{center}
 \end{figure}

\end{document}